\documentclass[prd,preprint,preprintnumbers,floatfix,showpacs,
secnumarabic,nofootinbib,amsmath,amsfonts,amssymb]{revtex4}
\usepackage{graphicx}
\usepackage{bm}
\usepackage{pxfonts}
\usepackage[colorlinks=true,citecolor=blue,linkcolor=red,
hyperfootnotes=false]{hyperref}
\newcommand{\SU}{\mathrm{SU}}

\begin{document}
\title{Quark-lepton mass unification at TeV scales}
\author{Mehrdad~Adibzadeh}
\email{mehrdad@mailaps.org}
\affiliation{Department of Physics, University of Virginia, P.O.Box 400714,
Charlottesville, VA 22904, USA}
\author{P.~Q.~Hung}
\email{pqh@virginia.edu}
\affiliation{Department of Physics, University of Virginia, P.O.Box 400714,
Charlottesville, VA 22904, USA}

\date{May 2007}

\begin{abstract}
A scenario combining a model of early (TeV) unification of quarks and leptons
with the physics of large extra dimensions provides a natural mechanism linking
quark and lepton masses at TeV scale. This has been dubbed as \textit{early
quark-lepton mass unification} by one of us (PQH) in one of the two models of
early quark-lepton unification, which are consistent with data, namely
$\SU(4)_{\mathrm {PS}} \otimes \SU(2)_{L} \otimes \SU(2)_{R} \otimes
\SU(2)_{H}$. In particular, it focused on the issue of naturally light Dirac
neutrino. The present paper will focus on similar issues in the other model,
namely $\SU(4)_{\mathrm {PS}} \otimes \SU(3)_L \otimes \SU(3)_{H}$.
\end{abstract}
\maketitle

\section{Introduction}
Could quark and lepton masses be related at TeV scales? Not long ago, one of us
explored this possibility in the framework of the so-called \textit{early
quark-lepton mass unification}~\cite{Hung2005}. The idea was to combine two TeV
scale scenarios, namely one of the two petite unification models
$\mathrm{PUT}_1=\SU(4)_{\mathrm {PS}} \otimes \SU(2)_{L} \otimes \SU(2)_{R}
\otimes \SU(2)_{H}$, and TeV scale large extra dimensions~\cite{LED,Arkani1998}.

The Petite Unification Theories (PUT's)~\cite{Hung1982,Buras2003} are
quark-lepton unification models, which occur at TeV scales and have the gauge
group structure $G = G_{S}(g_S) \otimes G_{W}(g_W)$. Both PUT  models propose
unusually charged heavy quarks and leptons, in addition to the fermion content
of the Standard Model (SM).

The model in Ref.~\cite{Hung2005} made use of the mechanism of wave function
overlap along the large extra dimension~\cite{Arkani1998,Antoniadis1998}, which
was originally employed to justify the smallness of Dirac neutrino
mass~\cite{Arkani2000,Hung2003a,SNM}. The mechanism connects the strengths of
the couplings in the mass terms of the fermions in four dimensions, as
\textit{effective} Yukawa couplings, to the magnitudes of wave function overlaps
between the corresponding left- and right-handed fermionic zero modes along the
large extra dimension~\cite{Arkani2000,Hung2003a}.

In this framework, therefore, the shapes of the wave functions of left- and
right-handed fermions plus distances between those wave functions in the extra
dimension determine the strengths of the mass terms in four dimensions.

The geometry of the fermionic zero modes along the extra dimension was
systematically set in Ref.~\cite{Hung2005} by breaking the symmetries of the
model in the extra dimension down to that of the Standard Model, which was the
approach originally suggested in Ref.~\cite{Hung2003a}.
As a result, Ref.~\cite{Hung2005} obtained \textit{early quark-lepton mass
unification}, within which the four-dimensional (4D) Yukawa couplings of the
chiral fermions of the model related to each other and a light Dirac neutrino
was made possible.

The present work intends to build a model based on the marriage of the other
petite unification model, ${\mathrm {PUT}}_2=\SU(4)_{\mathrm {PS}} \otimes
\SU(3)_L \otimes \SU(3)_{H}$, and the physics of large extra dimension in the
context of ``brane world" picture, in order to explore its implications. Similar
to the work in Ref.~\cite{Hung2005}, we make use of the idea of wave function
overlaps along the extra dimension and set the geometry of the zero modes by
symmetry breakings.

Historically, questions on quark-lepton mass relation were addressed in a
quark-lepton unification scenario, e.g., Grand Unified Theories
(GUT's)~\cite{GUT}. A well-known example of this is the equality of
$\tau$-lepton and bottom-quark masses~\cite{Buras1978} at $M_{GUT}$ in $\SU(5)$
scenario. A TeV scale quark-lepton mass relation differs from a GUT one in the
amount of ``running\footnote{including both coupling constants and masses.}" one
needs to be concerned about if one attempts to explore the implications at lower
energies, say $M_Z$.

On another front, the present work assumes a Dirac neutrino, which will turn out
light in a direct correlation with the masses of heavy unconventional fermions.
Such connection between a light Dirac neutrino and TeV-scale physics is in
contrast with the traditional seesaw mechanism~\cite{Seesaw}, where its scale is
limited perhaps only by Planck mass. Very recently, however, a TeV scale
scenario for seesaw mechanism~\cite{Hung2007} has been put forward, which
broadens the implications on TeV-scale physics to both Dirac and Majorana light
neutrinos. Of course, the final word on the nature of neutrino, whether it is a
Majorana or Dirac particle, must come from experiment, in particular those
regarding lepton number violation.

The outline of the paper is as follows. First, we go over the idea of petite
unification theories briefly followed by a review on the group structure and the
particle content of $\mathrm{PUT}_2$ scenario. Then, we present a five
dimensional model based on $\mathrm{PUT}_2$ scenario plus a short review on  the
wave function overlap mechanism. Afterward, we set the geometry of the zero mode
wave functions of chiral fermions by systematic symmetry breakings in the extra
dimension. In subsequent sections, we move toward the computation of chiral
fermion mass scales by relating them to the magnitudes of applicable overlaps in
the extra dimension. A numerical analysis concludes the mass scale computation,
which substantiates the notion of early quark-lepton mass unification. Then, we
examine the validity of our model by computing the electroweak oblique parameter
$S$ and the lifetimes of heavy chiral fermions.
\section{Petite unification of quarks and leptons}\label{sec:PUT}
Petite unification models \cite{Hung1982} were built around the idea of unifying
quarks and leptons at  an energy scale not too much higher than the electroweak
scale. They have the gauge group structure of $G = G_{S}(g_S) \otimes
G_{W}(g_W)$ with two independent couplings $g_S$ and $g_W$, which must contain
the SM fields. The first PUT model was constructed based on the knowledge of the
low-energy $\sin^{2} \theta_W$ value and known fermion representations at the
time. With the $\SU(4)_{\mathrm{PS}}$ group of Pati and Salam \cite{Pati1974}
chosen for $G_S$ and the constraint from the experimental value of
$\sin^{2}\theta_{W}(M_{Z}^{2})$, known at the time, the gauge group
$\mathrm{PUT}_0=\SU(4)_{\mathrm {PS}} \otimes [\SU(2)]^4$ with unification scale
of several hundreds of TeV emerged and was proposed in Ref.~\cite{Hung1982}.

Later precise measurements of $\sin^{2}\theta_{W}(M_{Z}^{2})$ plus renewed
interest in TeV scale physics, however, resulted in a thorough re-examination of
the PUT idea \cite{Buras2003}, yielding three favorable PUT models:
$\mathrm{PUT}_0$ and $\mathrm{PUT}_{1,2}$, where
\begin{equation}
{\mathrm {PUT}}_1=\SU(4)_{\mathrm {PS}} \otimes \SU(2)_{L} \otimes
\SU(2)_{R} \otimes \SU(2)_{H} ,
\end {equation}
and 
\begin{equation}
{\mathrm {PUT}}_2=\SU(4)_{\mathrm {PS}} \otimes \SU(3)_L \otimes \SU(3)_{H} .
\end {equation}

The new measured value of $\sin^{2}\theta_{W}(M_{Z}^{2})$, which was higher than
its old value, lowered the unification scale down to a few-TeV region. This
lower scale rules out $\mathrm{PUT}_0$ scenario due to problems with the decay
rate of $K_L \rightarrow \mu e$ at tree level. The remaining two models,
$\mathrm{PUT}_{1}$ and $\mathrm{PUT}_{2}$, however, are found to naturally avoid
the violation of the upper bound on the $K_L \rightarrow \mu e$ rate at tree
level. The SM gauge group with three couplings, $\SU(3)_c (g_3)\otimes \SU(2)_L
(g_2) \otimes \mathrm{U}(1)_Y (g_1)$, is assumed to be embedded into  the PUT
groups with two couplings. The symmetry breaking scheme of PUT scenarios is
given by\footnote{The gauge symmetry breakdown of PUT scenarios down to that of
the SM with an additional discrete $\mathcal{Z}$ symmetry and its implications
on monopoles is discussed in Ref.~\cite{Zubkov2007}.}
\begin{subequations}\label{equ:sb}
\begin{equation}
G \stackrel{\textstyle M}{\longrightarrow} G_1 
\stackrel{\textstyle \tilde{M}}{\longrightarrow} G_2
\stackrel{\textstyle M_Z}{\longrightarrow} \SU(3)_c \otimes \mathrm{U}(1)_{EM} ,
\end{equation}
where
\begin{equation}
G_1 = \SU(3)_{c}(g_3) \otimes \mathrm{U}(1)_S(\tilde{g}_S) \otimes
G_{W}(g_W) \, ,
\end {equation}
and
\begin{equation}
G_2 = \SU(3)_{c}(g_3) \otimes \SU(2)_{L}(g_2) \otimes
\mathrm{U}(1)_{Y}(g^\prime)\ ,
\end{equation}
\end{subequations}
with $M_Z < \tilde{M} \leq M$. The two PUT scenarios have three new generations
of unconventional quarks and leptons, in addition to the three standard
generations of quarks and leptons. The magnitude of the charges of these new
particles can reach up to $4/3$ (for ``quarks'') and 2 (for ``leptons''). The
horizontal groups $\SU(2)_H$ and $\SU(3)_H$ connect the standard fermions to the
unconventional ones, as well as the gauge bosons of
$\SU(4)_{\mathrm{PS}}/\left[\SU(3)_c \otimes \mathrm{U}(1)_S \right]$.

In both PUT models the $\SU(4)_{\mathrm {PS}}$ quartets contain either
``unconventional quark and the SM lepton'' or ``SM quark and unconventional
lepton.'' As a result, there is no tree-level transition between ordinary quarks
and leptons mediated by the $\SU(4)_{\mathrm{PS}}/\left[\SU(3)_c \otimes
\mathrm{U}(1)_S \right]$ gauge bosons. This important property prevents rare
decays such as $K_L \rightarrow \mu e$ from acquiring large rates, since it can
only occur through one-loop processes which can be made small enough to comply
with the experimental bound.

Another property of PUT scenarios is the existence of new contributions to
flavor changing neutral current (FCNC) processes, involving standard quarks and
leptons, which are mediated by the horizontal $\SU(2)_H$ and $\SU(3)_H$ weak
gauge bosons and the new unconventional quarks and leptons. Nonetheless, they
appear at one-loop level and can be made consistent with the existing
experimental bounds. A thorough analysis of $\mathrm{PUT}_1$ was carried out by
the authors of Ref.~\cite{Buras2004}.
\section{\texorpdfstring{$\text{PUT}_2$}{PUT2} model}\label{sec:PUT2}
In this scenario the weak gauge group is $G_W=\SU(3)_{L} \otimes \SU(3)_{H}$,
where the SM's $\SU(2)_L$ is the subgroup of its $\SU(3)_L$. The gauge symmetry
breaking of $\mathrm{PUT}_2$ follows the scheme given in Eqs.~(\ref{equ:sb}).

Within such symmetry breaking, the strong $\mathrm{U}(1)_S$ group corresponds to
the unbroken diagonal generator of $\SU(4)_{\mathrm{PS}}$, i.e., $\hat Y_S$. The
weak hypercharge $\mathrm{U}(1)_Y$ group emerges from $\mathrm{U}(1)_S$ and
$G_W$ breaking, whose generator $\hat{Y}_W$ can be written as $\hat Y_W  = C_S
\hat T_{15\mathrm{PS}}  + C_L \hat T_{8L}  + C_{1H} \hat T_{8H}  + C_{2H} \hat
T_{3H}$ where $\hat T$'s are the diagonal generators of $\tilde{G}_S$,
$\SU(3)_L$ and $\SU(3)_H$ symmetries. The SM's $\hat T_{3L}$ generator is simply
the third generator of $\SU(3)_L$, which goes into the unbroken $\SU(2)_L$
subgroup. Note that this is all in the ``unlocked standard model'' picture of
Ref.~\cite{Hung1982}, where the generators of $\SU(2)_L$ are the unbroken
generators of $G_W$. The $C_i$ coefficients in $\hat{Y}_W$ define the embedment
of the SM's weak hypercharge group $\mathrm{U}(1)_Y$ into $G_1$.

The two symmetry breaking scales $M$ and $\tilde M$ were determined in
Ref.~\cite{Buras2003} by renormalization group (RG) evolution combined with the
very precise experimental value of $\sin ^2 \theta _W \left( {M_Z^2 } \right)$.
The values could differ by up to an order of magnitude, roughly $3 \leqslant M
\leqslant 10{\text{ TeV}}$ and $0.8 \leqslant \tilde M \leqslant 3{\text{
TeV}}$.

The charge operator in PUT scenarios is defined as $\hat Q = \hat Q_W  + C_S
\hat T_{15\mathrm{PS}}$, where $\hat Q_W$ is the weak charge given by $\hat Q_W 
= \hat T_{3L}  + C_L \hat T_{8L}  + C_{1H} \hat T_{8H}  + C_{2H} \hat T_{3H}$.
The weak charge $Q_W $, as shown in Ref.~\cite{Hung1982}, is related to $\sin ^2
\theta _W^0 $ defining the charge distribution of the relevant representations
of PUT scenarios. For $\mathrm{PUT}_2$ model, $C_S^2  = {8 \mathord{\left/
{\vphantom {8 3}} \right. \kern-\nulldelimiterspace} 3}$ and the important group
theoretical factor $\sin ^2 \theta _W^0 $ is given by $\sin ^2 \theta _W^0  = {1
\mathord{\left/ {\vphantom {1 {\left( {1 + C_W^2 } \right)}}} \right.
\kern-\nulldelimiterspace} {\left( {1 + C_W^2 } \right)}} = {3 \mathord{\left/
{\vphantom {3 8}} \right. \kern-\nulldelimiterspace} 8}$ , where $C_W^2  = C_L^2
 + C_{1H}^2  + C_{2H}^2  = {5 \mathord{\left/ {\vphantom {5 3}} \right.
\kern-\nulldelimiterspace} 3}$.

For the model in question, the fermion representations, which together are
anomaly-free, are $(4,3,\bar{3})$ and $(4,\bar{3},3)$. The charge distribution
of the fermion content of $(4,3,\bar{3})$ representation is
\begin{equation}\label{equ:charge1}
\mathcal{Q}_1  = \Biggl( {\left[ {\left( {\frac{1}
{3},\frac{4}
{3},\frac{4}
{3}} \right),\left( { - 1,0,0} \right)} \right],\left[ {\left( { - \frac{2}
{3},\frac{1}
{3},\frac{1}
{3}} \right),\left( { - 2, - 1, - 1} \right)} \right],\left[ {\left( { -
\frac{2}
{3},\frac{1}
{3},\frac{1}
{3}} \right),\left( { - 2, - 1, - 1} \right)} \right]} \Biggr) \, ,
\end{equation}
Similarly, for $(4,\bar{3},3)$ the charge distribution is given by
\begin{equation}\label{equ:charge2}
\mathcal{Q}_2  = \Biggl( {\left[ {\left( {\frac{1}
{3}, - \frac{2}
{3}, - \frac{2}
{3}} \right),\left( { - 1, - 2, - 2} \right)} \right],\left[ {\left( {\frac{4}
{3},\frac{1}
{3},\frac{1}
{3}} \right),\left( {0, - 1, - 1} \right)} \right],\left[ {\left( {\frac{4}
{3},\frac{1}
{3},\frac{1}
{3}} \right),\left( {0, - 1, - 1} \right)} \right]} \Biggr) \, ,
\end{equation}
In terms of $\SU(2)_{L}$ doublets and singlets, one can write $(4,3,\bar{3})$ as
\begin{equation}\label{equ:psi1}
\Psi _{1,L}  = \Biggl( {\bigg[ {\Big( {\bm{\psi}^{Q*} ,D^c } \Big),\Big( {\psi
^l ,\nu ^c } \Big)} \bigg],\bigg[ {\Big( {\bm{\psi} ^{q*} ,d^c } \Big),\Big(
{\bm{\psi}^{L*} ,l_{d}^c } \Big)} \bigg],\bigg[ {\Big( {\tilde {\bm{\psi}} ^{q*}
,\tilde{\mathtt{d}}^{*} } \Big),\Big( {\tilde{\bm{\psi}}^{L*}
,\tilde{\mathtt{l}}^{*} } \Big)} \bigg]} \Biggr)_L ,
\end{equation}
and  $(4,\bar{3},3)$ as
\begin{equation}\label{equ:psi2}
\Psi _{2,L}  = \Biggl( {\bigg[ {\Big( {\tilde {\bm{\psi}} ^{q,c} ,u^c }
\Big),\Big( {\tilde{\bm{\psi}}^{L,c} ,l_{u}^c } \Big)} \bigg],\bigg[ {\Big(
{\tilde{\bm{\psi}}^{Q,c} ,U^c } \Big),\Big( {\tilde \psi ^{l,c*} ,l^{c*} }
\Big)} \bigg],\bigg[ {\Big( {\tilde{\bm{\psi}}^{Q*} ,\tilde{\mathtt{d}}^{c} }
\Big),\Big( {\tilde \psi ^l ,\tilde{\mathtt{l}}^{c} } \Big)} \bigg]} \Bigg)_L . 
\end{equation}

Before we identify the $\SU(2)_{L}$ doublets and singlets appearing in
Eqs.~(\ref{equ:psi1} and \ref{equ:psi2}), let us first  point out that in
Eqs.~(\ref{equ:psi1} and \ref{equ:psi2}) the right-handed fields are written in
terms of the left-handed charge conjugates; so that the whole representation is
left handed, e.g., $\nu_L^c$ or $u_L^c$. Besides, to match the charge
distributions of Eqs.~(\ref{equ:charge1} and \ref{equ:charge2}), some
$\SU(2)_{L}$ doublets, in Eqs.~(\ref{equ:psi1} and \ref{equ:psi2}), appear in
italic-boldface typeset. To explain this notation, consider an arbitrary doublet
\begin {equation}
\psi _{L,R}  = \left( {\begin{array}{*{20}c}
   {\psi _u }  \\
   {\psi _d }  \\
 \end{array} } \right)_{L,R} ,
\end {equation}
then $\bm{\psi} _{L,R}$, the rotated doublet in $\SU(2)$ space by $\pi$ about
the second axis, is defined as
\begin {equation}\label{equ:cpsi}
\bm{\psi} _{L,R}  \equiv i\tau _2 \psi _{L,R}  = \left( {\begin{array}{*{20}c}
   {\psi _d }  \\
   { - \psi _u }  \\
 \end{array} } \right)_{L,R}.
\end{equation}
The $\SU(2)_L$ doublets and singlets present in $(4,3,\bar{3})$ are\footnote{As
a convention, the fields presented by tilded letters are vector-like (i.e., not
chiral).}
\begin{subequations}\label{rep1}
\begin{equation}
\psi^{q}_{L} = \left(
\begin{array}{c}
u(2/3)\\ 
d(-1/3)
\end{array}
\right)_L \,; \,\, d^{c}_L(1/3)=C \bar{d}^{\,T}_{R}\, ,
\end{equation}
\begin{equation}
\psi^{l}_{L} = \left(
\begin{array}{c}
\nu(0) \\ 
l(-1)
\end{array}
\right)_L \, ; \,\, \nu^{c}_L= C \bar{\nu}_{R}^{T} \, ,
\end{equation}
\begin{equation}
\psi^Q_{L} = \left(
\begin{array}{c}
U(-1/3)\\ 
D(-4/3)
\end{array}
\right)_L \,; \,\, D^{c}_{L}(4/3) = C \bar{D}^{T}_{R} \, ,
\end{equation}
\begin{equation}
\psi^L_{L} = \left(
\begin{array}{c}
l_{u}(2)\\ 
l_{d}(1)
\end{array}
\right)_L \,; \,\, l^{c}_{d,L}(-1)= C \bar{l}^{\,T}_{d,R}\, ,
\end{equation}
\begin{equation}
\tilde{\psi}^L_{L} = \left(
\begin{array}{c}
\tilde{l}_{u}(2)\\ 
\tilde{l}_{d}(1)
\end{array}
\right)_L \, ; \,\, \tilde{\mathtt{l}}_L(+1) \, ,
\end{equation}
\begin{equation}
\tilde{\psi}^{q}_{L} = \left(
\begin{array}{c}
\tilde{u}(2/3)\\ 
\tilde{d}(-1/3)
\end{array}
\right)_L \, ; \,\, \tilde{\mathtt{d}}_L(-1/3) \, . 
\end{equation}
\end{subequations}
In the above list, one notices normal quarks and leptons, and those with unusual
electric charges. On the other hand, the $\SU(2)$ doublets and singlets of
$(4,\bar{3},3)$ are
\begin{subequations}\label{rep2}
\begin{equation}
\tilde{\psi}^{l}_{L,R} = \left(
\begin{array}{c}
\tilde{\nu}(0)\\ 
\tilde{l}(-1)
\end{array}
\right)_{L,R} \, ; \,\,\, l^{c}_{L}(+1) = C \bar{l}^{\,T}_{R} \, ,
\end{equation}
\begin{equation}
\tilde{\mathtt{l}}_R (+1)\ \, ;\ \, u_L^c \left( {{{ - 2} \mathord{\left/
 {\vphantom {{ - 2} 3}} \right.
 \kern-\nulldelimiterspace} 3}} \right)=C \bar{u}^{T}_{R} \,,
\end{equation}
\begin{equation}
\tilde{\psi}^{q}_{R} = \left(
\begin{array}{c}
\tilde{u}(+2/3)\\ 
\tilde{d}(-1/3)
\end{array}
\right)_R\ \, ; \,\, \tilde{\mathtt{d}} _R (-1/3) \, ,
\end{equation}
\begin{equation}
\tilde{\psi}^L_{R} = \left(
\begin{array}{c}
\tilde{l}_{u}(2)\\ 
\tilde{l}_{d}(1)
\end{array}
\right)_{R}\ \, ; \,\,
l^{c}_{u,L}(-2)= C \bar{l}^{\,T}_{u,R} \, ,
\end{equation}
\begin{equation}
\tilde{\psi}^Q_{L,R} = \left(
\begin{array}{c}
\tilde{U}(-1/3)\\ 
\tilde{D} (-4/3)
\end{array}
\right)_{L,R}\ \,; \;\, U^{c}_{L}(1/3) = C 
\bar{U}^{T}_{R} \, .
\end{equation}
\end{subequations}

One notices two types of families with SM transformation property in both
$\Psi_1$ and $\Psi_2$. This means left-handed doublets and right-handed singlets
for each family. One family includes SM quarks and leptons (normal fermions) and
the other contains unconventional quarks and leptons, i.e., those with unusual
charges. These unconventional particles are $\psi^Q_L$, $D^c_L$, $U^c_L$, and
$\psi^L_{L}$, $l^c_{d,L}$, $l^c_{u,L}$. The normal and unconventional quarks and
leptons will receive mass through their couplings with the SM Higgs field.

In addition, the fermion content of $\mathrm{PUT}_2$ includes two vector-like
$\SU(2)_L$ doublets of quarks and leptons $(\tilde\psi^q,\tilde\psi^l)_{L,R}$
and $(\tilde{\psi}^Q,\tilde{\psi}^L)_{L,R}$, with normal and unusual charges,
and two vector-like $\SU(2)_L$ singlets $\tilde{\mathtt{l}}_{L,R}$ and
$\tilde{\mathtt{d}}_{L,R}$. These vector-like particles can obtain large bare
masses as mentioned in Ref.~\cite{Buras2003}.

Let us write the two representations in terms of quartets and triplets of the
corresponding gauge symmetry groups. For $\Psi _{1L}$, we have the following
multiplets:
\begin{itemize}
\item
$\SU(4)_{\mathrm{PS}}$ quartets
\begin{subequations}\label{equ:quartets1}
\begin{equation}
\begin{array}{*{20}c}
   {\left( {\begin{array}{*{20}c}
   {D^* _L  \left( {{4 \mathord{\left/
 {\vphantom {4 3}} \right.
 \kern-\nulldelimiterspace} 3}} \right)}  \\
   {\nu _L \left( 0 \right)}  \\
 \end{array} } \right)} & ; & {\left( {\begin{array}{*{20}c}
   {-U^* _L  \left( {{1 \mathord{\left/
 {\vphantom {1 3}} \right.
 \kern-\nulldelimiterspace} 3}} \right)}  \\
   {l_L \left( { - 1} \right)}  \\
 \end{array} } \right)} & ; & {\left( {\begin{array}{*{20}c}
   {D_L^c \left( {{4 \mathord{\left/
 {\vphantom {4 3}} \right.
 \kern-\nulldelimiterspace} 3}} \right)}  \\
   {\nu _L^c \left( 0 \right)}  \\
 \end{array} } \right)} \,,  \\
 \end{array} 
\end{equation}
\begin{equation}
\begin{array}{*{20}c}
   {\left( {\begin{array}{*{20}c}
   {d^* _L \left( {{1 \mathord{\left/
 {\vphantom {1 3}} \right.
 \kern-\nulldelimiterspace} 3}} \right)}  \\
   {l^* _{d,L} \left( { - 1} \right)}  \\
 \end{array} } \right)} & ; & {\left( {\begin{array}{*{20}c}
   {-u^* _L  \left( {{{ - 2} \mathord{\left/
 {\vphantom {{ - 2} 3}} \right.
 \kern-\nulldelimiterspace} 3}} \right)}  \\
   {-l^* _{u,L} \left( { - 2} \right)}  \\
 \end{array} } \right)} & ; & {\left( {\begin{array}{*{20}c}
   {d_L^c \left( {{1 \mathord{\left/
 {\vphantom {1 3}} \right.
 \kern-\nulldelimiterspace} 3}} \right)}  \\
   {l_{d,L}^c \left( { - 1} \right)}  \\
 \end{array} } \right)}\,,  \\
 \end{array} 
\end{equation}
\begin{equation}
\begin{array}{*{20}c}
   {\left( {\begin{array}{*{20}c}
   {\tilde{d}^* _L \left( {{{ - 2} \mathord{\left/
 {\vphantom {{ - 2} 3}} \right.
 \kern-\nulldelimiterspace} 3}} \right)}  \\
   {\tilde{l}^* _{d,L} \left( { - 1} \right)}  \\
 \end{array} } \right)} & ; & {\left( {\begin{array}{*{20}c}
   {-\tilde{u}^* _L \left( {{{ - 2} \mathord{\left/
 {\vphantom {{ - 2} 3}} \right.
 \kern-\nulldelimiterspace} 3}} \right)}  \\
   {-\tilde{l}^* _{u,L} \left( { - 2} \right)}  \\
 \end{array} } \right)} & ; & {\left( {\begin{array}{*{20}c}
   {\tilde{\mathtt{d}}_L^{*} \left( {{1 \mathord{\left/ 
 {\vphantom {1 3}} \right. \kern-\nulldelimiterspace} 3}} \right)}  \\
   {\tilde{\mathtt{l}}_L^{*} \left( { - 1} \right)}  \\
 \end{array} } \right)}\,,  \\
 \end{array} 
\end{equation}
\end{subequations}
\item
$\SU(3)_L$ triplets
\begin{subequations}\label{equ:tripletsL1}
\begin{equation}
\begin{array}{*{20}c}
   {\left( {\begin{array}{*{20}c}
   {D^* _L \left( {{4 \mathord{\left/
 {\vphantom {4 3}} \right.
 \kern-\nulldelimiterspace} 3}} \right)}  \\
   {-U^* _L  \left( {{1 \mathord{\left/
 {\vphantom {1 3}} \right.
 \kern-\nulldelimiterspace} 3}} \right)}  \\
   {D_L^c \left( {{4 \mathord{\left/
 {\vphantom {4 3}} \right.
 \kern-\nulldelimiterspace} 3}} \right)}  \\
 \end{array} } \right)} & ; & {\left( {\begin{array}{*{20}c}
   {\nu _L \left( 0 \right)}  \\
   {l_L \left( { - 1} \right)}  \\
   {\nu _L^c \left( 0 \right)}  \\
 \end{array} } \right)} & ; & {\left( {\begin{array}{*{20}c}
   {d^* _L \left( {{1 \mathord{\left/
 {\vphantom {1 3}} \right.
 \kern-\nulldelimiterspace} 3}} \right)}  \\
   {-u^* _L  \left( {{{ - 2} \mathord{\left/
 {\vphantom {{ - 2} 3}} \right.
 \kern-\nulldelimiterspace} 3}} \right)}  \\
   {d_L^c \left( {{1 \mathord{\left/
 {\vphantom {1 3}} \right.
 \kern-\nulldelimiterspace} 3}} \right)}  \\
 \end{array} } \right)}\,,  \\
 \end{array}
\end{equation}
\begin{equation}\label{equ:tripletsL12}
\begin{array}{*{20}c}
   {\left( {\begin{array}{*{20}c}
   {l^* _{d,L} \left( { - 1} \right)}  \\
   {-l^* _{u,L} \left( { - 2} \right)}  \\
   {l_{d,L}^c \left( { - 1} \right)}  \\
 \end{array} } \right)} & ; & {\left( {\begin{array}{*{20}c}
   {\tilde{d}^* _L \left( {{1 \mathord{\left/
 {\vphantom {1 3}} \right.
 \kern-\nulldelimiterspace} 3}} \right)}  \\
   {-\tilde{u}^* _L \left( {{{ - 2} \mathord{\left/
 {\vphantom {{ - 2} 3}} \right.
 \kern-\nulldelimiterspace} 3}} \right)}  \\
   {\tilde{\mathtt{d}}_L^{*} \left( {{1 \mathord{\left/
 {\vphantom {1 3}} \right.
 \kern-\nulldelimiterspace} 3}} \right)}  \\
 \end{array} } \right)} & ; & {\left( {\begin{array}{*{20}c}
   {\tilde{l}^* _{d,L} \left( { - 1} \right)}  \\
   {-\tilde{l}^* _{u,L} \left( { - 2} \right)}  \\
   {\tilde{\mathtt{l}}_L^{*} \left( { - 1} \right)}  \\
 \end{array} } \right)}\,,  \\
 \end{array}
\end{equation}
\end{subequations}
\item
$\SU(3)_H$ antitriplets
\begin{subequations}\label{equ:atripletsH1}
\begin{equation}
\begin{array}{*{20}c}
   {\left( {\begin{array}{*{20}c}
   {d^* _L \left( {{1 \mathord{\left/
 {\vphantom {1 3}} \right.
 \kern-\nulldelimiterspace} 3}} \right)}  \\
   {D^* _L \left( {{4 \mathord{\left/
 {\vphantom {4 3}} \right.
 \kern-\nulldelimiterspace} 3}} \right)}  \\
   {\tilde{d}^* _L \left( {{1 \mathord{\left/
 {\vphantom {1 3}} \right.
 \kern-\nulldelimiterspace} 3}} \right)}  \\
 \end{array} } \right)} & ; & {\left( {\begin{array}{*{20}c}
   {-u^* _L \left( {{{ - 2} \mathord{\left/
 {\vphantom {{ - 2} 3}} \right.
 \kern-\nulldelimiterspace} 3}} \right)}  \\
   {-U^* _L \left( {{1 \mathord{\left/
 {\vphantom {1 3}} \right.
 \kern-\nulldelimiterspace} 3}} \right)}  \\
   {-\tilde{u}^* _L  \left( {{{ - 2} \mathord{\left/
 {\vphantom {{ - 2} 3}} \right.
 \kern-\nulldelimiterspace} 3}} \right)}  \\
 \end{array} } \right)} & ; & {\left( {\begin{array}{*{20}c}
   {d_L^c \left( {{1 \mathord{\left/
 {\vphantom {1 3}} \right.
 \kern-\nulldelimiterspace} 3}} \right)}  \\
   {D_L^c \left( {{4 \mathord{\left/
 {\vphantom {4 3}} \right.
 \kern-\nulldelimiterspace} 3}} \right)}  \\
   {\tilde{\mathtt{d}}_L^{*} \left( {{1 \mathord{\left/
 {\vphantom {1 3}} \right.
 \kern-\nulldelimiterspace} 3}} \right)}  \\
 \end{array} } \right)}\,,  \\
 \end{array}
\end{equation}
\begin{equation}
\begin{array}{*{20}c}
   {\left( {\begin{array}{*{20}c}
   {l^* _{d,L} \left( { - 1} \right)}  \\
   {\nu _L \left( 0 \right)}  \\
   {\tilde{l}^* _{d,L}  \left( { - 1} \right)}  \\
 \end{array} } \right)} & ; & {\left( {\begin{array}{*{20}c}
   {-l^* _{u,L} \left( { - 2} \right)}  \\
   {l_L \left( { - 1} \right)}  \\
   {-\tilde{l}^* _{u,L} \left( { - 2} \right)}  \\
 \end{array} } \right)} & ; & {\left( {\begin{array}{*{20}c}
   {l_{d,L}^c \left( { - 1} \right)}  \\
   {\nu _L^c \left( 0 \right)}  \\
   {\tilde{\mathtt{l}}_L^{*} \left( { - 1} \right)}  \\
 \end{array} } \right)}\,,  \\
 \end{array}
\end{equation}
\end{subequations}
\end{itemize}
For $\Psi _{2L} $, on the other hand, the corresponding multiplets are:
\begin{itemize}
\item
$\SU(4)_{\mathrm{PS}}$ quartets
\begin{subequations}\label{equ:quartets2}
\begin{equation}
\begin{array}{*{20}c}
   {\left( {\begin{array}{*{20}c}
   {\tilde d_L^{c} \left( {{1 \mathord{\left/
 {\vphantom {1 3}} \right.
 \kern-\nulldelimiterspace} 3}} \right)}  \\
   {\tilde{l}_{d,L}^{c} \left( { - 1} \right)}  \\
 \end{array} } \right)} & ; & {\left( {\begin{array}{*{20}c}
   {-\tilde u_L^{c} \left( {{{ - 2} \mathord{\left/
 {\vphantom {{ - 2} 3}} \right.
 \kern-\nulldelimiterspace} 3}} \right)}  \\
   {-\tilde{l}_{u,L}^{c} \left( { - 2} \right)}  \\
 \end{array} } \right)} & ; & {\left( {\begin{array}{*{20}c}
   {u_L^c \left( {{{ - 2} \mathord{\left/
 {\vphantom {{ - 2} 3}} \right.
 \kern-\nulldelimiterspace} 3}} \right)}  \\
   {l_{u,L}^c \left( { - 2} \right)}  \\
 \end{array} } \right)} \,, \\
 \end{array}
\end{equation}
\begin{equation}
\begin{array}{*{20}c}
   {\left( {\begin{array}{*{20}c}
   {\tilde{D}_L^{c} \left( {{4 \mathord{\left/
 {\vphantom {4 3}} \right.
 \kern-\nulldelimiterspace} 3}} \right)}  \\
   {\tilde \nu _L^{c*} \left( 0 \right)}  \\
 \end{array} } \right)} & ; & {\left( {\begin{array}{*{20}c}
   {-\tilde{U}_L^{c} \left( {{1 \mathord{\left/
 {\vphantom {1 3}} \right.
 \kern-\nulldelimiterspace} 3}} \right)}  \\
   {\tilde l_L^{\,c*} \left( { - 1} \right)}  \\
 \end{array} } \right)} & ; & {\left( {\begin{array}{*{20}c}
   {U_L^c \left( {{1 \mathord{\left/
 {\vphantom {1 3}} \right.
 \kern-\nulldelimiterspace} 3}} \right)}  \\
   {l_L^{c*} \left(-1 \right)}  \\
 \end{array} } \right)} \,, \\
 \end{array}
\end{equation}
\begin{equation}
\begin{array}{*{20}c}
   {\left( {\begin{array}{*{20}c}
   {\tilde{D}_L^{*}  \left( {{4 \mathord{\left/
 {\vphantom {4 3}} \right.
 \kern-\nulldelimiterspace} 3}} \right)}  \\
   {\tilde \nu _L \left( 0 \right)}  \\
 \end{array} } \right)} & ; & {\left( {\begin{array}{*{20}c}
   {-\tilde{U}_L^{*}  \left( {{1 \mathord{\left/
 {\vphantom {1 3}} \right.
 \kern-\nulldelimiterspace} 3}} \right)}  \\
   {\tilde l_L \left( { - 1} \right)}  \\
 \end{array} } \right)} & ; & {\left( {\begin{array}{*{20}c}
   {\tilde{\mathtt{d}}_L^{c} \left( {{1 \mathord{\left/
 {\vphantom {1 3}} \right.
 \kern-\nulldelimiterspace} 3}} \right)}  \\
   {\tilde{\mathtt{l}}_L^{c} \left( { - 1} \right)}  \\
 \end{array} } \right)} \,, \\
 \end{array}
\end{equation}
\end{subequations}
\item
$\SU(3)_L$ antitriplets
\begin{subequations}\label{equ:atripletsL2}
\begin{equation}
\begin{array}{*{20}c}
   {\left( {\begin{array}{*{20}c}
   {-\tilde u_L^{c} \left( {{{ - 2} \mathord{\left/
 {\vphantom {{ - 2} 3}} \right.
 \kern-\nulldelimiterspace} 3}} \right)}  \\
   {\tilde d_L^{c} \left( {{1 \mathord{\left/
 {\vphantom {1 3}} \right.
 \kern-\nulldelimiterspace} 3}} \right)}  \\
   {u_L^c \left( {{{ - 2} \mathord{\left/
 {\vphantom {{ - 2} 3}} \right.
 \kern-\nulldelimiterspace} 3}} \right)}  \\
 \end{array} } \right)} & ; & {\left( {\begin{array}{*{20}c}
   {-\tilde{l}_{u,L}^{c} \left( { - 2} \right)}  \\
   {\tilde{l}_{d,L}^{c} \left( { - 1} \right)}  \\
   {l_{u,L}^c \left( { - 2} \right)}  \\
 \end{array} } \right)} & ; & {\left( {\begin{array}{*{20}c}
   {-\tilde{U}_L^{c} \left( {{1 \mathord{\left/
 {\vphantom {1 3}} \right.
 \kern-\nulldelimiterspace} 3}} \right)}  \\
   {\tilde{D}_L^{c} \left( {{4 \mathord{\left/
 {\vphantom {4 3}} \right.
 \kern-\nulldelimiterspace} 3}} \right)}  \\
   {U_L^c \left( {{1 \mathord{\left/
 {\vphantom {1 3}} \right.
 \kern-\nulldelimiterspace} 3}} \right)}  \\
 \end{array} } \right)} \,, \\
 \end{array} 
\end{equation}
\begin{equation}\label{equ:atripletsL22}
\begin{array}{*{20}c}
   {\left( {\begin{array}{*{20}c}
   {\tilde l_L^{c*} \left( { - 1} \right)}  \\
   {\tilde \nu _L^{c*} \left( 0 \right)}  \\
   {l_L^{c*} \left( { - 1} \right)}  \\
 \end{array} } \right)} & ; & {\left( {\begin{array}{*{20}c}
   {-\tilde{U}_L^{*}  \left( {{1 \mathord{\left/
 {\vphantom {1 3}} \right.
 \kern-\nulldelimiterspace} 3}} \right)}  \\
   {\tilde{D}_L^{*}  \left( {{4 \mathord{\left/
 {\vphantom {4 3}} \right.
 \kern-\nulldelimiterspace} 3}} \right)}  \\
   {\tilde{\mathtt{d}}_L^{c} \left( {{1 \mathord{\left/
 {\vphantom {1 3}} \right.
 \kern-\nulldelimiterspace} 3}} \right)}  \\
 \end{array} } \right)} & ; & {\left( {\begin{array}{*{20}c}
   {\tilde l_L \left( { - 1} \right)}  \\
   {\tilde \nu _L \left( 0 \right)}  \\
   {\tilde{\mathtt{l}}_L^{c} \left( { - 1} \right)}  \\
 \end{array} } \right)} \,, \\
 \end{array}
\end{equation}
\end{subequations}
\item
$\SU(3)_H$ triplets
\begin{subequations}\label{equ:tripletsH2}
\begin{equation}
\begin{array}{*{20}c}
   {\left( {\begin{array}{*{20}c}
   {\tilde{D}_L^{c} \left( {{4 \mathord{\left/
 {\vphantom {4 3}} \right.
 \kern-\nulldelimiterspace} 3}} \right)}  \\
   {\tilde d_L^{c} \left( {{1 \mathord{\left/
 {\vphantom {1 3}} \right.
 \kern-\nulldelimiterspace} 3}} \right)}  \\
   {\tilde{D}_L^{*}  \left( {{4 \mathord{\left/
 {\vphantom {4 3}} \right.
 \kern-\nulldelimiterspace} 3}} \right)}  \\
 \end{array} } \right)} & ; & {\left( {\begin{array}{*{20}c}
   {-\tilde{U}_L^{c} \left( {{1 \mathord{\left/
 {\vphantom {1 3}} \right.
 \kern-\nulldelimiterspace} 3}} \right)}  \\
   {-\tilde u_L^{c} \left( {{{ - 2} \mathord{\left/
 {\vphantom {{ - 2} 3}} \right.
 \kern-\nulldelimiterspace} 3}} \right)}  \\
   {-\tilde{U}_L^{*}  \left( {{1 \mathord{\left/
 {\vphantom {1 3}} \right.
 \kern-\nulldelimiterspace} 3}} \right)}  \\
 \end{array} } \right)} & ; & {\left( {\begin{array}{*{20}c}
   {U_L^c \left( {{1 \mathord{\left/
 {\vphantom {1 3}} \right.
 \kern-\nulldelimiterspace} 3}} \right)}  \\
   {u_L^c \left( {{{ - 2} \mathord{\left/
 {\vphantom {{ - 2} 3}} \right.
 \kern-\nulldelimiterspace} 3}} \right)}  \\
   {\tilde{\mathtt{d}}_L^{c} \left( {{1 \mathord{\left/
 {\vphantom {1 3}} \right.
 \kern-\nulldelimiterspace} 3}} \right)}  \\
 \end{array} } \right)} \,, \\
 \end{array}
\end{equation}
\begin{equation}
\begin{array}{*{20}c}
   {\left( {\begin{array}{*{20}c}
   {\tilde \nu _L^{c*} \left( 0 \right)}  \\
   {\tilde{l}_{d,L}^{c} \left( { - 1} \right)}  \\
   {\tilde \nu _L \left( 0 \right)}  \\
 \end{array} } \right)} & ; & {\left( {\begin{array}{*{20}c}
   {\tilde l_L^{c*} \left( { - 1} \right)}  \\
   {-\tilde{l}_{u,L}^{c} \left( { - 2} \right)}  \\
   {\tilde l_L \left( { - 1} \right)}  \\
 \end{array} } \right)} & ; & {\left( {\begin{array}{*{20}c}
   {l_L^{c*} \left( { - 1} \right)}  \\
   {l_{u,L}^c \left( { - 2} \right)}  \\
   {\tilde{\mathtt{l}}_L^{c} \left( { - 1} \right)}  \\
 \end{array} } \right)} \,, \\
 \end{array}
\end{equation}
\end{subequations}
\end{itemize}

Before we end this section, it is worth mentioning that all left-handed SM-type
fermions are in $\Psi _1 $. Plus, four of the corresponding right-handed fields
are in $\Psi _1 $ (i.e., $d_L^c$, $D_L^c$, $l_{d,L}^c$, $\nu _L^c$) and the
other four in $\Psi _2 $ (i.e., $u_L^c$, $U_L^c$, $l_L^{c*}$, $l_{u,L}^c$). The
right-handed fields, in both representations, are the third components of the
$\SU\left( 3 \right)_L $ triplets.
\section{Early quark-lepton unification in five dimensions}
Generalization to five-dimensional (5D) space is simply done by introducing an
extra spatial dimension, $y$. It is well known that 5D fermions are of Dirac
type and not chiral. As we would like the SM-type fermion content of our five
dimensional model to mimic the \textit{chiral} spectrum of the 4D SM-type
fermions; we compactify the extra dimension on an ${{\mathrm{S}_1 }
\mathord{\left/ {\vphantom {{S_1 } {\mathbb{Z}_2 }}} \right.
\kern-\nulldelimiterspace} {\mathbb{Z}_2 }}$ orbifold with a TeV-scale size.
That means the size of the extra dimension for our model is about the inverse of
the partial unification scale ($M \sim 3.3 - 10$ TeV).

In the ``brane world" picture, however, such chiral fermions are assumed to be
trapped onto a three-dimensional (3D) sub-manifold (``brane'' or ``domain wall''
\cite{Rubakov1983}) as zero modes. The localization of fermions into brane is
achievable by coupling the fermionic field to a background scalar field with a
kink solution.

In addition to localization, the shapes of zero-mode wave functions are to be
set. For doing that, we follow the idea in Ref.~\cite{Hung2003a} for which a
short review is given here. 

\begin{figure*}
\includegraphics{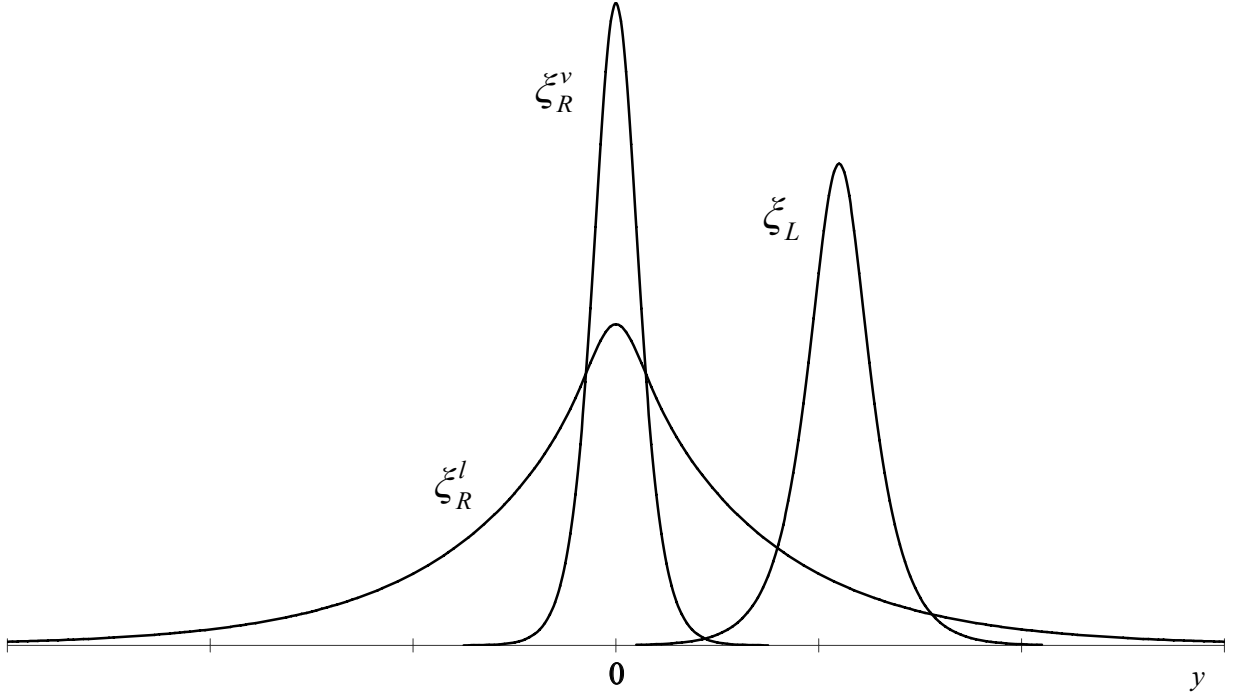}
\caption{\label{fig:lroverlaps} Schematic depiction of left-right overlaps for
neutrino and charged lepton in the extra dimension: $\xi^{\nu.l}_R$, zero-mode
wave function for right-handed neutrino and charged lepton; $\xi_L$, zero-mode
wave function for left-handed leptonic doublet.}
\end{figure*}

In Ref.~\cite{Hung2003a} a 5D left-right symmetric model was considered. After
localizing the right-handed fermions of a given doublet at the same point, the
$\SU(2)_R$ symmetry was spontaneously broken along the extra dimension via the
kink solution of a triplet scalar field. The outcome of such symmetry breaking
is significant in the sense that one element of the right-handed doublet obtains
a narrow, while the other element acquires a broad wave function along the extra
dimension. With left handed doublet localized at some other point along the
extra dimension, two very different left-right overlaps are resulted. An
exaggerated depiction of such overlaps is shown in Fig.~\ref{fig:lroverlaps} for
a leptonic doublet, $\nu$ and $l$. Fermionic Dirac mass terms involve left- and
right-handed fields and when the extra dimension is integrated out, the Yukawa
coupling in 4D space will be proportional to the corresponding left-right
overlaps in the extra dimension. The spirit of the work presented in
Ref.~\cite{Hung2003a} is that when zero-mode wave functions of the right-handed
fields overlap with the left-handed wave function (common for both $\nu$ and
$l$) there will be a large difference between the effective Yukawa couplings of
neutrino and charged lepton.

The objective in our 5D model is to localize the SM-type fermions of our model
on 3D slices and break the relevant symmetries along the extra dimension, which
in turn define the geometry of zero modes and ultimately will determine the
effective Yukawa couplings in the 4D theory.

The localization and symmetry breakings along the extra dimension involve Yukawa
couplings, e.g.,  in the form $f\bar \Psi _1 \Phi \Psi _1  + f\bar \Psi _2 \Phi
\Psi _2 $, where $\Psi _1 $ and $\Psi _2 $ couple to the same scalar field with
the same coupling constant to localize at the same point or shift position with
the same amount. This suggests an $\SU(2)_G$ global symmetry among $\Psi _1 $
and $\Psi _2 $ in the extra-dimensional Yukawa sector. For the Yukawa sector in
the extra dimension, therefore, the symmetry group of the theory can be written
as the product of global and gauge groups, $G=G_{\text{gauge}} \otimes
G_{\text{global}}$. Although the $\SU(2)_G$ global symmetry is limited to the
extra-dimensional Yukawa sector, there is an analogous, however implicit, global
$\SU(2)$ symmetry among $\Psi_1$ and $\Psi_2$ if only the strong
$\SU(4)_{\mathrm{PS}}$ quartets are looked at, i.e., weak group neglected. The
fact that the weak group representations of $\Psi_1$ and $\Psi_2$ differ means
that such extensive symmetry is explicitly broken by $G_W$.

The fermion representation of the model for couplings with scalar fields in the
extra dimension can be written as
\begin {equation}
\Psi \left( {x^\mu  ,y} \right) = \left( {\begin{array}{*{20}c}
   {\begin{array}{*{20}c}
   {4,3,\bar 3}  \\
   {4,\bar 3,3}  \\

 \end{array} } & ; & 2  \\

 \end{array} } \right) = \left( {\begin{array}{*{20}c}
   {\Psi _1 \left( {x^\mu  ,y} \right)}  \\
   {\Psi _2 \left( {x^\mu  ,y} \right)}  \\
 \end{array} } \right),
\end {equation}
where we used the notation $\left( {\begin{array}{*{20}c}{\mathrm{Gauge}} & ; &
{\mathrm{Global}}  \\ \end{array} } \right)$ to articulate the multiplet
structure of $\Psi$ with respect to the gauge and global groups.

To find out the appropriate group representations of the background scalar
fields, needed for localization and symmetry breakings, we should examine the
bilinear form of $\Psi$ under $\SU(4)_{\mathrm {PS}} \otimes \SU(3)_L \otimes
\SU(3)_{H} \otimes \SU(2)_{G}$, explicitly
\begin{equation}\label{equ:bilinear}
\bar \Psi \left( {x,y} \right)\Psi \left( {x,y} \right) = \left(
{\begin{array}{*{20}c}
   {15 \oplus 1,8 \oplus 1,8 \oplus 1} & ; & {3 \oplus 1}  \\
\end{array} } \right).
\end{equation}

From Eq.~(\ref{equ:bilinear}), one can pick suitable scalar fields to
\begin{enumerate}
\item
Localize the right-handed and left-handed fermions in the extra dimension at
different locations,
\item
Give different profiles to up and down sectors of the right-handed fermions,
\item
Differentiate between normal and unconventional fermions, also quarks and
leptons.
\end{enumerate}
In the following sections, we shall carry out these tasks one by one.
\subsection{Localization of fermions}\label{sec:localization}
To localize the SM-type fermions as chiral zero modes, we first note that we
wrote the fermion representations of $\mathrm{PUT}_2$ as left-handed multiplets
(see section~\ref{sec:PUT2}). Therefore, by choosing a chiral ${{\mathrm{S}_1 }
\mathord{\left/ {\vphantom {{S_1 } {\mathbb{Z}_2 }}} \right.
\kern-\nulldelimiterspace} {\mathbb{Z}_2 }}$ orbifold and positive couplings in
the localization process \cite{Georgi2001},  we can assign zero modes to all
left-handed fields of the representation including the right handed fields which
are written as charge conjugates. This way, the relevant 5D Dirac spinors
transform as left-handed SM fermions. Symbolically, the 5D representation of the
model as chiral zero mode can be imagined as
\begin {equation}
\Psi_L \left( {x ,y} \right) = \left( {\begin{array}{*{20}c}
   {\Psi _{1L} \left( x\right) \xi_1 \left( y \right)}  \\
   {\Psi _{2L} \left( x \right) \xi_2 \left( y \right)}  \\
 \end{array} } \right),
\end {equation}
where $\xi$'s take on the appropriate zero-mode wave functions for each
constituent field of the multiplet. Throughout this work and for clarity, we
denote the zero-mode wave functions of the left- and right-handed fields with
subscripts $L$ and $R$, respectively.

For localization, consider a singlet scalar field $\Phi _S  = \left(
{\begin{array}{*{20}c}{1,1,1} & ; & 1  \\ \end{array} } \right)$. The gauge- and
global-invariant Yukawa coupling of such scalar field with fermions looks like
\begin{equation}\label{equ:singcoup}
\mathcal{L}_S  = f_S \bar \Psi \Phi _S \Psi  = f_S \left( {\bar \Psi _1 \Phi _S
\Psi _1  + \bar \Psi _2 \Phi _S \Psi _2 } \right),
\end{equation}
where $f_S  > 0$. To localize at some non-zero point, $y \ne0$, let the kink
solution of $\Phi _S$ be in the form
\begin{equation}
\left\langle {\Phi _S } \right\rangle  = h_S \left( y \right) +v_S \, .
\end{equation}
The equation of motion for the zero-mode wave functions of the left- and
right-handed SM-type fermions is then given by
\begin{equation}\label{equ:localized}
\partial _y \xi _{L,R}  + \left[ {f_S h_S \left( y \right) + f_S v_S } \right]
\xi _{L,R}  = 0 \, .
\end{equation}
However, if one wants to have left-right overlaps between the zero modes, one
needs to separate the zero-mode wave functions of the left- and right-handed
fields along the extra dimension. This can be done by moving the left- and
right- handed zero-modes asymmetrically. To do this, we need to couple fermions
to a background scalar field that would only acquire a minimum energy solution
and \textit{not} a kink solution. We introduce a scalar field $\Phi _O = \left(
{\begin{array}{*{20}c} {1,8 ,1} & ; & 1  \\ \end{array} } \right)$, whose
coupling with fermions takes the form
\begin{equation}\label{equ:Octcoup}
\mathcal{L}_O  = - f_O \bar \Psi \Phi _O \Psi = - f_O \left( \bar \Psi_1 \Phi _O
\Psi_1 + \bar \Psi_2 \Phi _O \Psi_2 \right),
\end{equation}
where $f_O  > 0$. The minimum energy solution of $\Phi _O$ (which leaves
$\SU(2)_L$ unbroken) for such asymmetrical shift can simply take on the eighth
direction of $\SU(3)_L$, i.e.,
\begin{equation}\label{equ:octVEV}
\left\langle {\Phi _O } \right\rangle  =  \delta \left( {\begin{array}{*{20}c}
   1 & 0 & 0  \\
   0 & 1 & 0  \\
   0 & 0 & { - 2}  \\

 \end{array} } \right),
\end{equation}
where $\delta$ is the vacuum expectation value (VEV) of $\Phi _O$. The coupling
in Eq.~(\ref{equ:Octcoup}), when $\Phi _O$ develops VEV, shifts the position of
the left- and right-handed zero modes along the extra dimension differently,
which is obvious from their equations of motion
\begin{subequations}\label{equ:lrheq0}
\begin{equation}\label{equ:lheq0}
\partial _y \xi _L \left( y \right) + \left[ {f_S h_S \left( y \right)  + f_S
v_S - f_O \delta } \right]\xi _L \left( y \right) = 0,
\end{equation}
\begin{equation}\label{equ:rheq0}
\partial _y \xi _R \left( y \right) + \left[ {f_S h_S \left( y \right)  + f_S
v_S + 2f_O \delta} \right] \xi _R \left( y \right) = 0.
\end{equation}
\end{subequations}
with $\delta  \ne 0$. The possibility of $\delta = 0$ will be discussed later.
We remind ourselves that the left-handed zero-mode wave functions $\xi _L$, are
$\SU(2)_L$ doublets, while the right-handed zero-mode wave functions $\xi _R$,
are just singlets.
\subsection{Distinguishing the up and down sectors of the right-handed
fermions}\label{sec:uddis}
Since one sector of the right-handed SM-type fields are in $\Psi_1$ and the
other in $\Psi_2$, distinguishing these two sectors along the extra dimension
demands a coupling which differentiates between them in the extra dimension.
Looking at Eq.~(\ref{equ:bilinear}), we consider two $\SU(2)_G$ triplet fields
$\Phi _T = \left( {\begin{array}{*{20}c} {1,1 ,1} & ; & 3  \\ \end{array} }
\right)$ and $\Phi' _T = \left( {\begin{array}{*{20}c} {1,8,1} & ; & 3  \\
\end{array} } \right)$, for an asymmetrical profile changing. The Yukawa
couplings with fermions would be
\begin{equation}\label{equ:Gcoupling}
\mathcal{L}_T  = f_T \bar \Psi \Phi _T \Psi + f'_T \bar \Psi \Phi' _T \Psi,
\end{equation}
where $f_T , f'_T > 0$. To alter the shapes of the right-handed zero-mode wave
functions, these two triplet fields must attain kink solutions, they are
\begin{equation}
\left\langle {\Phi _{T} } \right\rangle  = h_T \left( y \right) \left(
{\begin{array}{*{20}c}
   1 & 0  \\
   0 & { - 1}  \\
 \end{array} } \right),
\end{equation}
and
\begin{equation}
\left\langle {\Phi' _{T} } \right\rangle  = h'_T \left( y \right) \left(
{\begin{array}{*{20}c}
   1 & 0 & 0  \\
   0 & 1 & 0  \\
   0 & 0 & { - 2}  \\
 \end{array} } \right) \otimes \left( {\begin{array}{*{20}c}
   1 & 0  \\
   0 & { - 1}  \\
 \end{array} } \right),
\end{equation}
where $h_T \left( y \right)$ and $h'_T \left( y \right)$ are the kink solutions
of $\Phi _T$ and $\Phi' _T$, respectively.
The equations of motion for the zero-mode wave functions now read
\begin{subequations}\label{equ:eomTT}
\begin{equation}\label{equ:eomTTl0}
\partial _y \xi _L \left( y \right) + \left[ {f_S h_S \left( y \right) + f_T h_T
\left( y \right) + f'_T h'_T \left( y \right)  + f_S v_S - f_O \delta}
\right]\xi _L \left( y \right) = 0 \, ,
\end{equation}
\begin{equation}\label{equ:eomTTru0}
\partial _y \xi _R^{up} \left( y \right) + \left[ {f_S h_S \left( y \right) +
\left({ f_T h_T \left( y \right) -2f'_T h'_T \left( y \right) } \right) + f_S
v_S - 2f_O \delta }\right]\xi _R^{up} \left( y \right) = 0 \, ,
\end{equation}
\begin{equation}\label{equ:eomTTrd0}
\partial _y \xi _R^{up} \left( y \right) + \left[ {f_S h_S \left( y \right) -
\left({ f_T h_T \left( y \right) -2f'_T h'_T \left( y \right) } \right) + f_S
v_S - 2f_O \delta }\right]\xi _R^{up} \left( y \right) = 0 \, ,
\end{equation}
\end{subequations}
where $\xi _R^{up}$ and $\xi _R^{down}$ refer to the right-handed zero-mode wave
functions of $\Psi_1$ (i.e., those of $d$, $D$, $l_{d}$, and $\nu$) and $\Psi_2$
(i.e., those of $u$, $U$, $l_{u}$, and $l$), respectively. The doublet $\xi _L$
still refers to both normal and unconventional left-handed fermion zero-mode
wave functions, which means the left-handed fermions of $\Psi_1$. 

It can be seen, from Eqs.~(\ref{equ:eomTTru0} and \ref{equ:eomTTrd0}) that the
profiles of the right-handed zero-mode wave functions of $\Psi_1$ and $\Psi_2$,
which we denote by $\xi _R^{up}$ and $\xi _R^{down}$, are now different: a broad
wave function for $\xi _R^{down}$ and a narrow wave function for $\xi _R^{up}$.
This disparity between the profiles of the two sectors of right-handed zero
modes may become more clear in section~\ref{sec:separation}. Let us define
\begin{subequations}\label{equ:hs}
\begin{equation}
h_{sym} \left( y \right) \equiv f_S h_S \left( y \right) + \left({ f_T h_T
\left( y \right) - 2f'_T h'_T \left( y \right)} \right) \, ,
\end{equation}
\begin{equation}
h_{asym} \left( y \right) \equiv f_S h_S \left( y \right) - \left({ f_T h_T
\left( y \right) - 2f'_T h'_T \left( y \right)} \right) \, ,
\end{equation}
\end{subequations}
for future compactness of equations.
\subsection{Distinguishing normal and unconventional fermions, quarks and
leptons}\label{sec:undis}
As the geometry of the zero-mode wave functions in the extra dimension
determines the overlaps and therefore the effective Yukawa couplings, one would
like to differentiate between the zero-mode wave functions of normal and
unconventional fermions, also between those of quarks and leptons. Since these
fermions are mixed by groups $\SU(4)_\mathrm{PS}$ and $\SU(3)_H$, breaking those
symmetries along the extra dimension seems plausible.
The desired symmetry breaking can be achieved by four scalar fields, which only
develop VEV's and not kink solutions. The scalar fields are $\Sigma  = \left(
{\begin{array}{*{20}c}   {15,8,1} & ; & 1  \\ \end{array} } \right)$, $\Sigma' 
= \left({\begin{array}{*{20}c}   {15,1,1} & ; & 1  \\ \end{array} } \right)$,
$\Omega  = \left( {\begin{array}{*{20}c}   {15,8,8} & ; & 1  \\ \end{array} }
\right)$, and $\Omega' = \left( {\begin{array}{*{20}c}   {15,1,8} & ; & 1  \\
\end{array} } \right)$, with Yukawa couplings in the form
\begin{equation}
\mathcal{L}_Y  = \bar \Psi \left( {f_{\Sigma } \Sigma  + f_{\Sigma' } \Sigma'  +
f_{\Omega } \Omega  + f_{\Omega' } \Omega' } \right)\Psi ,
\end{equation}
where $f_{\Sigma }, \, f_{\Sigma'}, \, f_{\Omega}, \, f_{\Omega'}  > 0$. The
minimum energy solutions of these fields are taken as
\begin{subequations}\label{equ:sigomegvev}
\begin{equation}
\left\langle {\Sigma } \right\rangle  = \sigma \left( {\begin{array}{*{20}c}
   1 & 0 & 0 & 0  \\
   0 & 1 & 0 & 0  \\
   0 & 0 & 1 & 0  \\
   0 & 0 & 0 & { - 3}  \\

 \end{array} } \right) \otimes \left( {\begin{array}{*{20}c}
   1 & 0 & 0  \\
   0 & 1 & 0  \\
   0 & 0 & { - 2}  \\

 \end{array} } \right),
\end{equation}
\begin{equation}
\left\langle {\Sigma'} \right\rangle  = \sigma' \left( {\begin{array}{*{20}c}
   1 & 0 & 0 & 0  \\
   0 & 1 & 0 & 0  \\
   0 & 0 & 1 & 0  \\
   0 & 0 & 0 & { - 3}  \\

 \end{array} } \right),
\end{equation}
\begin{equation}
\left\langle {\Omega} \right\rangle  = \left( {\begin{array}{*{20}c}
   1 & 0 & 0 & 0  \\
   0 & 1 & 0 & 0  \\
   0 & 0 & 1 & 0  \\
   0 & 0 & 0 & { - 3}  \\

 \end{array} } \right) \otimes \left( {\begin{array}{*{20}c}
   1 & 0 & 0  \\
   0 & 1 & 0  \\
   0 & 0 & { - 2}  \\

 \end{array} } \right) \otimes \left( {\begin{array}{*{20}c}
   {\omega} & 0 & 0  \\
   0 & { - \omega} & 0  \\
   0 & 0 & 0  \\

 \end{array} } \right),
\end{equation}
\begin{equation}
\left\langle {\Omega' } \right\rangle  = \left( {\begin{array}{*{20}c}
   1 & 0 & 0 & 0  \\
   0 & 1 & 0 & 0  \\
   0 & 0 & 1 & 0  \\
   0 & 0 & 0 & { - 3}  \\

 \end{array} } \right) \otimes \left( {\begin{array}{*{20}c}
   {\omega'} & 0 & 0  \\
   0 & { - \omega'} & 0  \\
   0 & 0 & 0  \\

 \end{array} } \right).
\end{equation}
\end{subequations}

Similar to $\Phi_O$'s role in section \ref{sec:localization}, the role of these
scalar fields is to shift the positions of the zero-mode wave functions of
normal and unconventional fermions, even those of quarks and leptons along the
extra dimension. That means different left-right separations for each one of
those classes, which would indicate different overlaps and therefore effective
Yukawa couplings.

Let us start with the left-handed zero-mode wave functions. Their equation of
motion, Eq.~(\ref{equ:eomTTl0}), now splits into four different equations
\begin{subequations}\label{equ:lheom0}
\begin{equation}
\partial _y \xi _L^q \left( y \right) + \left[ {f_S h_S \left( y \right) + f_T
h_T \left( y \right) + f'_T h'_T \left( y \right)  + f_S v_S - f_O \delta  + 
X_L^q} \right]\xi _L^q \left( y \right) = 0 \, ,
\end{equation}
\begin{equation}
\partial _y \xi _L^Q \left( y \right) + \left[ {f_S h_S \left( y \right) + f_T
h_T \left( y \right) + f'_T h'_T \left( y \right)  + f_S v_S - f_O \delta +
X_L^Q} \right]\xi _L^Q \left( y \right) = 0 \, ,
\end{equation}
\begin{equation}
\partial _y \xi _L^l \left( y \right) + \left[ {f_S h_S \left( y \right) + f_T
h_T \left( y \right) + f'_T h'_T \left( y \right)  + f_S v_S - f_O \delta  -
3X_L^q} \right]\xi _L^l \left( y \right) = 0 \, ,
\end{equation}
\begin{equation}
\partial _y \xi _L^L \left( y \right) + \left[ {f_S h_S \left( y \right) + f_T
h_T \left( y \right) + f'_T h'_T \left( y \right)  + f_S v_S - f_O \delta - 3
X_L^Q} \right]\xi _L^L \left( y \right) = 0 \, ,
\end{equation}
\end{subequations}
where
\begin{subequations}\label{equ:xls}
\begin{equation}
 X_L^q = f_{\Sigma} \sigma + f_{\Sigma'} \sigma' + f_{\Omega} \omega +
f_{\Omega'} \omega' \, ,
\end{equation}
\begin{equation}
 X_L^Q = f_{\Sigma } \sigma + f_{\Sigma'} \sigma' - f_{\Omega} \omega -
f_{\Omega'} \omega' \, ,
\end{equation}
\end{subequations}

In Eqs.~(\ref{equ:lheom0}), the superscripts $q$, $l$, $Q$, and $L$, correspond
to normal quark, normal lepton, unconventional quark and unconventional lepton,
respectively.

The two equations of motion for right-handed zero-mode wave functions,
Eqs.~(\ref{equ:eomTTru0} and \ref{equ:eomTTrd0}), also split into eight
equations for those of quarks and leptons, unconventional and normal. For $\xi
_R^{up}$, we obtain
\begin{subequations}\label{equ:rheom0}
\begin{equation}
\partial _y \xi _R^{q,up} \left( y \right) + \left[ {h_{sym} \left( y \right) +
f_S v_S + 2f_O \delta + X_R^q} \right]\xi _R^q \left( y \right) = 0 \, ,
\end{equation}
\begin{equation}
\partial _y \xi _R^{Q,up} \left( y \right) + \left[ {h_{sym} \left( y \right) +
f_S v_S + 2f_O \delta + X_R^Q} \right]\xi _R^Q \left( y \right) = 0 \, ,
\end{equation}
\begin{equation}
\partial _y \xi _R^{l,up} \left( y \right) + \left[ {h_{sym} \left( y \right) +
f_S v_S + 2f_O \delta -3X_R^q} \right]\xi _R^l \left( y \right) = 0 \, ,
\end{equation}
\begin{equation}
\partial _y \xi _R^{L,up} \left( y \right) + \left[ {h_{sym} \left( y \right) +
f_S v_S + 2f_O \delta -3X_R^Q} \right]\xi _R^L \left( y \right) = 0 \, ,
\end{equation}
\end{subequations}
where
\begin{subequations}\label{equ:xrs}
\begin{equation}
 X_R^q = -2 f_{\Sigma } \sigma + f_{\Sigma'} \sigma' -2 f_{\Omega} \omega +
f_{\Omega'} \omega' \, ,
\end{equation}
\begin{equation}
 X_R^Q = -2 f_{\Sigma} \sigma + f_{\Sigma'} \sigma' +2 f_{\Omega} \omega -
f_{\Omega'} \omega' \, ,
\end{equation}
\end{subequations}
while for $\xi _R^{down}$ the same equations are valid with $h_{sym} \left( y
\right) \rightarrow h_{asym} \left( y \right)$ of Eqs.~(\ref{equ:hs}). From
Eqs.~(\ref{equ:lheom0} and \ref{equ:rheom0}), it is clear that due to strong and
horizontal symmetry breaking, each type of left- and right-handed zero mode is
localized at different point in the extra dimension. Therefore, the left-right
separations which determine the overlaps would be different for each type, as we
desired.
Although Eqs.~(\ref{equ:lheom0} and \ref{equ:rheom0}) seem to suggest that the
displacements due to strong and horizontal symmetry breakings are expressed in
terms of four parameters $X_L^q$, $X_L^Q$, $X_R^q$, and $X_R^Q$, there are only
two independent parameters involved. For example, since
\begin{subequations}\label{equ:reduction}
 \begin{equation}
  X_L^q + X_L^Q = 2f_{\Sigma} \sigma + 2f_{\Sigma'} \sigma' \, ,
 \end{equation}
 \begin{equation}
  X_R^q + X_R^Q = -4f_{\Sigma} \sigma + 2f_{\Sigma'} \sigma' \, ,
 \end{equation}
\end{subequations}
once one fixes the two coupling constants and vacua on the right hand side of
Eqs.~(\ref{equ:reduction}), the $X$'s can be expressed in terms of each other.
That means two of these $X$'s are indeed arbitrary and can be viewed as
references for the other two.

Hence, let us set $X_R^q=X_R^Q=0$ and let $X_L^q$, $X_L^Q$ be the two
independent parameters of strong and horizontal symmetry breakings, they become
\begin{subequations}\label{equ:xlsn}
\begin{equation}
 X_L^q = 3f_{\Sigma} \sigma + 3f_{\Omega} \omega \, ,
\end{equation}
\begin{equation}
 X_L^Q = 3f_{\Sigma} \sigma - 3f_{\Omega} \omega \, .
\end{equation}
\end{subequations}
At this stage and to differentiate the normal fermions from the unconventional
ones, we demand the important phenomenological constraint
\begin{equation}\label{equ:const6}
f_{\Sigma}  \sigma  =  f_{\Omega}  \omega \, .
\end{equation}

This assumption separates the zero-mode wave functions of normal and
unconventional SM-type fermions in a fashion that results in stronger left-right
overlaps for unconventional fermions and consequently higher mass scales. That
is what we expect, since the unconventional fermions have not been
experimentally detected yet. With the constraint of Eq.~(\ref{equ:const6}), the
two independent distances $X_L^q$ and $X_L^Q$ become
\begin{subequations}\label{equ:XLf}
\begin{equation} 
 X_L^q = 6f_\Omega  \omega \, ,
\end{equation}
\begin{equation}
 X_L^Q = 0 \, ,
\end{equation}
\end{subequations}
Therefore, the zero-mode wave functions of left-handed SM-type fermions satisfy
\begin{subequations}\label{equ:lheomj}
\begin{equation}
\partial _y \xi _L^q \left( y \right) + \left[ {f_S h_S \left( y \right) + f_T
h_T \left( y \right) + f'_T h'_T \left( y \right)  + f_S v_S - f_O \delta  +
6f_\Omega  \omega } \right]\xi _L^q \left( y \right) = 0 \, ,
\end{equation}
\begin{equation}\label{equ:QL0}
\partial _y \xi _L^Q \left( y \right) + \left[ {f_S h_S \left( y \right) + f_T
h_T \left( y \right) + f'_T h'_T \left( y \right)  + f_S v_S - f_O \delta }
\right]\xi _L^Q \left( y \right) = 0 \, ,
\end{equation}
\begin{equation}
\partial _y \xi _L^l \left( y \right) + \left[ {f_S h_S \left( y \right) + f_T
h_T \left( y \right) + f'_T h'_T \left( y \right)  + f_S v_S - f_O \delta  -
18f_\Omega  \omega } \right]\xi _L^l \left( y \right) = 0 \, ,
\end{equation}
\begin{equation}\label{equ:LL0}
\partial _y \xi _L^L \left( y \right) + \left[ {f_S h_S \left( y \right) + f_T
h_T \left( y \right) + f'_T h'_T \left( y \right)  + f_S v_S - f_O \delta }
\right]\xi _L^L \left( y \right) = 0 \, ,
\end{equation}
\end{subequations}
Looking at  Eqs.~(\ref{equ:QL0} and \ref{equ:LL0}), one notices that $\xi
_L^Q=\xi _L^L$. On the other hand, the zero-mode wave functions for the
right-handed SM-type fermions still obey Eqs.~(\ref{equ:eomTTru0} and
\ref{equ:eomTTrd0}).
\subsection{Simplification of numerical algorithm}
So far, we have localized SM-type fermions at different points and given
different shapes to the right-handed zero mode wave functions by symmetry
breakings along the extra dimension. The equations of motion for left- and
right-handed zero mode wave functions can be simplified considerably, for
numerical ease, however without affecting the values of left-right overlaps.
To begin with, let us assume
\begin{equation}\label{equ:const1}
f_S v_S = - 2 f_O\delta \, ,
\end{equation}	
which preserves the distance between the left- and right-handed zero modes,
however places the right-handed zero modes at the origin. As the distance and
profiles of the zero modes are the only important factors in determining the
overlaps, such assumption only simplifies numerical procedure.

On the other hand, in analogy with the idea presented in
Refs.~\cite{Hung2003a,Hung2005}, where the difference in profiles for the up and
down sectors of the right-handed zero-mode wave functions is sufficient to
describe the sizes of corresponding overlaps, we may also consider
\begin{equation}\label{equ:const2}
 f'_T h'_T \left( y \right) = - f_T h_T \left( y \right) \, ,
\end{equation}
which simplifies the left-handed zero-mode wave functions. The dissimilar
(narrow and broad) profiles of the right-handed zero-mode wave functions $\xi
_R^{up}$ and $\xi _R^{down}$ remain in place regardless of the condition of
Eq.~(\ref{equ:const2}) and since that difference in shapes is what matters (see
Fig.~\ref{fig:lroverlaps}), the numerical value of left-right overlaps will not
change. With these simplifications, the equations of motion for zero-mode wave
functions read
\begin{subequations}\label{equ:lheom}
\begin{equation}
\partial _y \xi _L^q \left( y \right) + \left[ {f_S h_S \left( y \right) - 3f_O
\delta  + 6f_\Omega  \omega } \right]\xi _L^q \left( y \right) = 0 \, ,
\end{equation}
\begin{equation}\label{equ:QL}
\partial _y \xi _L^Q \left( y \right) + \left[ {f_S h_S \left( y \right) - 3f_O
\delta } \right]\xi _L^Q \left( y \right) = 0 \, ,
\end{equation}
\begin{equation}
\partial _y \xi _L^l \left( y \right) + \left[ {f_S h_S \left( y \right)  - 3f_O
\delta  - 18f_\Omega  \omega } \right]\xi _L^l \left( y \right) = 0 \, ,
\end{equation}
\begin{equation}\label{equ:LL}
\partial _y \xi _L^L \left( y \right) + \left[ {f_S h_S \left( y \right) - 3f_O
\delta } \right]\xi _L^L \left( y \right) = 0 \, ,
\end{equation}
\end{subequations}
and
\begin{subequations}\label{equ:rheom}
\begin{equation}
\partial _y \xi _R^{up} \left( y \right) + \left[ {f_S h_S \left( y \right) +
3f_T h_T \left( y \right) } \right] \xi _R^{up} \left( y \right) = 0 \, ,
\end{equation}
\begin{equation}
\partial _y \xi _R^{down} \left( y \right) + \left[ {f_S h_S \left( y \right) -
3f_T h_T \left( y \right) } \right] \xi _R^{down} \left( y \right) = 0 \, .
\end{equation}
\end{subequations}
In mass scale calculations, we find out that the distances between localized
left-handed and right-handed zero-mode wave functions along the extra dimension
are needed. Estimating those separations is the subject of the next section.
\subsection{Left-right separations along the extra
dimension}\label{sec:separation}
The localization process of SM-type fermions involved scalar fields, with
classical kink solutions. The kink solutions, however, yet to be specified. In
order to estimate the left-right separations, we give a Gaussian shape to zero
mode wave functions. Let us consider a linear approximation for the kink
solutions, explicitly
\begin{subequations}\label{equ:linapp}
\begin{equation}
h_S \left( y \right) \approx \mu _S^2 y \, ,
\end{equation}
\begin{equation}
h_T \left( y \right) \approx \mu _T^2 y \, .
\end{equation}
\end{subequations}

In this linear approximation, the equations of motion for the right-handed
zero-mode wave functions, Eqs.~(\ref{equ:rheom}), become
\begin{subequations}\label{equ:aprheom1}
\begin{equation}
\partial _y \xi _R^{up} \left( y \right) + \left( {f_S \mu _S^2 + 3f_T \mu _T^2}
\right)y\xi _R^{up} \left( y \right) = 0 \, ,
\end{equation}
\begin{equation}
\partial _y \xi _R^{down} \left( y \right) + \left( {f_S \mu _S^2 - 3f_T \mu
_T^2} \right) y \xi _R^{down} \left( y \right) = 0 \, ,
\end{equation}
\end{subequations}
The Gaussians defined by Eqs.~(\ref{equ:aprheom1}) are clearly localized at
$y=0$, meaning
\begin{equation}\label{equ:rhlocs}
y_R^{up} =y_R^{down}= 0 \,,
\end{equation}
where $y_R^{up}$ corresponds to the location of the right-handed zero modes of
$\Psi_1$ (i.e., those of $d$, $D$, $l_{d}$, and $\nu$) and $y_R^{down}$ refers
to the location of the right-handed zero modes of $\Psi_2$ (i.e., those of $u$,
$U$, $l_{u}$, and $l$). From Eqs.~(\ref{equ:aprheom1}), one clearly sees the
shape notion of narrow $\xi _R^{up}$ and wide $\xi _R^{down}$. The locations of
the left-handed zero-mode wave functions, on the other hand, can be determined
from their differential equations, Eqs.~(\ref{equ:lheom}). Those equations, in
the linear approximation scheme, now read
\begin{subequations}\label{equ:aplheom}
\begin{equation}
\partial _y \xi _L^q \left( y \right) + \left( {f_S \mu _{S}^2 y - 3f_O \delta 
+ 6f_\Omega  \omega} \right)\xi _L^q \left( y \right) = 0 \, ,
\end{equation}
\begin{equation}
\partial _y \xi _L^Q \left( y \right) + \left( {f_S \mu _{S}^2 y - 3f_O \delta }
\right)\xi _L^Q \left( y \right) = 0 \, ,
\end{equation}
\begin{equation}\partial _y \xi _L^l \left( y \right) + \left( {f_S \mu _{S}^2 y
- 3f_O \delta  - 18f_\Omega  \omega} \right)\xi _L^l \left( y \right) = 0 \, ,
\end{equation}
\begin{equation}
\partial _y \xi _L^L \left( y \right) + \left( {f_S \mu _{S}^2 y - 3f_O \delta }
\right)\xi _L^L \left( y \right) = 0 \, .
\end{equation}
\end{subequations}
For our future convenience, let us define
\begin{equation}\label{equ:w}
  w \equiv \frac{3f_O \delta }{f_\Omega \omega} \, .
\end{equation}
With this definition, The locations of localized left-handed zero-mode wave
functions can be written as
\begin{subequations}\label{equ:lhloc}
\begin{align}
y_L^q  &= \frac{f_\Omega \omega }{f_S \mu _{S}^2}\left( {w - 6} \right)\, ,\\
y_L^Q  &= \frac{f_\Omega \omega }{f_S \mu _{S}^2} w \, ,\\
y_L^l  &= \frac{f_\Omega \omega }{f_S \mu _{S}^2} \left( {w + 18} \right) \, ,\\
y_L^L  &= \frac{f_\Omega \omega }{f_S \mu _{S}^2} w\, .
\end{align}
\end{subequations}

The superscripts on $y$'s in Eqs.~(\ref{equ:lhloc}) have the same meanings
explained in sections \ref{sec:uddis} and \ref{sec:undis} for $\xi$'s. Each
location given in Eqs.~(\ref{equ:lhloc}) is applicable to both components of the
left-handed zero-mode wave function doublet to which it refers. Since the mass
terms involve left- and right-handed fields, the relevant wave function
separations are those between the left- and right-handed ones. Using the
locations we already found, those left-right separations can be computed easily.
They are
\begin{itemize}
\item
For normal quarks
\begin{equation}\label{equ:qseps}
\left| {\Delta y^{q} } \right| = \left| {y_R - y_L^q } \right| = \left|
{\frac{f_\Omega \omega }{f_S \mu _{S}^2}\left( {6 - w} \right)} \right|,
\end{equation}
\item
For normal leptons
\begin{equation}\label{equ:lseps}
\left| {\Delta y^{l} } \right| = \left| {y_R - y_L^l } \right| = \left|
{\frac{f_\Omega \omega }{f_S \mu _{S}^2} \left( {w + 18} \right)} \right|,
\end{equation}
\item
For unconventional quarks and leptons
\begin{equation}\label{equ:QLseps}
\left| {\Delta y^{L} } \right| = \left| {\Delta y^{Q} } \right| = \left| {y_R  -
y_L^L } \right| = \left| {y_R  - y_L^Q } \right| = \left| {\frac{f_\Omega \omega
}{f_S \mu _{S}^2} w} \right|.
\end{equation}
\end{itemize}

In these left-right separations, $\Delta y$'s refer to both up and down sectors
of each flavor doublet and $y_R = y_R^{up} = y_R^{down}$. The identical
left-right separations of unconventional quarks and leptons,
Eq.~(\ref{equ:QLseps}), imply similar mass scales. Obviously, the magnitude of
such mass scale can be large and remains to be explored. Alternatively, and with
the help of Eqs.~(\ref{equ:qseps} and \ref{equ:lseps}), the relation between the
wave function separations of quarks and leptons can be found, i.e.,
\begin{equation}\label{equ:lqsep}
\left| {\Delta y^l } \right| = 3\left| {\frac{{{w \mathord{\left/
 {\vphantom {w {18}}} \right.
 \kern-\nulldelimiterspace} {18}} + 1}}
{{1 - {w \mathord{\left/
 {\vphantom {w 6}} \right.
 \kern-\nulldelimiterspace} 6}}}} \right|\left| {\Delta y^q } \right|
\end{equation}
We can also find relationship between the left-right separations of
unconventional fermions and ordinary quarks, i.e.,
\begin{equation}\label{equ:unqsep}
\left| {\Delta y^{Q/L} } \right| = \left|{\frac{w}{6 - w}}\right| \left|{\Delta
y^{q} } \right| = \frac{1}{4}\left( {\frac{\left| {\Delta y^{l} }
\right|}{\left| {\Delta y^{q} } \right|} - 3} \right)\left| {\Delta y^{q} }
\right|.
\end{equation}

Since the left-right separations are determinant factors in mass scale
computations, Eq.~(\ref{equ:lqsep}) implies relationship between the mass scales
of ordinary quarks and leptons in one generation, as fixing one would restrain
the other.

This can also be extended to unconventional fermions, as Eq.~({\ref
{equ:unqsep}) relates the left-right separations of unconventional fermions to
those of ordinary fermions. Thus the masses of unconventional fermions cannot
just be heavy enough to escape detection; they must yield meaningful masses for
ordinary fermions, as the known physics is concerned.

We have then arrived at a point where the masses of unconventional fermions not
only should comply with the existent experimental check on ordinary quarks and
charged lepton's masses but they could in principle restrain the mass scales for
the neutrino sector of ordinary leptons, as their left-right separations in the
extra dimension restrain the left-right separations of ordinary leptons
including those of neutrinos.

Let us now discuss the possibility of having $\delta  = 0$, which we left aside
in section~\ref{sec:localization}. Obviously, $\delta  = 0$ corresponds to $w =
0$, which would mean $\left| {\Delta y^{l} } \right| = 3\left| {\Delta y^{q} }
\right|$ and $ \left| {\Delta y^{Q/L} } \right| = 0$. Phenomenologically, we
prefer $\delta  \ne 0$ for the reason that will be clear when we give numerical
results for the mass scales. With a minimum at $w = 0$, Eq.~(\ref{equ:lqsep})
can be also written as $\left| {\Delta y^l } \right| \ge 3\left| {\Delta y^q }
\right|$, which clearly indicates that lepton's wave function overlaps can be
potentially weaker than those of quarks. This seems plausible knowing the
profound differences between the mass scales of quarks and leptons.
\section{Return to four dimensions: The mass scales}
Speaking of SM-type fermion mass terms and mass scales implies that the gauge
symmetry is reduced to that of the SM and is going to break further down by the
SM Higgs vacuum. 
A complete analysis of the gauge symmetry breakdown of the model is rather
lengthy and is not consistent with the flow of the paper at this point. However,
the necessary scalar fields for the gauge symmetry breakdown and the mixing of
charged gauge bosons are crucial to our analyses in sections~\ref{sec:EWP} and
\ref{sec:decays}. For that reason and completeness, a detailed gauge symmetry
breakdown is given in Appendix~\ref{app:gsb}.

As we only concentrate on the mass scales, we therefore will not discuss issues
such as the fermion mixings in the mass matrix~\cite{Split}. We follow the mass
scale calculations with some rough numerical analysis.
\subsection{Effective Yukawa couplings and the mass scales}
Dirac mass terms for chiral fermions involve couplings of left-handed and
right-handed fields with a Higgs field, which acquires VEV and breaks the SM
symmetry as well. The minimal SM symmetry breakdown of our model can occur
through a Higgs multiplet transforming as $\Theta = \left( 1,8,8 \right)$. The
decomposition of $\Theta$'s $\SU(3)_L$ octet in terms of $\SU(2)_L \otimes
\mathrm{U}(1)_Y$ multiplets or quantum numbers
\begin{equation}
\left[ 8 \right]_{\SU\left( 3 \right)_L}  = \left( 3,0 \right) \oplus \left( 2,1
\right) \oplus \left( 2,-1 \right) \oplus \left( 1,0 \right),
\end{equation}
shows that $\Theta$ indeed possesses a SM Higgs field, which we denote by
$H=\left( {2,1} \right)$. Thus, $\Theta$ can break the SM symmetry and give mass
to chiral fermions by developing a VEV in $H$. The Yukawa couplings between the
left- and right-handed SM-type fermions can be written in the form
\begin{equation}\label{equ:mcouplings}
\mathcal{L}_{\mathrm{mass}}  = \kappa _1 \Psi_1^T \Theta C \Psi_2^\ast  + \kappa
_2 \Psi_1^T \tilde \Theta  C \Psi_1^\ast  + h.c.\, .
\end{equation}

In the above couplings, $\kappa_1$, $\kappa_2$ can be different in general,
$\tilde \Theta= i \hat \lambda_{2L} \Theta^*$, and $C = i \gamma^{2}
\gamma^{0}$. The mass terms in Eq.~(\ref{equ:mcouplings}) seem compact but they
can be expanded very easily. For example, they yield
\begin{equation}\label{equ:qmass}
\mathcal{L}_{\mathrm{mass}}^{q} =\kappa _1 \frac{v}{{\sqrt 2 }}\bar u_L u_R  -
\kappa _2 \frac{v}{{\sqrt 2 }}\bar d_L d_R  + h.c. \;,
\end{equation}
for normal quarks. These mass expressions have been worked out for transparency
in Appendix~\ref{app:gsb}. Similar expressions for other SM-type fermions can be
obtained easily. 

We assume a delocalized Higgs field along the extra dimension and use its lowest
KK mode, which entirely depends on 4D coordinates. This means that the zero mode
of the Higgs field is independent of $y$, and can be written as, e.g., $H^0
\left( {x,y} \right) = K\phi \left( x \right)$. The zero mode of the SM Higgs
field $\phi$, then obtains VEV in the usual form $\bigl( \begin{smallmatrix} 0
\\ {{v \mathord{\left/ {\vphantom {v {\sqrt 2 }}} \right.
\kern-\nulldelimiterspace} {\sqrt 2 }}} \end{smallmatrix} \bigl)$.

The mass terms involve Yukawa couplings determining the magnitude of each mass
term. In our model and in four dimensional space, those couplings can be viewed
as ``effective'' Yukawa couplings whose strengths are determined by the geometry
of the zero-mode wave functions in the extra dimension. The reduction to 4D
space is simply done by integrating the extra dimension out, and that is how the
couplings in mass terms become ``effective'' 4D Yukawa couplings.

Mass scales can be computed from the mass terms in Eq.~(\ref{equ:mcouplings}).
To proceed, we define dimensionless couplings
\begin{equation}
g_{Y1,2}  = \kappa_{1,2} K \, .
\end{equation}
The relationship between the mass scales and the mass matrix is given by
\begin{equation}
  \mathcal{M} = \Lambda M \, ,
\end{equation}
where $M$ is a dimensionless matrix, whose form depends on the model for fermion
masses. In our case, we may write explicitly
\begin{equation}
\mathcal{M}_{u,d,\nu ,l,U,D,l_{u} ,l_{d} }  = \Lambda _{u,d,\nu ,l,U,D,l_{u}
,l_{d} } M_{u,d,\nu ,l,U,D,l_{u} ,l_{d} } \; ,
\end{equation}
where $\Lambda _{u,d,\nu ,l,U,D,l_{u} ,l_{d} }$ are the mass scales of interest
and the subscripts refer to the SM-type fermions of the theory. The mass scales
in 4D space are proportional to the 4D effective Yukawa couplings, which in turn
are proportional to the overlaps between the relevant left- and right-handed
zero-mode wave functions in the extra dimension. Therefore, they simply are
\begin{align}\label{equ:mscales}
 \Lambda _u  &= \frac{v}{\sqrt 2} g_{Y1} \int_0^L {dy\xi _L^q } \left( y
\right)\xi _R^{down} \left( y \right)
 \qquad ;  & 
 \Lambda _d  &= \frac{v}{\sqrt 2} g_{Y2} \int_0^L {dy\xi _L^q } \left( y
\right)\xi _R^{up} \left( y \right),
\\
 \Lambda _\nu  &= \frac{v}{\sqrt 2} g_{Y2} \int_0^L {dy\xi _L^l } \left( y
\right)\xi _R^{up} \left( y \right)
 \qquad \; \; \; ;  &
 \Lambda _l  &= \frac{v}{\sqrt 2} g_{Y1} \int_0^L {dy\xi _L^l } \left( y
\right)\xi _R^{down} \left( y \right),
\\
 \Lambda _U  &= \frac{v}{\sqrt 2} g_{Y1} \int_0^L {dy\xi _L^Q } \left( y
\right)\xi _R^{down} \left( y \right) 
 \qquad ;  &
 \Lambda _D  &= \frac{v}{\sqrt 2} g_{Y2} \int_0^L {dy\xi _L^Q } \left( y
\right)\xi _R^{up} \left( y \right),
 \\
 \Lambda _{l_{u}} &= \frac{v}{\sqrt 2} g_{Y1} \int_0^L {dy\xi _L^L } \left( y
\right)\xi _R^{down} \left( y \right) 
 \qquad ;  &
 \Lambda _{l_{d}} &= \frac{v}{\sqrt 2} g_{Y2} \int_0^L {dy\xi _L^L } \left( y
\right)\xi _R^{up} \left( y \right). 
 \end{align}

Note that the left-handed $\xi _L $'s appearing in Eqs.~(\ref{equ:mscales}) are
no longer doublets, but the relevant components of those doublets. The fact is
that the geometry of each flavor component is the same as that assigned to the
corresponding doublet. As we are only concerned with the geometry of zero-mode
wave functions, we do not introduce new notation for the flavor components, as
if they were the relevant doublets. There are two possibilities that one can
explore: $g_{Y1}  = g_{Y2} $ and $g_{Y1}  \ne g_{Y2} $. The relationships
between the mass scales may depend on those choices.
\begin{enumerate}
\item
$g_{Y1}  = g_{Y2}$ : One can write all sorts of ratios, which would only depend
on wave function overlaps. For example, we can write ratios relating mass scales
of two sectors of one family, or ratios involving mass scales from different
families. Some of those ratios are
\begin{subequations}\label{equ:ratios1}
\begin{equation}\label{equ:du}
\frac{{\Lambda _d }}
{{\Lambda _u }} = \frac{{\int_0^L {dy\xi _L^q } \left( y \right)\xi _R^{up}
\left( y \right)}}
{{\int_0^L {dy\xi _L^q } \left( y \right)\xi _R^{down} \left( y \right)}} \, ,
\end{equation}

\begin{equation}\label{equ:nl}
\frac{{\Lambda _\nu  }}
{{\Lambda _l }} = \frac{{\int_0^L {dy\xi _L^l } \left( y \right)\xi _R^{up}
\left( y \right)}}
{{\int_0^L {dy\xi _L^l } \left( y \right)\xi _R^{down} \left( y \right)}} \, ,
\end{equation}

\begin{equation}\label{equ:DU}
\frac{{\Lambda _D }}
{{\Lambda _U }} = \frac{{\int_0^L {dy\xi _L^Q } \left( y \right)\xi _R^{up}
\left( y \right)}}
{{\int_0^L {dy\xi _L^Q } \left( y \right)\xi _R^{down} \left( y \right)}} \, ,
\end{equation}

\begin{equation}\label{equ:nu}
\frac{{\Lambda _\nu  }}
{{\Lambda _u }} = \frac{{\int_0^L {dy\xi _L^l } \left( y \right)\xi _R^{up}
\left( y \right)}}
{{\int_0^L {dy\xi _L^q } \left( y \right)\xi _R^{down} \left( y \right)}} \, ,
\end{equation}

\begin{equation}\label{equ:lu}
\frac{{\Lambda _l }}
{{\Lambda _u }} = \frac{{\int_0^L {dy\xi _L^l } \left( y \right)\xi _R^{down}
\left( y \right)}}
{{\int_0^L {dy\xi _L^q } \left( y \right)\xi _R^{down} \left( y \right)}} \, ,
\end{equation}

\begin{equation}\label{equ:Du}
\frac{{\Lambda _D }}
{{\Lambda _u }} = \frac{{\int_0^L {dy\xi _L^Q } \left( y \right)\xi _R^{up}
\left( y \right)}}
{{\int_0^L {dy\xi _L^q } \left( y \right)\xi _R^{down} \left( y \right)}} \, ,
\end{equation}

\begin{equation}\label{equ:Ud}
\frac{{\Lambda _U }}
{{\Lambda _d }} = \frac{{\int_0^L {dy\xi _L^Q } \left( y \right)\xi _R^{down}
\left( y \right)}}
{{\int_0^L {dy\xi _L^q } \left( y \right)\xi _R^{up} \left( y \right)}}\, ,
\end{equation}

\begin{equation}\label{equ:nU}
\frac{{\Lambda _\nu  }}
{{\Lambda _U }} = \frac{{\int_0^L {dy\xi _L^l } \left( y \right)\xi _R^{up}
\left( y \right)}}
{{\int_0^L {dy\xi _L^Q } \left( y \right)\xi _R^{down} \left( y \right)}} \, ,
\end{equation}

\begin{equation}\label{equ:lD}
\frac{{\Lambda _l }}
{{\Lambda _D }} = \frac{{\int_0^L {dy\xi _L^l } \left( y \right)\xi _R^{down}
\left( y \right)}}
{{\int_0^L {dy\xi _L^Q } \left( y \right)\xi _R^{up} \left( y \right)}} \, .
\end{equation}
\end{subequations}
One notices that the ratios involving unconventional leptons are identical to
those of unconventional quarks, since they share the same left-handed wave
functions and the same separations.
\item
$g_{Y1}  \ne g_{Y2}$:
In this case, we may still find some ratios, depending only on wave function
overlaps. They are
\begin{subequations}\label{equ:ratios2}
\begin{equation}\label{equ:lu2}
\frac{{\Lambda _l }}
{{\Lambda _u }} = \frac{{\int_0^L {dy\xi _L^l } \left( y \right)\xi _R^{down}
\left( y \right)}}
{{\int_0^L {dy\xi _L^q } \left( y \right)\xi _R^{down} \left( y \right)}} \, ,
\end{equation}

\begin{equation}
\frac{{\Lambda _\nu }}
{{\Lambda _d }} = \frac{{\int_0^L {dy\xi _L^l } \left( y \right)\xi _R^{up}
\left( y \right)}}
{{\int_0^L {dy\xi _L^q } \left( y \right)\xi _R^{up} \left( y \right)}} \, ,
\end{equation}

\begin{equation}
\frac{{\Lambda _U }}
{{\Lambda _u }} = \frac{{\int_0^L {dy\xi _L^Q } \left( y \right)\xi _R^{down}
\left( y \right)}}
{{\int_0^L {dy\xi _L^q } \left( y \right)\xi _R^{down} \left( y \right)}} \, ,
\end{equation}

\begin{equation}
\frac{{\Lambda _D }}
{{\Lambda _d }} = \frac{{\int_0^L {dy\xi _L^Q } \left( y \right)\xi _R^{up}
\left( y \right)}}
{{\int_0^L {dy\xi _L^q } \left( y \right)\xi _R^{up} \left( y \right)}} \, ,
\end{equation}

\begin{equation}
\frac{{\Lambda _l }}
{{\Lambda _U }} = \frac{{\int_0^L {dy\xi _L^l } \left( y \right)\xi _R^{down}
\left( y \right)}}
{{\int_0^L {dy\xi _L^Q } \left( y \right)\xi _R^{down} \left( y \right)}} \, ,
\end{equation}

\begin{equation}
\frac{{\Lambda _\nu }}
{{\Lambda _D }} = \frac{{\int_0^L {dy\xi _L^l } \left( y \right)\xi _R^{up}
\left( y \right)}}
{{\int_0^L {dy\xi _L^Q } \left( y \right)\xi _R^{up} \left( y \right)}} \, .
\end{equation}
\end{subequations}
\end{enumerate}

So far, we have been able to find relationships between the mass scales of the
fermions of interest. However, there exist parameters in these relations, which
need to be determined in order to give numerical results. In the next section,
we attempt to express mass scales in terms of the mass scales of up- and
down-sectors of ordinary quarks and charged-sector of ordinary leptons by fixing
some of the parameters and deriving others.
\subsection{Numerical analysis}
To obtain numerical values for the mass scales of neutrino and unconventional
fermions, we first need to specify the analytical expressions for the zero mode
wave functions involved in the overlap integrals. To start, let us consider the
general case of $w \ne 0$. 
For the left- and right-handed zero-mode wave functions, we employ the same
expressions as those used in Refs.~\cite{Hung2003a, Hung2005, Georgi2001}. The
left-handed zero mode wave functions are
\begin{equation}\label{equ:lhexp}
\xi _L^i \left( y \right) = N^i_L \exp \Big[ { - C_S \ln \Big( {\cosh \left(
{\mu _S \left( {y - y_i } \right)} \right)} \Big)} \Big],
\end{equation}
where $i=q,l,Q,L$, and $N_L $'s are normalization factors, $C_{S} = f_{S}
\sqrt{{2 \mathord{\left/ {\vphantom {2 {\lambda _{S} }}} \right.
\kern-\nulldelimiterspace} {\lambda _{S} }}}$, and $y_i$'s are the positions of
the left-handed zero modes along the extra dimension. The right-handed zero mode
wave functions, which are slightly more complicated, are expressed in the form
\begin{subequations}\label{equ:rhexp}
\begin{equation}
\xi _R^{up} \left( y \right) = N_R^{up} \exp \Big[ { - \Big( {C_S \ln \left(
{\cosh \mu _S y} \right) + C_T \ln \left( {\cosh \mu _T y} \right)} \Big)}
\Big],
\end{equation}
and
\begin{equation}
\xi _R^{down} \left( y \right) = N_R^{down} \exp \Big[ { - \Big( {C_S \ln \left(
{\cosh \mu _S y} \right) - C_T \ln \left( {\cosh \mu _T y} \right)} \Big)}
\Big],
\end{equation}
\end{subequations}
where $N_R $'s are normalization factors and $C_T = 3f_T \sqrt{{2
\mathord{\left/ {\vphantom {2 {\lambda _{T} }}} \right.
\kern-\nulldelimiterspace} {\lambda _{T} }}}$. Note that $C_S $ and $C_T $
contain factors from both the Yukawa coupling with fermions  $f_{S,T} $, and the
scalar field self-interaction $\lambda _{S,T} $.

To calculate the mass scales of interest, we note that the mass scale ratios of
ordinary quarks and leptons may be estimated from the experimental values for
mass eigenstates (we shall elucidate this issue momentarily). Therefore, we can
use the \textit{estimated} value of ${{\Lambda _d } \mathord{\left/ {\vphantom
{{\Lambda _d } {\Lambda _u }}} \right. \kern-\nulldelimiterspace} {\Lambda _u
}}$ in conjunction with the relevant mass scale ratio of Eq.~(\ref{equ:du}) to
obtain the quark left-right separation, $\Delta y^q $. In addition, we may use
the \textit{estimated} value of ${{\Lambda _l } \mathord{\left/ {\vphantom
{{\Lambda _l } {\Lambda _d }}} \right. \kern-\nulldelimiterspace} {\Lambda _d
}}$ and the ratio in Eq.~(\ref{equ:lu}) to find the lepton left-right
separation, $\Delta y^l $, which in turn can determine the mass scale of Dirac
neutrino, say using Eqs.~(\ref{equ:nl},\ref{equ:nu}). Once  $\Delta y^q $ and
$\Delta y^l $ are known, we can find the unconventional fermion left-right
separation, using Eq.~(\ref{equ:unqsep}), in the linear approximation scheme of
section~\ref{sec:separation}. Consequently, we can estimate unconventional
fermion mass scales, say using Eqs.~(\ref{equ:Du},\ref{equ:Ud},\ref{equ:lD}).

The outlined numerical method makes use of the ratios of
Eqs.~(\ref{equ:ratios1}), which are obtained assuming that $g_{Y1} = g_{Y2}$. It
turns out that $g_{Y1} \ne g_{Y2}$ case gives the same mass scales, however with
a bit different numerical approach. We shall explain this at the closing of this
section.

To evaluate the mass scales, we need to fix some of the parameters in zero-mode
wave function expressions, Eqs.~(\ref{equ:lhexp} and \ref{equ:rhexp}), and vary
some. Since the difference between $C_S \ln \left( {\cosh \mu _S y} \right)$ and
$C_T \ln \left( {\cosh \mu _T y} \right)$, in Eqs.~(\ref{equ:rhexp}), is what
matters, we choose $C_S  = C_T  = 1$, set $\mu _S  = 1$  (in some units) and let
$\mu _T $ vary. Therefore, for a given $\mu _T $ we may find the quark and
lepton left-right separations that satisfy the phenomenological constraints and
use those separations to estimate the Dirac neutrino and unconventional fermion
mass scales. Technically speaking, varying $\mu_T$ means varying the width of
the right-handed zero-mode wave functions; therefore we look for
width-separation combinations that would satisfy the estimated mass scale
ratios.

To estimate the phenomenological constraints on the mass scale ratios, we need
to make an assumption concerning the nature of mass matrices of up- and
down-quark sectors and charged-lepton sector of ordinary fermions. The mass
matrix $\mathcal{M}$, is related to the mass scale $\Lambda$, through the
expression
\begin{equation}
  \mathcal{M} = \Lambda M \, ,
\end{equation}
where $M$ is a dimensionless matrix. Obviously, mass scale $\Lambda$ is a common
factor in the mass matrix and $M$, which determines the flavor mixings and
masses, is to be specified by the model describing the mass issues. We shall not
engage in discussing mass matrices here, as the subject itself is rich and well
beyond the scope of this work. Nevertheless, to relate the mass scales of
up-quark, down-quark and charged-lepton sectors of ordinary fermions to
experimentally measured mass eigenvalues, a general case could be considered,
where the relevant mass scales lie within two bounds, namely
\begin{equation}\label{equ:scalebounds}
\begin{gathered}
  \frac{{m_t }}
{3} \leqslant \Lambda _u  \leqslant m_t , \hfill \\
  \frac{{m_b }}
{3} \leqslant \Lambda _d  \leqslant m_b , \hfill \\
  \frac{{m_\tau  }}
{3} \leqslant \Lambda _l  \leqslant m_\tau , \hfill \\
\end{gathered} 
\end{equation}
where $m_t$, $m_b$, and $m_{\tau}$ are the largest eigenvalues of up-quark,
down-quark and charged-lepton mass matrices, respectively. The lower bounds
correspond to pure democratic mass matrices \cite{Dem}, which are impractical
since they cannot replicate proper mass spectrum and CKM matrix. The upper
bounds, on the other hand, refer to ``highly hierarchical'' mass
matrices\footnote{There have been a lot of works done on hierarchical mass
matrices, which span from phenomenological to superstring theory inspired
models. See Ref.~\cite{Fritzsch2000} for a mini review and references therein.},
where the largest eigenvalues are approximately equal to the mass scales, i.e.,
\begin{equation}\label{equ:eigenmass}
\begin{gathered}
  \Lambda_u \approx m_t \,, \hfill \\
  \Lambda_d \approx m_b \, , \hfill \\
  \Lambda_l \approx m_{\tau} \, . \hfill \\ 
\end{gathered} 
\end{equation}

To carry out the mass scale calculations, we consider this highly hierarchical
scheme. We will come back to Eqs.~(\ref{equ:scalebounds}) and the mass scales
within the two bounds, which do not correspond to pure democratic mass matrices.
\begin{table}
\caption{\label{tab:parameters} Values of $\mu_T$, $\Delta y^l$, $\Delta y^q$,
and $\Delta y^{L,Q}$ that give meaningfull results in accordance with the
phenomenological constraints of Eqs.~(\ref{equ:expconst}). Each set of values is
labeled with a roman letter.}
\begin{ruledtabular}
\begin{tabular}{ccccc}
&	$\mu_T$	&	$\left|{\Delta y^l}\right|$&	$\left|{\Delta
y^q}\right|$&	$\left|{\Delta y^{Q/L}}\right|$	\\
\hline
a	&	0.81	&	31.360	&	6.940	&	2.635	\\
b	&	0.80	&	30.200	&	7.000	&	2.300	\\
c	&	0.79	&	29.170	&	7.070	&	1.990	\\
d	&	0.75	&	24.715	&	7.530	&	0.531	\\
e	&	0.73	&	24.115	&	7.690	&	0.211	\\
f	&	0.70	&	23.285	&	7.815	&	0.040	\\
\end{tabular}
\end{ruledtabular}
\end{table}

We employ the masses of top and bottom quarks and tau lepton at $M_Z$ for $m_t$,
$m_b$, and $m_{\tau}$, and to simplify our numerical computations ignore any
running between $M_Z$ and the early unification scale. That seems plausible as
the early unification scale is not much higher than $M_Z$, meaning that there
would not be much of a ``running.'' We use 
\begin{equation}
\begin{gathered}
  m_t \left( {M_Z } \right) = 181{\text{ GeV,}} \hfill \\
  m_b \left( {M_Z } \right) = 3{\text{ GeV,}} \hfill \\
  m_\tau  \left( {M_Z } \right) = 1.747{\text{ GeV}}{\text{.}} \hfill \\ 
\end{gathered} 
\end{equation}
Therefore, the phenomenological constraints on the mass scale ratios can be
written as
\begin{subequations}\label{equ:expconst}
\begin{equation}\label{equ:duratio}
\frac{{\Lambda _d }}
{{\Lambda _u }} \approx \frac{{m_b \left( {M_Z } \right)}}
{{m_t \left( {M_Z } \right)}} \approx 0.0166,
\end{equation}
and 
\begin{equation}\label{equ:luratio}
\frac{{\Lambda _l }}
{{\Lambda _u }} \approx \frac{{m_\tau  \left( {M_Z } \right)}}
{{m_t \left( {M_Z } \right)}} \approx 0.00965.
\end{equation}
\end{subequations}

With the mass scale ratios of Eqs.~(\ref{equ:expconst}), the left-right
separations of normal quarks and leptons are at grab, which then lead us to the
left-right separation for unconventional fermions and finally the mass scales
for neutrino and unconventional fermions.

It turns out that there are a few width-separation combinations that satisfy the
phenomenological constraints. Consequently, there will be a few sets of mass
scales for neutrino and unconventional fermions which in turn imply a
relationship between the masses. The possible values of $\mu_T$, $\Delta y^l$,
$\Delta y^q$, and $\Delta y^{L,Q}$, which satisfy the phenomenological
conditions are listed in Table~\ref{tab:parameters} for completeness. The
left-right separations of Table~\ref{tab:parameters} demonstrate a hierarchy in
the form $\Delta y^l > \Delta y^q > \Delta y^{L,Q}$, which means a hierarchy in
overlaps where the largest is that of  unconventional fermions and the smallest
belongs to ordinary leptons.

With the values of Table~\ref{tab:parameters}, the left- and right-handed zero
mode wave functions are specified and finally the mass scales of Dirac neutrino
and unconventional fermions for each allowed case can be determined. Those mass
scales are listed in Table~\ref{tab:massscales} for each allowed set of
parameters.
\begin{table}
\caption{\label{tab:massscales} Predicted mass scales for Dirac neutrino
$\Lambda_\nu$, and unconventional quarks $\Lambda_U$, $\Lambda_D$, and leptons
$\Lambda_{l_{u}}$, $\Lambda_{l_{d}}$. Note that $\Lambda_U \approx
\Lambda_{l_{u}}$ and  $\Lambda_D \approx \Lambda_{l_{d}}$ and that each label in
the left column refers to the corresponding set of parameters in
Table~\ref{tab:parameters}.}
\begin{ruledtabular}
\begin{tabular}{cccc}
&$\Lambda_\nu \,\, \text{(eV)} \approx$  & $\Lambda_U \,\, \text{(GeV)} \approx$
& $\Lambda_D  \,\, \text{(GeV)} \approx$ \\
\hline
a	&	0.065	&	406	&	181 \\
b	&	0.23	&	456	&	252 \\
c	&	0.67	&	513	&	336 \\
d	&	23	&	802	&	791 \\
e	&	87	&	988	&	1053 \\
f	&	486	&	1321	&	1435 \\
\end{tabular}
\end{ruledtabular}
\end{table}

Looking at Table~\ref{tab:massscales}, it is obvious that the mass scales of
neutrino and unconventional fermions increase monotonically together. One can
argue that there is a correlation between the masses of neutrino and those of
unconventional fermions, such that the mass of one can set a bound on the mass
of the other. For instance, we could start with a mass scale for unconventional
fermions and find the corresponding left-right separation, which together with
quark left-right separation would determine that of lepton and therefore the
mass scale of neutrino. Such relationship can also be seen, however naively, by
looking at Eq.~(\ref{equ:nU}) where the mass scales of one sector can set a
bound on another.

The neutrino oscillation data provide mass differences between the neutrinos of
different families. The most recent data~\cite{PDBook} on neutrino mass
differences indicate $\Delta m_{21}^2 = \left( {8.0_{ - 0.3}^{ + 0.4} } \right)
\times 10^{ - 5} {\text{ eV}}^2 $ and $\left| {\Delta m_{32}^2 } \right| =
1.9\text{ to }3.0 \times 10^{ - 3} {\text{ eV}}^2 $.

The neutrino mass scales of of Table~\ref{tab:massscales} increase with those of
unconventional fermions. One could see two distinct possibilities by looking at
$\Delta m_{21}^2$ and $\left| {\Delta m_{32}^2 } \right|$, namely:
\begin{enumerate}
\item 
For the lightest unconventional fermions, i.e., mass scales not smaller than 180
GeV, the neutrino sector is very light, about 0.065 eV. That corresponds to
either quasi-degenerate or hierarchical mass matrix for neutrinos.
\item 
For heavier unconventional fermions, i.e., mass scales between 250 and 500 GeV,
the neutrino sector is light, ranging between 0.2 and 0.7 eV. In this case the
neutrino mass matrix ought to be quasi-degenerate in order to satisfy the
neutrino oscillation data.
\end{enumerate}

The mass scales in Table~\ref{tab:massscales} are similar to those obtained in
Ref.~\cite{Hung2005} based on $\mathrm{PUT}_1$ scenario. This similarity is
mainly due to the common strong $\SU(4)_{\mathrm{PS}}$ group. The breaking of
this symmetry in the extra dimension yields similar relations between the
left-right distances of quarks and leptons. Nevertheless, the actual masses for
unconventional fermions in each scenario can be different in principal, as the
mass matrices can be different.

In obtaining the mass scales of Table~\ref{tab:massscales}, we assumed highly
hierarchical mass matrices for up-quark, down-quark and charged-lepton sectors
of ordinary fermions, which resulted in phenomenological conditions of
Eqs.~(\ref {equ:expconst}). However, between the two bounds defined by
Eqs.~(\ref{equ:scalebounds}), the corresponding mass matrices are no longer
purely democratic \cite{NonDem}. There are models (e.g., those in
Refs.~\cite{HungSeco}) where the mass matrices deviate enough from pure
democratic case that can generate suitable mass spectrum and CKM matrix. In
those models, the mass scales of interest can be taken nearly as low as half of
the largest eigenvalues. In such regime, we end up with mass scales at least
half of those given in Table~\ref{tab:massscales}, which make the degeneracy of
neutrino sector for heavier  (250~-~500 GeV) unconventional fermions seem less
reflective.

Let us talk about the possible mass scales that can be computed when $w=0$.
Although such case simplifies the relation between the left-right separations of
ordinary quarks and leptons, it yields ordinary charged lepton with mass scale
in order of 11 GeV. That alone is sufficient to dismiss the $w=0$ case, as
Ref.~\cite{Hung2005} also suggests.

Now that we know $w=0$ leads to unphysical mass scales, we may explain the
numerical method for computing mass scales if we were to use the ratios in
Eqs.~(\ref{equ:ratios2}) when $g_{Y1} \ne g_{Y2}$, which would only make sense
if $w\ne0$. Similar to $g_{Y1} = g_{Y2}$ case, the ratios of
Eqs.~(\ref{equ:ratios2}) should also comply with the corresponding
\textit{estimated} ratios. That, nevertheless, requires adjusting the left-right
separations of quarks and leptons accordingly (e.g., to fix the ratio in
Eq.~(\ref{equ:lu2})), which means varying one more parameter and that is $w$.
Once $\Delta y^l$ and $\Delta y^q$ are known, $\Delta y^{Q/L}$ can be estimated
and then the mass scales of interest can be evaluated.

A few remarks are in order here. If the unconventional fermions are very heavy,
the neutrinos are quasi-degenerate. That would imply that the mixing angles in
PMNS matrix~\cite{PMNS} will be mainly determined by the angles of the charged
lepton sector. If the unconventional fermions are lighter, it would imply that
the mixing angles could come from both charged lepton and neutrino sectors,
since the neutrino sector could also be hierarchical in this case.
\section{Constraints from precision electroweak measurements} \label{sec:EWP}
The oblique corrections to the SM are best presented in terms of the so-called
electroweak oblique parameters $S$, $T$, and $U$ \cite{Peskin}. They are
primarily defined for sorts of new physics that have no or insignificant direct
couplings to the SM particle content and have mass scales larger than $M_Z$.

Of these parameters, $U$ plays a relatively minor role and is not linked to any
precision measurement but that of $M_W$. The other two, however, are strongly
correlated and important in limiting the type of new physics that could couple
to the SM. To give a conceptual sense, $S$ measures the momentum dependence of
the vacuum polarization and $T$ measures the custodial isospin violation.

The new physics corrections to oblique parameters in our model come from the
SM-type unconventional fermions, and the scalars of the theory, since the
vector-like fermions decouple for large vector-like masses (\textit{decoupling
theorem}).

The experimental values of oblique parameters refer to the allowed contributions
from new physics with respect to the SM reference point. The latest experimental
values of oblique parameters are \cite{PDBook}
\begin{subequations}\label{equ:expoblique}
 \begin{equation}\label{equ:expS}
   S =  - 0.13 \pm 0.10\left( { - 0.08} \right),
 \end{equation}
 \begin{equation}\label{equ:expT}
   T =  - 0.13 \pm 0.11\left( { - 0.09} \right),
 \end{equation}
 \begin{equation}\label{equ:expU}
  U = 0.20 \pm 0.12\left( { + 0.01} \right),
 \end{equation} 
\end{subequations}
where the central values assume the SM Higgs mass $M_H = 117$ GeV, and the
values in parentheses show the change for $M_H = 300$ GeV.

The custodial isospin symmetry constraint presented by $T$, forbids too much of
a difference between the masses of $U$, $l_{u}$ and $D$, $l_{d}$ respectively.
This will have implications on the decay modes of unconventional fermions, as it
constrains the phase space for decays such as $U \to D + W_L$ or $D \to U + W_L$
happening in real $W_L$'s.

The $S$ parameter, on the other hand, can be estimated for our model. For
fermionic contribution to $S$, we note that the mass scales give the maximum
masses for unconventional fermions and therefore their maximum contributions to
$S$. The total $S$ from one extra generation of fermions can be estimated that
way, i.e.,
\begin{equation}\label{equ:Sfermions}
S = \frac{1}
{{6\pi }}\left[ {16x_U  + 32x_D  + 2\ln \frac{{x_U }}
{{x_D }} + \left( {4x_U  - 1} \right)G\left( {x_U } \right) + \left( {8x_D  + 1}
\right)G\left( {x_D } \right)} \right],
\end{equation}
with
\[
x_{U}  = \frac{{M_U }}
{{M_Z }} \approx \frac{{M_{l_{u} } }}
{{M_Z }} \,\,\,{\text{   and   }}\,\,\,x_{D}  = \frac{{M_D }}
{{M_Z }} \approx \frac{{M_{l_{d} } }}
{{M_Z }}\,\,.
\]

This expression obviously depends only on the masses of unconventional fermions
for which we use the calculated mass scales. The derivation of
Eq.~(\ref{equ:Sfermions}) is given in Appendix~\ref{app:Scalc}, for
completeness. The minimum and maximum fermionic contributions to the electroweak
$S$ parameter, obtained using the computed mass scales as maximum masses, are
given in Table~\ref{tab:Sferm}. The $S$ values in Table~\ref{tab:Sferm} indicate
$0.391\geq S \geq 0.195$, form one generation of unconventional fermions.
\begin{table}
\caption{\label{tab:Sferm} Minimum and maximum values of electroweak $S$
parameter from one generation of unconventional fermions for given masses.}
\begin{ruledtabular}
\begin{tabular}{ccc}
$M_U$(GeV)	&	$M_D$(GeV)	&	$S$	\\
\hline
406	&	181	&	0.391	\\
1321	&	1435	&	0.195	\\
\end{tabular}
\end{ruledtabular}
\end{table}
If we assume generational mass degeneracy among the three generations of
unconventional fermions, the total fermionic correction to $S$ can reach up to
three times those values.

Correction to electroweak $S$ parameter from scalar fields, generally, takes
negative sign~\cite{Georgi1991,Dugan1991}. The scalar contributions to $S$ come
from the Higgs multiplets responsible for the gauge symmetry breakdown (see
Appendix~\ref{app:gsb}).
The gauge symmetry breaking of our model involves many scalar multiplets.
However, only those with $\SU(2)_L$ quantum number can contribute to $S$. Those
are $\Phi _L  = \left( {1,8,1} \right)$, $\Phi _H^{\left( 2 \right)}  = \left(
{4,3,3} \right)$, and $\Theta = \left( {1,8,8} \right)$. In terms of their
$\SU(2)_L$ multiplets, the scalar fields which carry $\SU(2)_L$  quantum number
consist of 9 triplets and 18 doublets. The computations of scalar corrections to
$S$, in the paradigm of Ref.~~\cite{Dugan1991}, are given in detail in 
Appendix~\ref{app:Scalc}. The $S$ parameter due to an $\SU(2)_L$ doublet with
mass $m$ and mass splitting parameter $m' \geq 0$ is
\begin{equation}\label{equ:Sdoublet}
S_{\mathrm{doublet}}  = \frac{1}
{\pi }\int_0^1 {dx{\text{ }}x\left( {1 - x} \right)\ln \left( {x + \zeta \left(
{1 - x} \right)} \right)} \, ,
\end{equation}
where $\zeta  = {{\left( {1 - {{3\beta ^2 } \mathord{\left/ {\vphantom {{3\beta
^2 } 2}} \right. \kern-\nulldelimiterspace} 2}} \right)} \mathord{\left/
{\vphantom {{\left( {1 - {{3\beta ^2 } \mathord{\left/ {\vphantom {{3\beta ^2 }
2}} \right. \kern-\nulldelimiterspace} 2}} \right)} {\left( {1 + {{\beta ^2 }
\mathord{\left/ {\vphantom {{\beta ^2 } 2}} \right. \kern-\nulldelimiterspace}
2}} \right)}}} \right. \kern-\nulldelimiterspace} {\left( {1 + {{\beta ^2 }
\mathord{\left/ {\vphantom {{\beta ^2 } 2}} \right. \kern-\nulldelimiterspace}
2}} \right)}}$ with $\beta  = {{m'} \mathord{\left/ {\vphantom {{m'} m}} \right.
\kern-\nulldelimiterspace} m}$. For an $\SU(2)_L$ triplet with mass $m$ and mass
splitting parameter $m' \geq 0$, contribution to $S$ is
\begin{equation}\label{equ:Striplet}
S_{\mathrm{triplet}}  = \frac{2}
{{9\pi }}\left\{ {\frac{1}
{3}\ln \zeta + 8\int_0^1 {dx{\text{ }}x\left( {1 - x} \right)\ln \left( {x +
\zeta \left( {1 - x} \right)} \right)} } \right\},
\end{equation}
where $\zeta  = {{\left( {1 - 2\beta ^2 } \right)} \mathord{\left/ {\vphantom
{{\left( {1 - 2\beta ^2 } \right)} {\left( {1 + \beta ^2 } \right)}}} \right.
\kern-\nulldelimiterspace} {\left( {1 + \beta ^2 } \right)}}$ with $\beta  =
{{m'} \mathord{\left/ {\vphantom {{m'} m}} \right. \kern-\nulldelimiterspace}
m}$. The integrals in Eqs.~(\ref{equ:Sdoublet} and \ref{equ:Striplet}) can be
easily computed, which yield $S$ parameters that depend only on $\beta$ for each
scalar multiplet. Figure~\ref{fig:S} shows the dependence of those $S$
parameters on $\beta$.
\begin{figure*}
\includegraphics{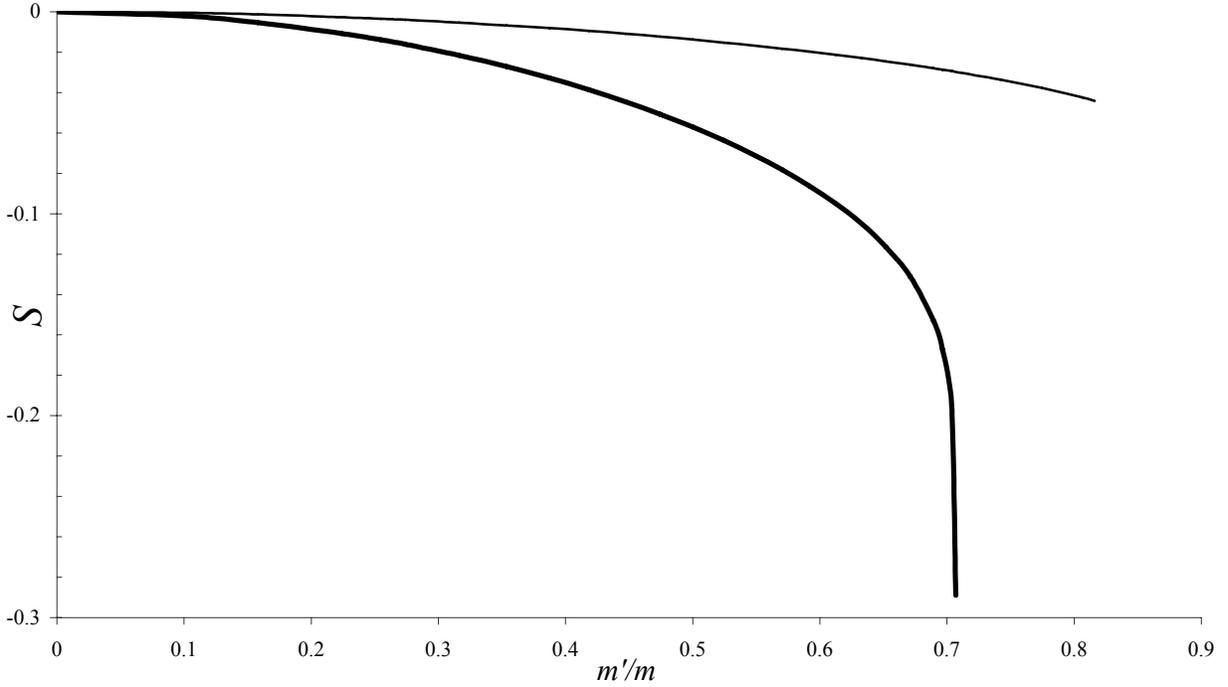}
\caption{\label{fig:S} Electroweak $S$ parameter from left-handed doublet and
triplet scalars: thick solid line, left-handed triplet; thin solid line,
left-handed doublet.}
\end{figure*}
Since the $S$ parameters in Fig.~\ref{fig:S} are bounded from above, at
$\beta=0$, the contribution from the scalars of the theory is net negative. The
total contribution from the scalar fields to $S$, however, is a sum of all
scalar contributions. As the $\beta$ parameter for each scalar multiplet is not
known, we may only speak of the bounds the total scalar contribution should lie
within, namely
\begin{equation}\label{equ:Ss}
0 \geqslant S_{\mathrm{scalars}}  \geqslant  - 3.13 \,\, .
\end{equation}

To be inclusive, we may also consider the contribution from a heavy SM model
Higgs\footnote{By heavy SM Higgs, we mean heavier than the 300 GeV Higgs for
which the experimental \protect $S$ value is provided in Eq.~(\ref{equ:expS}).}
to $S$, which can be positive depending on its mass. This contribution, however,
is relatively small even for an exotic Higgs with 1 TeV mass, where it can reach
up to 0.06. With such contribution, if heavy SM Higgs exists at all, it would
not change the maximum negative $S$ provided by other scalars, significantly. On
the other hand, the fermionic contribution to $S$ from three generations of
unconventional fermions is within
\begin{equation}\label{equ:Sf}
1.17 \geqslant S_{\mathrm{fermions}}  \geqslant 0.58 \,\, .
\end{equation}

The positive contribution from unconventional fermions obviously violates the
experimental bounds. Nonetheless, the negative scalar contribution to $S$ has
the potential to bring the total $S$ in agreement with the experimental
constraint on new physics, given in Eq.~(\ref{equ:expS}). Therefore, the notion
of three extra generations of heavy fermions can, in principal, be accommodated
within the model.
\section{The decay of unconventional fermions}\label{sec:decays}
Although unconventional fermions can have transitions involving ordinary
fermions via $\SU(4)_{\mathrm{PS}}$ and $\SU(3)_{H}$ mediated processes, the
lightest unconventional fermion cannot decay into light ordinary fermions, even
through weak channels. That poses an alarming danger:  a stable unconventional
fermion. There are stringent constraints on heavy stable fermions (quarks or
leptons) from cosmology (e.g., nucleosynthesis) and earth-based experiments.
Those constraints are discussed in length in Ref.~\cite{Frampton2000} and the
references therein.

Fortunately, the decay of the lightest unconventional fermion is possible via
the mixing among the charged gauge bosons. Such mixing is possible through
$\Theta$'s VEV, which mixes the charged gauge bosons $W_{\mu H}^ \pm $ and
$W_{\mu L}^ \pm $, since it carries $\SU(2)_L$ and $\SU(2)_H$ quantum numbers.
This mixing has been discussed in Appendix~\ref{app:gsb}, where its rather long
expressions are given for completeness.

The mixing between the charged gauge bosons corresponds to an equivalent mixing
among the relevant currents as well. If we denote gauge eigenstates of charged
bosons by $W_{\mu H}^ \pm $ and $W_{\mu L}^ \pm $, and mass eigenstates by
$\tilde W_{\mu H}^ \pm $ and $\tilde W_{\mu L}^ \pm $, we may write (to first
approximation)
\begin{subequations}
\begin{equation}
\tilde W_{\mu L}^ \pm = W_{\mu L}^ \pm  + \mathcal{O}\left({\frac{{v^2}}
{{v_H^{'2}}}} \right) W_{\mu H}^ \pm \, ,
\end{equation}
\begin{equation}
\tilde W_{\mu H}^ \pm = W_{\mu H}^ \pm  + \mathcal{O}\left({\frac{{v^2}}
{{v_H^{'2}}}} \right) W_{\mu L}^ \pm \, ,
\end{equation}
\end{subequations}
where $v$ and $ v'_H $ are the VEV's of the SM Higgs $H$, and the horizontal
breaking Higgs $\Phi _H^{\left( 2 \right)}$, respectively. Let us also denote
currents coupled to gauge eigenstates by $J_{\mu H}$ and $J_{\mu L}$, and those
coupled to mass eigenstates by $\tilde J_{\mu H}$ and $\tilde J_{\mu L}$. The
same matrix that connects the gauge and mass eigenstate charged bosons relates
the corresponding currents to each other as well. Therefore, we have
\begin{subequations} 
\begin{align}
\tilde J_{\mu L} &= J_{\mu L}  + \mathcal{O}\left({\frac{{v^2}}{{ v_H^{'2}}}}
\right) J_{\mu H} \, ,\\
\tilde J_{\mu H} &= J_{\mu H} + \mathcal{O}\left({\frac{{v^2}}{{ v_H^{'2}}}}
\right) J_{\mu L} \, .
\end{align}
\end{subequations}
The consequent interaction terms are then given by
\begin{subequations}\label{equ:ints}
\begin{equation}\label{equ:Lint}
\mathcal{L}_L  = g_W \tilde J_L^\mu  \tilde W_{\mu L}  = g_W \left[ {J_L^\mu   +
\mathcal{O}\left( {\frac{{v^2 }}
{{v_H^{'2} }}} \right)J_H^\mu  } \right]\tilde W_{\mu L} ,
\end{equation}
and
\begin{equation}\label{equ:Hint}
\mathcal{L}_H  = g_W \tilde J_H^\mu  \tilde W_{\mu H}  = g_W \left[ {J_H^\mu   +
\mathcal{O}\left( {\frac{{v^2 }}
{{v_H^{'2} }}} \right)J_L^\mu  } \right]\tilde W_{\mu H} .
\end{equation}
\end{subequations}

The interaction term of Eq.~(\ref{equ:Lint}) portrays how the lightest
unconventional fermion, in $J_{\mu H}$, can couple to $\tilde W_{\mu L}^ \pm $
and therefore decay into ordinary fermions. The decay mechanism falls within one
of the two possibilities:
\begin{enumerate}
\item
The mass of the lightest unconventional fermion is large enough to decay into a
real $\tilde W_{ L}$ and a regular fermion, according to $g_W \mathcal{O}\left(
{{{v^2 } \mathord{\left/ {\vphantom {{v^2 } {v_H^{'2} }}} \right.
\kern-\nulldelimiterspace} {v_H^{'2} }}} \right)J_H^\mu  \tilde W_{\mu L}$.
Although the $\mathcal{O}\left( {{{v^2 } \mathord{\left/ {\vphantom {{v^2 }
{v_H^{'2} }}} \right. \kern-\nulldelimiterspace} {v_H^{'2} }}} \right)$ factor
is small, the corresponding decay rate can be sizeable, since the unconventional
fermion decays into a real $\tilde W_{ L}$.
\item
The mass of the lightest unconventional fermion is not large enough to decay
into a real $\tilde W_{ L}$ and a regular fermion. In that case, $\tilde W_{ L}$
would be virtual and the interaction involves a $\tilde W_{ L}$ propagator,
i.e.,
\[
\mathcal{L}_{\mathrm{int}} \sim g_W^2 \mathcal{O}\left( {\frac{{v^2 }}
{{v_H^{'2} }}} \right)J_L^{\mu \dagger} \frac{{1}}
{{p^2 - M_{\tilde W_ L}^2 }} J_{\mu H} .
\]
\end{enumerate}

The fermions appearing in $J_{\mu L}$ and $J_{\mu H}$ should be expressed in
terms of mass eigenstates. This means new mixing angels, which are totally
different from the known CKM matrix elements. Thus, the computation of the
lifetimes of unconventional fermions involves unknown mixing angels; one could
only have a rational estimate for.

To estimate the lifetime and decay length of the lightest unconventional
fermion, which is the long-lived one, we note that the dominant decay is that
into a real $\tilde W_L$. For illustration purposes, let us assume that the
lightest unconventional quark and lepton are $D$ and $l_{d} $, respectively.
Their dominant decay modes would be
\begin{subequations}
\begin{equation}
D\left( -{\frac{4}{3}} \right) \to d^j \left( -{\frac{1}{3}} \right) + \tilde
W_L^ -  ,
\end{equation}
\begin{equation}
l_{d} \left( 1 \right) \to \nu ^k \left( 0 \right) + \tilde W_L^ +  ,
\end{equation}
\end{subequations}
where $j = d,s,b$, and $k=e,\mu,\tau$, and all fermions ought to be mass
eigenstates. The decay widths can be found easily. For $m_{d^j ,\nu ^k }  \ll
m_{D,l_{d} } $
, the decay widths are simply given by
\begin{subequations}\label{equ:widths1}
\begin{equation}
\Gamma _D  = \frac{{G_F m_D^3 }}
{{8\pi \sqrt 2 }}\left| {\mathcal{O}\left( {\frac{{v^2 }}
{{v_H^{'2} }}} \right)} \right|^2 \left| {V_Q } \right|^2 \left( {1 +
\frac{{2M_{\tilde W_L }^2 }}
{{m_D^2 }}} \right)\left( {1 - \frac{{M_{\tilde W_L }^2 }}
{{m_D^2 }}} \right)^2 ,
\end{equation}
\begin{equation}
\Gamma _{l_{d} }  = \frac{{G_F m_{l_{d} }^3 }}
{{8\pi \sqrt 2 }}\left| {\mathcal{O}\left( {\frac{{v^2 }}
{{v_H^{'2} }}} \right)} \right|^2 \left| {V_{L} } \right|^2 \left( {1 +
\frac{{2M_{\tilde W_L }^2 }}
{{m_{l_{d} }^2 }}} \right)\left( {1 - \frac{{M_{\tilde W_L }^2 }}
{{m_{l_{d} }^2 }}} \right)^2 .
\end{equation}
\end{subequations}

The factors $V_Q$ and $V_{L}$ are the relevant elements of matrices
$V_Q=\mathrm{U}_D^{-1}\mathrm{U}_U$ and 
$V_{L}=\mathrm{U}_{l_{d}}^{-1}\mathrm{U}_{l_{u}}$, which describe the mixings
among the unconventional quarks and leptons, respectively. To be precise,
$\mathrm{U}_U$, $\mathrm{U}_D$ and $\mathrm{U}_{l_{u}}$, $\mathrm{U}_{l_{d}}$
are matrices which diagonalize the up-, down-sector of unconventional quarks and
leptons, respectively. To obtain an estimate, let us take a rational value for
the masses of $D$ and $l_{d} $, namely $m_D \approx m_{l_{d}} \approx 250$ GeV
and apply realistic assumption $\mathcal{O}\left( {{{v^2 } \mathord{\left/
{\vphantom {{v^2 } {v_H^{'2} }}} \right. \kern-\nulldelimiterspace} {v_H^{'2}
}}} \right) \approx 10^{ - 2} $. Then typical lifetimes for the lightest
unconventional fermions can be estimated
\begin{subequations}\label{equ:LT}
\begin{equation}
\tau _D  \approx 1.3 \times 10^{ - 21} \left| {V_Q } \right|^{ - 2} \,\,
\mathrm{s} \, ,
\end{equation}
\begin{equation}
\tau _{l_{d} }  \approx 1.3 \times 10^{ - 21} \left| {V_{L} }  \right|^{ - 2}
\,\, \mathrm{s} \, .
\end{equation}
\end{subequations}

These lifetimes are obviously short, which indicate that unconventional fermions
decay fast and therefore pose no cosmological problems, unless the mixing
factors are peculiarly small. A typical decay length for the lightest
unconventional quark and lepton can also be estimated from the lifetimes of
Eqs.~(\ref{equ:LT}), they are
\begin{subequations}\label{equ:DL}
\begin{equation}
l _D  \approx 400 \left| {V_Q } \right|^{ - 2} \,\, \mathrm{fm} \, ,
\end{equation}
\begin{equation}
l _{l_{d} }  \approx 400 \left| {V_{L} }  \right|^{ - 2} \,\, \mathrm{fm} \, ,
\end{equation}
\end{subequations}
which also depend on the mixings $V_Q$ and $V_{L}$. To summarize, we showed that
the lightest of unconventional quarks or leptons will not be stable and can
decay through the mixing among the horizontal and left-handed charged gauge
bosons. Very short lifetimes (for reasonably small mixings) are possible for the
longest-lived unconventional fermions, which alleviate cosmological concerns on
heavy stable fermions.
\section{Summary}
We examined the idea of \textit{early quark-lepton mass unification} through one
of petite unification models $\mathrm{PUT}_2=\SU(4)_{\mathrm {PS}} \otimes
\SU(3)_L \otimes \SU(3)_{H}$. 

For this we embedded $\mathrm{PUT}_2$ into a 5D model, in brane world picture.
The petite unification scenario calls for new chiral fermions with
unconventional charges. The philosophy was that the sizes of 4D couplings in
chiral fermion mass terms are controlled by the left-right overlaps between the
corresponding localized fermions along the extra dimension. The magnitudes of
these overlaps are set by the geometry of the localized zero modes in the extra
dimension. We chose to establish such geometry by reducing the symmetry of the
model to that of the SM. This way the quark-lepton unification structure
translates into the geometry of the localized zero modes and yields a
\textit{quark-lepton mass unification structure}. This idea therefore sets the
symmetry breakings along the extra dimension in motion systematically.

As a result, the effective Yukawa couplings of quarks and leptons even those of
unconventional fermions and the SM fermions relate to each other. A numerical
estimation of mass scales showed that the unconventional fermion mass scales can
set bounds on the mass scales of Dirac neutrino and vice versa. For example,
light unconventional fermions (up to 500 GeV mass scales) set 1 eV bound on
neutrino mass scale, which imply a near-degenerate mass matrix for neutrino
sector. On the other hand, lighter unconventional fermions (as light as 180 GeV)
yield light (less than 0.1 eV) neutrino sector, which corresponds to a
hierarchical or near-degenerate mass matrix for neutrinos.

The mass scales obtained in this model are similar to those of $\mathrm{PUT}_1$.
The strong $\SU(4)_{\mathrm {PS}}$ is the common group of PUT scenarios. The
unconventional quarks and leptons are connected to their normal siblings through
the quartets of this group in both scenarios. Breaking $\SU(4)_{\mathrm {PS}}$
along the extra dimension results in similar relations between the left-right
separations of quarks and leptons and that translates into similar mass scales
for both models.

We computed the contributions of extra heavy fermions and scalars of the model
to the electroweak oblique parameter S, and showed that the extra generations of
heavy fermions may not violate the experimental bounds on new physics, in
principle.

The issue of the decay of the lightest unconventional fermion was also
discussed. We showed that the lightest unconventional fermion is indeed
unstable, as it decays to ordinary fermions through the mixing of charged gauge
bosons. The estimated lifetimes for the lightest unconventional quark and lepton
also appeared to be small enough to comply with the cosmological bounds on
stable heavy fermions.

In addition, we discussed the gauge symmetry breaking of the model in length,
where the mixings of neutral and charged gauge bosons of the model were
explained explicitly. 
\begin{acknowledgments}
This work was supported, in part, by the U.S. Department of Energy under grant
No. DE-A505-89ER40518.
\end{acknowledgments}

\appendix

\section{The gauge symmetry breakdown of the model}\label{app:gsb}
In this appendix, we describe the gauge symmetry breaking of our model in 4D
space down to $\SU(3)_c \otimes \mathrm{U}(1)_{EM}$. We follow the gauge
symmetry breaking pattern given in section \ref{sec:PUT}, Eqs.~(\ref{equ:sb}),
and describe the symmetry breakings, as usual, by scalar Higgs fields with
non-zero VEV's.
\subsection{Strong breakdown}
The strong breakdown of $\mathrm{PUT}_2$, at energy scale $M$, can be
accomplished by a Higgs-field multiplet, which we denote by $\Phi_{\mathrm{PS}}
= \left( {15,1,1} \right)$. The VEV of this Higgs field, must leave the QCD
gauge group unbroken, decouple the color quark triplet from the lepton color
singlet in the $\SU(4)_{\mathrm{PS}}$ fundamental representation, and give mass
to six intermediate (lepto-quark) gauge bosons of $\SU(4)_{\mathrm{PS}}$
symmetry. The lepto-quark gauge bosons in terms of $\SU(4)_{\mathrm{PS}}$ gauge
bosons $A_{\mu S}^j $ are
\begin{equation}
\begin{gathered}
  X_{\mu 1}^ \pm   = \frac{1}
{{\sqrt 2 }}\left( {A_{\mu S}^9  \mp iA_{\mu S}^{10} } \right) \, , \hfill \\
  X_{\mu 2}^ \pm   = \frac{1}
{{\sqrt 2 }}\left( {A_{\mu S}^{11}  \mp iA_{\mu S}^{12} } \right) \, , \hfill \\
  X_{\mu 3}^ \pm   = \frac{1}
{{\sqrt 2 }}\left( {A_{\mu S}^{13}  \mp iA_{\mu S}^{14} } \right) \, . \hfill
\\ 
\end{gathered} 
\end{equation}
They carry electric charge $ \pm {4 \mathord{\left/ {\vphantom {4 3}} \right.
\kern-\nulldelimiterspace} 3}$ and receive mass from the VEV of $\Phi
_{\mathrm{P}S} $. The covariant derivative of $\SU(4)_{\mathrm{PS}} \otimes
\SU(3)_L \otimes \SU(3)_H$ is
\begin{equation}
\hat D_\mu   = \partial _\mu   + ig_S \hat A_{\mu S}  + ig_W \hat A_{\mu L}  +
ig_W \hat A_{\mu H} \, ,
\end{equation}
where $\hat A_{\mu S}  = \sum\limits_{j = 1}^{15} {\hat T_j } A_{\mu S}^j$ and
$\hat A_{\mu L,H}  = \sum\limits_{j = 1}^8 {\hat F_j } A_{\mu L,H}^j$ with $\hat
T_j$ and $\hat F_j$ being the generators of $\SU(4)$ and $\SU(3)$ algebras. The
lepto-quark mass terms come from the kinetic energy of the Higss field $\Phi
_{\mathrm{PS}}$, i.e.,
\begin{equation}
\mathcal{L}_{\Phi _{\mathrm{PS}} ,\mathrm{kin}}  = \mathrm{Tr}\left( {\left|
{D_\mu  \Phi _{\mathrm{PS}} } \right|^2 } \right) = \frac{1}
{2} \mathrm{Tr} \left( {\left| {\partial _\mu  \Phi _{\mathrm{PS}}  + ig_S
\left[ {\hat A_{\mu S} ,\Phi _{\mathrm{PS}} } \right] } \right|^2 } \right).
\end{equation}
Once $\Phi _{\mathrm{PS}} $ attains VEV
\begin{equation}
\left\langle {\Phi _{PS} } \right\rangle  = v_{PS} \left( {\begin{array}{*{20}c}
   {\frac{1}
{3}} & 0 & 0 & 0  \\
   0 & {\frac{1}
{3}} & 0 & 0  \\
   0 & 0 & {\frac{1}
{3}} & 0  \\
   0 & 0 & 0 & { - 1}  \\
 \end{array} } \right) \,,
\end{equation}
the lepto-quark gauge bosons receive mass in the form
\begin{equation}
\mathcal{L}_{X,\mathrm{mass}}  = \frac{4}
{9}g_S^2 v_{PS}^2 \sum\limits_{j = 1}^3 {\left| {X_{\mu j} } \right|^2 }  =
M_X^2 \sum\limits_{j = 1}^3 {\left| {X_{\mu j} } \right|^2 } ,
\end{equation}
where $M_X  = {{2g_S v_{PS} } \mathord{\left/ {\vphantom {{2g_S v_S } 3}}
\right. \kern-\nulldelimiterspace} 3}$ and $j$ refers to the color degree of
freedom.
\subsection{Weak breakdown}
Right above the next symmetry breaking scale $\tilde{M}$, the gauge group that
needs to be broken is $\mathrm{U}(1) _ S \otimes  \SU(3) _ L \otimes \SU(3) _
H$. The breaking of this group should furnish the generator of the SM's weak
hypercharge group $\hat Y_W  = C_S \hat T_{15\mathrm{PS}}  + C_L \hat T_{8L}  +
C_{1H} \hat T_{8H}  + C_{2H} \hat T_{3H}$. This means that at least one of the
Higgs fields for such breaking must have non-vanishing $\mathrm{U}(1) _ S$
quantum number. The breaking of $\SU(3)_L$ into $\SU(2)_L \otimes
\mathrm{U}(1)_L$ can be done by the Higgs field $\Phi _L  = \left( {1,8,1}
\right)$. The VEV of $\Phi _L$ should decouple the left-handed doublet from the
singlet in $\SU(3) _ L$ fundamental representation [see, e.g.,
Eqs.~(\ref{equ:tripletsL1})] and give mass to four intermediate gauge bosons of
$\SU(3)_L$ symmetry. Therefore, its VEV should take on the eighth direction of
the multiplet. The intermediate gauge bosons can be written in terms of 
$\SU(3)_L$ gauge fields $A_{\mu L}^j $, i.e.,
\begin{equation}
\begin{gathered}
  V_{\mu 1}^ \pm   = \frac{1}
{{\sqrt 2 }}\left( {A_{\mu L}^4  \mp iA_{\mu L}^5 } \right), \hfill \\
  V_{\mu 2}^ \pm   = \frac{1}
{{\sqrt 2 }}\left( {A_{\mu L}^6  \mp iA_{\mu L}^7 } \right). \hfill \\ 
\end{gathered} 
\end{equation}
The covariant derivative of $\SU(3)_c \otimes \SU(3)_L \otimes \SU(3)_H \otimes
\mathrm{U}(1)_S$ is
\begin{equation}
\hat D_\mu   = \partial _\mu   + ig_3 \hat G_\mu   + ig_W \hat A_{\mu L}  + ig_W
\hat A_{\mu H}  + i\tilde g_S \hat Y_S \tilde A_{\mu S}\, ,
\end{equation}
where $G_\mu ^j$ are gluon fields and $\tilde A_{\mu S}$ is the neutral gauge
boson of $\mathrm{U}(1)_S$. As usual, the mass terms of $V_{\mu j}^ \pm  $ come
from the kinetic energy of $\Phi _L $, which is in the form $\mathrm{Tr}\left(
{\left| {D_\mu  \Phi _L } \right|^2 } \right)$. With $\Phi _L $ taking VEV in
the form
\begin{equation}
\left\langle {\Phi _L } \right\rangle  = v_L \left( {\begin{array}{*{20}c}
   1 & 0 & 0  \\
   0 & 1 & 0  \\
   0 & 0 & { - 2}  \\
 \end{array} } \right) \, ,
\end{equation}
the mass of $V_{\mu j} $ fields easily turns out to be $M_V  = {{3g_W v_L }
\mathord{\left/ {\vphantom {{3g_W v_L } 2}} \right. \kern-\nulldelimiterspace}
2}$.

The symmetry breaking of $\SU(3)_H$, on the other hand, needs to have
non-vanishing $\mathrm{U}(1) _ S$ and $\mathrm{U}(1)_L$ quantum numbers, so that
at the end three of four neutral gauge bosons acquire masses and the weak
hypercharge gauge boson, $B_\mu$, emerges as a massless field. One economical
scenario for $\SU(3)_H$ symmetry breaking in two steps can consist of Higgs
field multiplets $\Phi^{(1)} _{H}  = \left( {1,1,8} \right)$ and $\Phi^{(2)}
_{H}  = \left( {4,3,3} \right)$. In the first step, $\Phi^{(1)} _{H}$ breaks
$\SU(3)_H$ into $\SU(2)_H \otimes \mathrm{U}(1)_H$ and then $\Phi^{(2)} _{H}$
destroys the horizontal symmetry completely. The VEV of $\Phi^{(1)} _{H}$ should
take on the eighth direction of the multiplet to decouple $\SU(2) _H$ and
$\mathrm{U}(1) _ H$ subgroups. In addition, $\Phi^{(1)} _{H}$'s VEV gives mass
to four intermediate gauge bosons of $\SU(3) _H$ symmetry. These are the gauge
bosons, which connect SM-type fermions to vector-like fermions. They can be
written in terms of the horizontal gauge bosons $A_{\mu H}^j $, as
\begin{equation}
\begin{gathered}
  U_{\mu 1}^ \pm   = \frac{1}
{{\sqrt 2 }}\left( {A_{\mu H}^4  \mp iA_{\mu H}^5 } \right), \hfill \\
  U_{\mu 2}^ \pm   = \frac{1}
{{\sqrt 2 }}\left( {A_{\mu H}^6  \mp iA_{\mu H}^7 } \right). \hfill \\ 
\end{gathered} 
\end{equation}
The VEV of $\Phi _H^{\left( 1 \right)} $ takes the same form as that of $\Phi _L
$, but in horizontal space. Therefore, the mass of $U_{\mu j} $ bosons simply
reads $M_U  = {{3g_W v_H } \mathord{\left/ {\vphantom {{3g_W v_H } 2}} \right.
\kern-\nulldelimiterspace} 2}$, where $v_H$ is the amplitude of $\Phi _H^{\left(
1 \right)} $'s VEV.

To complete horizontal symmetry breaking, $\Phi^{(2)} _{H}$ will develop VEV,
which must be on its colorless component to preserve QCD symmetry. In order to
keep $\SU(2) _L$ symmetry intact, $\Phi^{(2)} _{H}$'s VEV should also be on the
singlet component of its $\SU(3)_L$ triplet. Finally to destroy $\SU(2)_H$, the
horizontal triplet of $\Phi^{(2)} _{H}$ simply attains VEV in its doublet in the
form $\bigl( \begin{smallmatrix} 0 \\ {{{v'_H} \mathord{\left/ {\vphantom
{{v'_H} {\sqrt 2 }}} \right. \kern-\nulldelimiterspace} {\sqrt 2 }}}
\end{smallmatrix} \bigl)$. In the process of destroying $\SU(2)_H$ symmetry, the
four unbroken neutral gauge bosons will mix and as a result we end up with three
massive and one massless gauge bosons. The massless gauge boson $B_\mu  $,
corresponds to $\mathrm{U}(1)_Y$ symmetry. The $\SU(2)_H$ symmetry has three
gauge bosons: $W_{\mu H}^ \pm$ and $W_{\mu H}^3$. The charged gauge bosons
receive mass through the kinetic energy term of $\Phi _H^{\left( 2 \right)} $,
which turns out to be $M_{W_H }  = {{g_W v'_H } \mathord{\left/ {\vphantom {{g_W
v'_H } {\sqrt 2 }}} \right. \kern-\nulldelimiterspace} {\sqrt 2 }}$. The mixing
of the neutral gauge bosons $\tilde A_{\mu S} $, $\tilde A_{\mu L} $, $\tilde
A_{\mu H} $, $W_{\mu H}^3$ is through the VEV of $\Phi _H^{\left( 2 \right)}$,
i.e.,
\begin{equation}
\left\langle {\Phi _H^{\left( 2 \right)} } \right\rangle  = \left(
{\begin{array}{*{20}c}
   0  \\
   0  \\
   0  \\
   1  \\

 \end{array} } \right) \otimes \left( {\begin{array}{*{20}c}
   0  \\
   0  \\
   1  \\

 \end{array} } \right) \otimes \left( {\begin{array}{*{20}c}
   0  \\
   {\tfrac{{v'_H }}
{{\sqrt 2 }}}  \\
   0  \\

 \end{array} } \right).
\end{equation}
The kinetic energy of $\Phi _H^{\left( 2 \right)} $ is
\begin{equation}
\mathcal{L}_{\Phi _H^{\left( 2 \right)} ,\mathrm{kin}}  = \mathrm{Tr}\left[
{\left| {\left( {\partial _\mu  + ig_W \hat W_{\mu H}  - \frac{{ig_W }}
{{3\sqrt 3 }}\hat A_{\mu L}  + \frac{{ig_W }}
{{6\sqrt 3 }}\hat A_{\mu H}  - i\sqrt {\frac{3}
{8}} \tilde g_S \hat Y_S \tilde A_{\mu S} }\right) \Phi _H^{\left( 2 \right)}}
\right|^2 } \right],
\end{equation}
where $\hat W_{\mu H}  = \sum\limits_{j = 1}^3 {W_{\mu H}^j } \hat T_j$. The
squared mass matrix of neutral gauge bosons is obtained from the above trace,
i.e.,
\begin{equation}
M^2  = \left( {\begin{array}{*{20}c}
   {\frac{3}
{4}g_W^2 v'^2_H } & { - \frac{1}
{{12\sqrt 6 }}g_W^2 v'^2_H } & {\frac{1}
{{6\sqrt 6 }}g_W^2 v'^2_H } & {\frac{{\sqrt 3 }}
{8}g_W \tilde g_S v'^2_H }  \\
   { - \frac{1}
{{12\sqrt 6 }}g_W^2 v'^2_H } & {\frac{1}
{{216}}g_W^2 v'^2_H } & { - \frac{1}
{{108}}g_W^2 v'^2_H } & { - \frac{1}
{{24\sqrt 2 }}g_W \tilde g_S v'^2_H }  \\
   {\frac{1}
{{6\sqrt 6 }}g_W^2 v'^2_H } & { - \frac{1}
{{108}}g_W^2 v'^2_H } & {\frac{1}
{{54}}g_W^2 v'^2_H } & {\frac{1}
{{12\sqrt 2 }}g_W \tilde g_S v'^2_H }  \\
   {\frac{{\sqrt 3 }}
{8}g_W \tilde g_S v'^2_H } & { - \frac{1}
{{24\sqrt 2 }}g_W \tilde g_S v'^2_H } & {\frac{1}
{{12\sqrt 2 }}g_W \tilde g_S v'^2_H } & {\frac{3}
{{16}}\tilde g_S^2 v'^2_H }  \\

 \end{array} } \right).
\end{equation}
The mass terms would look like
\begin{equation}
\left( {\begin{array}{*{20}c}
   {W_{\mu H}^3 } & {\tilde A_{\mu H} } & {\tilde A_{\mu L} } & {\tilde A_{\mu
S} }  \\

 \end{array} } \right)M^2 \left( {\begin{array}{*{20}c}
   {W_{\mu H}^3 }  \\
   {\tilde A_{\mu H} }  \\
   {\tilde A_{\mu L} }  \\
   {\tilde A_{\mu S} }  \\

 \end{array} } \right).
\end{equation}
The squared mass matrix can be diagonalized, which leaves one massless gauge
boson, $B^\mu  $. The eigenvector corresponding to the zero eigenvalue is (after
normalization)
\begin{equation}
B_\mu   = \frac{1}
{{\sqrt {15g_W^2  + 453\tilde g_S^2 } }}\left( { - 4\sqrt 3 \tilde g_S W_{\mu
H}^3  + 
\frac{{27\tilde g_S }}
{{\sqrt 5 }}\tilde A_{\mu H}  + \frac{{36\tilde g_S }}
{{\sqrt 5 }}\tilde A_{\mu L}  + 4g_W \tilde A_{\mu S} } \right).
\end{equation}
\subsection{The SM breakdown and fermion masses}
To break $\mathrm{PUT}_2$'s gauge group further down to $\SU(3)_c \otimes
\mathrm{U}(1)_{EM}$, we need another Higgs field, which in addition to the
symmetry breaking is also responsible for giving mass to all SM-type fermions of
the model. The mass terms for SM-type fermions involve couplings of left-handed
and right-handed fields with a Higgs field, which develops VEV. To achieve this
in our model, we need to couple $\Psi_1$ to itself and also to $\Psi_2$ through
appropriate Higgs fields.

The SM Higgs field transforms as $(2,1)$ under $\SU(2)_L \otimes
\mathrm{U}(1)_Y$. Therefore, the $\SU(3) _L$ representation of a Higgs multiplet
that breaks the SM, must contain a $(2,1)$. The SM symmetry breakdown, then, is
through the VEV of a Higgs multiplet transforming as $\Theta = \left( 1,8,8
\right)$. The decomposition of $\Theta$'s $\SU(3)_L$ octet, in terms of
$\SU(2)_L \otimes \mathrm{U}(1)_Y$ multiplets or quantum numbers, yields
\begin{equation}
\left[ 8 \right]_{\SU\left( 3 \right)_L}  = \left( 3,0 \right) \oplus \left( 2,1
\right) \oplus \left( 2,-1 \right) \oplus \left( 1,0 \right),
\end{equation}
which in matrix form can be written as
\begin{equation}\label{equ:octetmatrixap}
\left( {\begin{array}{*{20}c}
   {\left( {3,0} \right)} &  \vline  & {\left( {2,1} \right)}  \\
    \hline
   {\left( {2, - 1} \right)} &  \vline  & {\left( {1,0} \right)}  \\

 \end{array} } \right).
\end{equation}

Therefore, $\Theta$ contains a SM Higgs field, which we denote by $H=(2,1)$, and
should develop a VEV in  $H$. The decomposition of $\Theta$'s horizontal octet
is essentially the same as that of the $\SU(3)_L$ octet, but in terms of
multiplets or quantum numbers of $\SU(2)_H \otimes \mathrm{U}(1)_H$. To avoid
vector-like fermions receiving mass through the SM's Higgs the horizontal octet
cannot contribute to $\Theta$'s VEV in its $(1,0)$ part and must have a
vanishing $\mathrm{U}(1)_H$ quantum number. This means that the horizontal
contribution to $\Theta$'s VEV can only exist in the $(3,0)$ part.

Through the kinetic energy term, $\Theta$'s VEV breaks the SM symmetry down to
$\SU(3)_c \otimes \mathrm{U}(1)_{EM}$, and gives mass to $W^ \pm$ and $Z^0$. In
addition, $\Theta$'s VEV mixes the charged gauge bosons, $W_{\mu H}^ \pm $ and
$W_{\mu L}^ \pm $, since it carries $\SU(2)_L$ and $\SU(2)_H$ quantum numbers.
To be precise, such mixing of ``gauge eigenstates'' yields new charged gauge
bosons which we may call ``mass eigenstates.'' Let us denote the mass
eigenstates of such mixing by $\tilde W_{\mu H}^ \pm $ and $\tilde W_{\mu L}^
\pm $. They can easily be expressed in terms of the gauge eigenstates, $W_{\mu
H}^ \pm $ and $W_{\mu L}^ \pm $ (and vice versa).
The squared mass matrix of charged gauge bosons, similar to neutral gauge boson
mixing, can be diagonalized by an orthogonal matrix, $R$. Such matrix connects
the mass eigenstates, $\tilde W_{\mu H}^ \pm$, $\tilde W_{\mu L}^ \pm $, and
gauge eigenstates, $W_{\mu H}^ \pm$, $W_{\mu L}^ \pm$, i.e.,
\begin{equation}
\left( {\begin{array}{*{20}c}
   {\tilde W_{\mu L}^ \pm  }  \\
   {\tilde W_{\mu H}^ \pm  }  \\

 \end{array} } \right) = R\left( {\begin{array}{*{20}c}
   {W_{\mu L}^ \pm  }  \\
   {W_{\mu H}^ \pm  }  \\

 \end{array} } \right),
\end{equation}
and vice versa
\begin{equation}
\left( {\begin{array}{*{20}c}
   {W_{\mu L}^ \pm  }  \\
   {W_{\mu H}^ \pm  }  \\

 \end{array} } \right) = R^T \left( {\begin{array}{*{20}c}
   {\tilde W_{\mu L}^ \pm  }  \\
   {\tilde W_{\mu H}^ \pm  }  \\

 \end{array} } \right).
\end{equation}
The unnormalized mass eigenvectors and their masses are given by
\begin{subequations}
\begin{equation}
\begin{aligned}
\begin{aligned}
  \tilde W_{\mu L}^ \pm   = &W_{\mu L}^ \pm   - \left[ {\frac{{\left( {g_W^2  -
g_2^2 } \right)}}
{{g_W g_2 }}\frac{{v^2 }}
{{v_H^2 }}} \right. \hfill \\
  &\left. { + \frac{{\sqrt {\left[ {\left( {g_W^2  - g_2^2 } \right)\frac{{v^2
}}
{{v_H^2 }} + g_W^2 } \right]^2  + 4\left[ {\left( {g_W g_2  - g_W^2 g_2^2 }
\right)\frac{{v^4 }}
{{v_H^4 }} - g_W^2 g_2^2 \frac{{v^2 }}
{{v_H^2 }}} \right]} }}
{{g_W g_2 }}} \right]W_{\mu H}^ \pm   \hfill \\ 
\end{aligned}
\end{aligned}
\end{equation}
with
\begin{equation}
M_{\tilde W_L }  = \left[ {\frac{{\left( {g_2^2  + g_W^2 } \right)v^2  + g_W^2
v'^2_H  + \sqrt {\left( {g_2^2  + g_W^2 } \right)^2 v^4  + g_W^4 v'^4_H  +
2\left( {g_W^2  - g_2^2 } \right)v^2 v'^2_H } }}
{4}} \right]^{\frac{1}
{2}}\, ,
\end{equation}
\end{subequations}
and
\begin{subequations}
\begin{equation}
\begin{aligned}
\begin{aligned}
  \tilde W_{\mu H}^ \pm   = &W_{\mu H}^ \pm   - \left[ {\frac{{\left( {g_W^2  -
g_2^2 } \right)}}
{{g_W g_2 }}\frac{{v^2 }}
{{v_H^2 }}} \right. \hfill \\
  &\left. { - \frac{{\sqrt {\left[ {\left( {g_W^2  - g_2^2 } \right)\frac{{v^2
}}
{{v_H^2 }} + g_W^2 } \right]^2  + 4\left[ {\left( {g_W g_2  - g_W^2 g_2^2 }
\right)\frac{{v^4 }}
{{v_H^4 }} - g_W^2 g_2^2 \frac{{v^2 }}
{{v_H^2 }}} \right]} }}
{{g_W g_2 }}} \right]W_{\mu L}^ \pm   \hfill \\ 
\end{aligned}
\end{aligned} 
\end{equation}
with
\begin{equation}
M_{\tilde W_H }  = \left[ {\frac{{\left( {g_2^2  + g_W^2 } \right)v^2  + g_W^2
v'^2_H  - \sqrt {\left( {g_2^2  + g_W^2 } \right)^2 v^4  + g_W^4 v'^4_H  +
2\left( {g_W^2  - g_2^2 } \right)v^2 v'^2_H } }}
{4}} \right]^{\frac{1}
{2}} .
\end{equation}
\end{subequations}

Dirac mass terms, involving the left- and right-handed SM-type fermions, can be
written in the form
\begin{equation}\label{equ:mcouplingsap}
\mathcal{L}_{\mathrm{mass}}  = \kappa _1 \Psi_1^T \Theta C \Psi_2^\ast  + \kappa
_2 \Psi_1^T \tilde \Theta  C \Psi_1^\ast  + h.c.\; ,
\end{equation} 
where $C$ is is the charge conjugation operator $C = i \gamma^{2} \gamma^{0}$;
$\kappa_1$ and $\kappa_2$ can be different in general, and $\tilde \Theta= i
\hat \lambda_{2L} \Theta^*$. One notices $i \hat \lambda_{2L}$ in $\tilde
\Theta$, which analogous to the presence of $i \hat \tau _2$ in the SM's quark
mass terms.

The mass terms in Eq.~(\ref{equ:mcouplingsap}) are rather compact. To convince
ourselves that they yield correct mass terms for SM-type fermions, let us expand
them for at least one flavor doublet. As explained, $\Theta$'s VEV has two
parts: an $\SU(3)_L$ octet, which contains the SM Higgs, and a horizontal octet.
The horizontal octet in $\Theta$'s VEV eliminates purely vector-like left-handed
triplets (e.g., those in Eqs.(\ref{equ:tripletsL12} and \ref{equ:atripletsL22}))
and therefore prevents them from receiving mass from the SM Higgs VEV. Thus, the
only left-handed triplets that stay in play are those which involve SM-type
fermions. Of those, let us consider $\Psi_1$'s 
\[
\left( {\begin{array}{*{20}c}
   {d_L^* }  \\
   { - u_L^* }  \\
   {d_L^c }  \\
 \end{array} } \right),
\]
and expand the mass terms in Eq.~(\ref{equ:mcouplingsap}) for normal quarks. The
SM Higgs doublet attains VEV in the usual form $\bigl( \begin{smallmatrix} 0 \\
{{v \mathord{\left/ {\vphantom {v {\sqrt 2 }}} \right.
\kern-\nulldelimiterspace} {\sqrt 2 }}} \end{smallmatrix} \bigl)$ and therefore
for normal quarks the mass terms in Eq.~(\ref{equ:mcouplingsap}) simply read
\begin{equation}\label{equ:mexpand2}
\begin{gathered}
\mathcal{L}_{\mathrm{mass}}^{\mathrm{quarks}}  = \kappa _1 \left(
{\begin{array}{*{20}c}
   {d_L^{*T} } & { - u_L^{*T} } & {d_L^{cT} }  \\
 \end{array} } \right)\left( {\begin{array}{*{20}c}
   0 & 0 & 0  \\
   0 & 0 & {\frac{v}
{{\sqrt 2 }}}  \\
   0 & 0 & 0  \\

 \end{array} } \right)C\left( {\begin{array}{*{20}c}
   { - \tilde u_L^{c*} }  \\
   {\tilde d_L^{c*} }  \\
   {u_L^{c*} }  \\
 \end{array} } \right) \hfill \\
  \;\quad\quad\quad+ \kappa _2 \left( {\begin{array}{*{20}c}
   {d_L^{*T} } & { - u_L^{*T} } & {d_L^{cT} }  \\
 \end{array} } \right)\left( {\begin{array}{*{20}c}
   0 & 0 & {\frac{v}
{{\sqrt 2 }}}  \\
   0 & 0 & 0  \\
   0 & 0 & 0  \\

 \end{array} } \right)C\left( {\begin{array}{*{20}c}
   {d_L }  \\
   { - u_L }  \\
   {d_L^{c*} }  \\
 \end{array} } \right) + h.c. \, , \hfill \\
\end{gathered} 
\end{equation}
which easily reduce to
\begin{equation}\label{equ:mexpand3}
\mathcal{L}_{\mathrm{mass}}^{\mathrm{quarks}}  =  - \kappa _1 \frac{v}
{{\sqrt 2 }}u_L^\dag  Cu_L^{c*}  + \kappa _2 \frac{v}
{{\sqrt 2 }}d_L^\dag  Cd_L^{c*}  + h.c. \; ,
\end{equation}
and finally
\begin{equation}\label{equ:mexpand5}
\mathcal{L}_{\mathrm{mass}}^{\mathrm{quarks}} =\kappa _1 \frac{v}{{\sqrt 2
}}\bar u_L u_R  - \kappa _2 \frac{v}{{\sqrt 2 }}\bar d_L d_R  + h.c. \; .
\end{equation}
\section{Contributions to electroweak \textit{S} parameter}\label{app:Scalc}
In this appendix, we show derivations of contributions to $S$ parameter from
unconventional fermions and scalar fields.
\subsection{Fermion contribution to \textit{S}}
For chiral fermions $\left( {\psi _1 ,\psi _2 } \right)$ with masses $\left(
{M_1 ,M_2 } \right)$ and hypercharge $Y$, one-loop contribution to $S$ is given
by Ref.~\cite{He2001}
\begin{equation}\label{equ:Sgeneral}
\begin{aligned}
  S = &\frac{{N_c }}
{{6\pi }}\left\{ {2\left( {4Y + 3} \right)x_1  + 2\left( { - 4Y + 3} \right)x_2 
- 2Y\ln \frac{{x_1 }}
{{x_2 }}} \right. \\ 
  &+\left[ {\left( {\frac{3}
{2} + 2Y} \right)x_1  + Y} \right]G\left( {x_1 } \right)  
  +\left. {\left[ {\left( {\frac{3}
{2} - 2Y} \right)x_2  - Y} \right]G\left( {x_2 } \right)} \right\}, 
\end{aligned} 
\end{equation}
where $x_i  = {{M_i } \mathord{\left/ {\vphantom {{M_i } {M_Z }}} \right.
\kern-\nulldelimiterspace} {M_Z }}$, $N_c $ is the color factor and $G\left( x
\right) =  - 4\sqrt {4x - 1} \arctan \left( {{1 \mathord{\left/ {\vphantom {1
{\sqrt {4x - 1} }}} \right. \kern-\nulldelimiterspace} {\sqrt {4x - 1} }}}
\right)$. For maximum contribution of unconventional fermions to $S$, we use
their mass scales as maximum masses. Since $\Lambda _{l_{u} }  \approx \Lambda
_U $ and $\Lambda _{l_{d} } \approx \Lambda _D $, we may write
\[
x_{1L}  = \frac{{M_{l_{u} } }}
{{M_Z }} \approx x_{1Q}  = \frac{{M_U }}
{{M_Z }}\,\,\,{\text{   and   }}\,\,\,x_{2L}  = \frac{{M_{l_{d} } }}
{{M_Z }} \approx x_{2Q}  = \frac{{M_D }}
{{M_Z }}\,\,.
\]
Let us employ a new notation $x_U $ and $x_D $ instead of $x_1 $ and $x_2 $,
where
\begin{equation}
x_U \equiv x_{1L} \approx x_{1Q} \,\,\,{\text{   and   }}\,\,\, x_D \equiv
x_{2L} \approx x_{2Q} \,\,.
\end{equation}
One loop fermionic contribution to $S$ for one generation of unconventional
fermions is
\begin{equation}
S = S_{leptons}  + S_{quarks} \,\,,
\end{equation}
where each $S$ can be calculated using Eq.~(\ref{equ:Sgeneral}). We easily find
\begin{equation}\label{equ:Sfermionsapp}
S = \frac{1}
{{6\pi }}\left[ {16x_U  + 32x_D  + 2\ln \frac{{x_U }}
{{x_D }} + \left( {4x_U  - 1} \right)G\left( {x_U } \right) + \left( {8x_D  + 1}
\right)G\left( {x_D } \right)} \right].
\end{equation}
\subsection{Scalar contribution to \textit{S}}
For a scalar multiplet, transforming as $\left( {j_L ,j_R } \right)$
 under $\SU(2)_L \otimes \SU(2)_R$, the $S$ parameter is given by
Ref.~\cite{Dugan1991}, i.e.,
\begin{equation}
S = \frac{1}
{{3\pi }}\sum\limits_{JJ'} {X_{JJ'} f\left( {m_J^2 ,m_{J'}^2 } \right)}\, ,
\end{equation}
where $X_{JJ'}$, $f$, and $m_J^2$  are explicitly defined in
Ref.~\cite{Dugan1991}.
For a scalar field, which transforms as $\left( {j_L ,j_R } \right) = \left( {{1
\mathord{\left/ {\vphantom {1 2}} \right. \kern-\nulldelimiterspace} 2},{1
\mathord{\left/ {\vphantom {1 2}} \right. \kern-\nulldelimiterspace} 2}}
\right)$, we find the group theoretical $X$ factors
\begin{equation}
X_{11}  = \frac{1}
{2}{\text{    }};{\text{   }}X_{00}  = 0{\text{   }};{\text{   }}X_{01}  =
X_{10}  =  - \frac{1}
{4} \; ,
\end{equation}
and the relevant $f$'s
\begin{subequations}
\begin{align}
 f\left( {m_1^2 ,m_1^2 } \right) &= - \ln \left( {\frac{{m_1^2 }}
{{\mu ^2 }}} \right)\,,\\ 
f\left( {m_1^2 ,m_0^2 } \right) &=f\left( {m_0^2 ,m_1^2 } \right) =  - 6\int_0^1
{dx{\text{ }}x\left( {1 - x} \right)\ln \left[ {\frac{{m_1^2 x + m_0^2 \left( {1
- x} \right)}}
{{\mu ^2 }}} \right]}\,.
\end{align}
\end{subequations}
With all this, the $S$ parameter for a scalar transforming as $ \left( {{1
\mathord{\left/ {\vphantom {1 2}} \right. \kern-\nulldelimiterspace} 2},{1
\mathord{\left/ {\vphantom {1 2}} \right. \kern-\nulldelimiterspace} 2}}
\right)$ reads
\begin{equation}
S = \frac{1}
{\pi }\int_0^1 {dx{\text{ }}x\left( {1 - x} \right)\ln \left( {x + \zeta \left(
{1 - x} \right)} \right)}\, ,
\end{equation}
where 
\begin{equation}
\zeta  = \frac{{m_0^2 }}
{{m_1^2 }} = \frac{{m^2  - \frac{3}
{2}m'^2 }}
{{m^2  + \frac{1}
{2}m'^2 }} = \frac{{1 - \frac{3}
{2}\beta ^2 }}
{{1 + \frac{1}
{2}\beta ^2 }} \,\,,
\end{equation}
$\beta  = {{m'} \mathord{\left/ {\vphantom {{m'} m}} \right.
\kern-\nulldelimiterspace} m}$, and $m'$ is the mass splitting parameter.
For a scalar field which transforms as $\left( {j_L ,j_R } \right) = \left(
{1,{1 \mathord{\left/ {\vphantom {1 2}} \right. \kern-\nulldelimiterspace} 2}}
\right)$, we find
\begin{equation}
X_{\tfrac{3}
{2}\tfrac{3}
{2}}  = \frac{{10}}
{9}{\text{    }};{\text{   }}X_{\tfrac{1}
{2}\tfrac{1}
{2}}  =  - \frac{2}
{9}{\text{   }};{\text{   }}X_{\tfrac{3}
{2}\tfrac{1}
{2}}  = X_{\tfrac{1}
{2}\tfrac{3}
{2}}  =  - \frac{4}
{9}\; ,
\end{equation}
and
\begin{subequations}
\begin{align}
  f\left( {m_{\tfrac{3}
{2}}^2 ,m_{\tfrac{3}
{2}}^2 } \right) &=  - \ln \left( {\frac{{m_{3 \mathord{\left/ {\vphantom {3 2}}
\right. \kern-\nulldelimiterspace} 2}^2 }}{{\mu ^2 }}} \right)\,, \\
f\left( {m_{\tfrac{1}
{2}}^2 ,m_{\tfrac{1}
{2}}^2 } \right) &=  - \ln \left( {\frac{{m_{1 \mathord{\left/ {\vphantom {1 2}}
\right. \kern-\nulldelimiterspace} 2}^2 }}{{\mu ^2 }}} \right)\,,\\
f\left( {m_{\tfrac{3}
{2}}^2 ,m_{\tfrac{1}
{2}}^2 } \right) &= f\left( {m_{\tfrac{1}
{2}}^2 ,m_{\tfrac{3}
{2}}^2 } \right)= - 6\int_0^1 {dx{\text{ }}x\left( {1 - x} \right)\ln \left[
{\frac{{m_{3 \mathord{\left/ {\vphantom {3 2}} \right.
\kern-\nulldelimiterspace} 2}^2 x + m_{1 \mathord{\left/ {\vphantom {1 2}}
\right. \kern-\nulldelimiterspace} 2}^2 \left( {1 - x} \right)}}
{{\mu ^2 }}} \right]}\, .
\end{align}
\end{subequations}
If we define
 \begin{equation}
\zeta  = \frac{{m_{1 \mathord{\left/ {\vphantom {1 2}} \right.
\kern-\nulldelimiterspace} 2}^2 }}
{{m_{3 \mathord{\left/ {\vphantom {3 2}} \right. \kern-\nulldelimiterspace} 2}^2
}} = \frac{{m^2  - 2m'^2 }}
{{m^2  + m'^2 }} = \frac{{1 - 2\beta ^2 }}
{{1 + \beta ^2 }}\,,
\end{equation}
with $\beta  = {{m'} \mathord{\left/ {\vphantom {{m'} m}} \right.
\kern-\nulldelimiterspace} m}$, the $S$ parameter for a scalar field
transforming as $\left( {1,{1 \mathord{\left/ {\vphantom {1 2}} \right.
\kern-\nulldelimiterspace} 2}} \right)$ is simply
\begin{equation}
S = \frac{2}
{{9\pi }}\left\{ {\frac{1}
{3}\ln \zeta + 8\int_0^1 {dx{\text{ }}x\left( {1 - x} \right)\ln \left( {x +
\zeta \left( {1 - x} \right)} \right)} } \right\}.
\end{equation}


\begin{thebibliography}{53}
\expandafter\ifx\csname natexlab\endcsname\relax\def\natexlab#1{#1}\fi
\expandafter\ifx\csname bibnamefont\endcsname\relax
  \def\bibnamefont#1{#1}\fi
\expandafter\ifx\csname bibfnamefont\endcsname\relax
  \def\bibfnamefont#1{#1}\fi
\expandafter\ifx\csname citenamefont\endcsname\relax
  \def\citenamefont#1{#1}\fi
\expandafter\ifx\csname url\endcsname\relax
  \def\url#1{\texttt{#1}}\fi
\expandafter\ifx\csname urlprefix\endcsname\relax\def\urlprefix{URL }\fi
\providecommand{\bibinfo}[2]{#2}
\providecommand{\eprint}[2][]{\url{#2}}

\bibitem[{\citenamefont{Hung}(2005)}]{Hung2005}
\bibinfo{author}{\bibfnamefont{P.~Q.} \bibnamefont{Hung}},
  \bibinfo{journal}{Nucl. Phys. B} \textbf{\bibinfo{volume}{720}},
  \bibinfo{pages}{89} (\bibinfo{year}{2005}),
  \href{http://arXiv.org/abs/arXiv:hep-ph/0412262v2}{{
  [arXiv:hep-ph/0412262v2]}}.

\bibitem[{\citenamefont{Antoniadis}(1990)}]{LED}
\bibinfo{author}{\bibfnamefont{I.}~\bibnamefont{Antoniadis}},
  \bibinfo{journal}{Phys. Lett. B} \textbf{\bibinfo{volume}{246}},
  \bibinfo{pages}{377} (\bibinfo{year}{1990});
\bibinfo{author}{\bibfnamefont{K.~R.} \bibnamefont{Dienes}},
  \bibinfo{author}{\bibfnamefont{E.}~\bibnamefont{Dudas}}, \bibnamefont{and}
  \bibinfo{author}{\bibfnamefont{T.}~\bibnamefont{Gherghetta}},
  \bibinfo{journal}{Nucl. Phys. B} \textbf{\bibinfo{volume}{537}},
  \bibinfo{pages}{47} (\bibinfo{year}{1999}),
  \href{http://arXiv.org/abs/arXiv:hep-ph/9806292v2}{{
  [arXiv:hep-ph/9806292v2]}}.

\bibitem[{\citenamefont{Arkani-Hamed et~al.}(1998)\citenamefont{Arkani-Hamed,
  Dimopoulos, and Dvali}}]{Arkani1998}
\bibinfo{author}{\bibfnamefont{N.}~\bibnamefont{Arkani-Hamed}},
  \bibinfo{author}{\bibfnamefont{S.}~\bibnamefont{Dimopoulos}},
  \bibnamefont{and} \bibinfo{author}{\bibfnamefont{G.}~\bibnamefont{Dvali}},
  \bibinfo{journal}{Phys. Lett. B} \textbf{\bibinfo{volume}{429}},
  \bibinfo{pages}{263} (\bibinfo{year}{1998}).

\bibitem[{\citenamefont{Hung et~al.}(1982)\citenamefont{Hung, Buras, and
  Bjorken}}]{Hung1982}
\bibinfo{author}{\bibfnamefont{P.~Q.} \bibnamefont{Hung}},
  \bibinfo{author}{\bibfnamefont{A.~J.} \bibnamefont{Buras}}, \bibnamefont{and}
  \bibinfo{author}{\bibfnamefont{J.~D.} \bibnamefont{Bjorken}},
  \bibinfo{journal}{Phys. Rev. D} \textbf{\bibinfo{volume}{25}},
  \bibinfo{pages}{805} (\bibinfo{year}{1982}).

\bibitem[{\citenamefont{Buras and Hung}(2003)}]{Buras2003}
\bibinfo{author}{\bibfnamefont{A.~J.} \bibnamefont{Buras}} \bibnamefont{and}
  \bibinfo{author}{\bibfnamefont{P.~Q.} \bibnamefont{Hung}},
  \bibinfo{journal}{Phys. Rev. D} \textbf{\bibinfo{volume}{68}},
  \bibinfo{pages}{035015} (\bibinfo{year}{2003}),
  \href{http://arXiv.org/abs/arXiv:hep-ph/0305238v1}{{
  [arXiv:hep-ph/0305238v1]}}.

\bibitem[{\citenamefont{Pati and Salam}(1974)}]{Pati1974}
\bibinfo{author}{\bibfnamefont{J.~C.} \bibnamefont{Pati}} \bibnamefont{and}
  \bibinfo{author}{\bibfnamefont{A.}~\bibnamefont{Salam}},
  \bibinfo{journal}{Phys. Rev. D} \textbf{\bibinfo{volume}{10}},
  \bibinfo{pages}{275} (\bibinfo{year}{1974}).

\bibitem[{\citenamefont{Antoniadis et~al.}(1998)\citenamefont{Antoniadis,
  Arkani-Hamed, Dimopoulos, and Dvali}}]{Antoniadis1998}
\bibinfo{author}{\bibfnamefont{I.}~\bibnamefont{Antoniadis}},
  \bibinfo{author}{\bibfnamefont{N.}~\bibnamefont{Arkani-Hamed}},
  \bibinfo{author}{\bibfnamefont{S.}~\bibnamefont{Dimopoulos}},
  \bibnamefont{and} \bibinfo{author}{\bibfnamefont{G.}~\bibnamefont{Dvali}},
  \bibinfo{journal}{Phys. Lett. B} \textbf{\bibinfo{volume}{436}},
  \bibinfo{pages}{257} (\bibinfo{year}{1998}),
  \href{http://arXiv.org/abs/arXiv:hep-ph/9804398v1}{{
  [arXiv:hep-ph/9804398v1]}}.

\bibitem[{\citenamefont{Arkani-Hamed and Schmaltz}(2000)}]{Arkani2000}
\bibinfo{author}{\bibfnamefont{N.}~\bibnamefont{Arkani-Hamed}}
  \bibnamefont{and} \bibinfo{author}{\bibfnamefont{M.}~\bibnamefont{Schmaltz}},
  \bibinfo{journal}{Phys. Rev. D} \textbf{\bibinfo{volume}{61}},
  \bibinfo{pages}{033005} (\bibinfo{year}{2000}),
  \href{http://arXiv.org/abs/arXiv:hep-ph/9903417v1}{{
  [arXiv:hep-ph/9903417v1]}}.

\bibitem[{\citenamefont{Hung}(2003)}]{Hung2003a}
\bibinfo{author}{\bibfnamefont{P.~Q.} \bibnamefont{Hung}},
  \bibinfo{journal}{Phys. Rev. D} \textbf{\bibinfo{volume}{67}},
  \bibinfo{pages}{095011} (\bibinfo{year}{2003}),
  \href{http://arXiv.org/abs/arXiv:hep-ph/0210131v1}{{
  [arXiv:hep-ph/0210131v1]}}.

\bibitem[{\citenamefont{Arkani-Hamed et~al.}(2001)\citenamefont{Arkani-Hamed,
  Dimopoulos, Dvali, and March-Russell}}]{SNM}
\bibinfo{author}{\bibfnamefont{N.}~\bibnamefont{Arkani-Hamed}},
  \bibinfo{author}{\bibfnamefont{S.}~\bibnamefont{Dimopoulos}},
  \bibinfo{author}{\bibfnamefont{G.}~\bibnamefont{Dvali}}, \bibnamefont{and}
  \bibinfo{author}{\bibfnamefont{J.}~\bibnamefont{March-Russell}},
  \bibinfo{journal}{Phys. Rev. D} \textbf{\bibinfo{volume}{65}},
  \bibinfo{pages}{024032} (\bibinfo{year}{2001});
\bibinfo{author}{\bibfnamefont{J.~M.} \bibnamefont{Frere}},
  \bibinfo{author}{\bibfnamefont{G.}~\bibnamefont{Moreau}}, \bibnamefont{and}
  \bibinfo{author}{\bibfnamefont{E.}~\bibnamefont{Nezri}},
  \bibinfo{journal}{Phys. Rev. D} \textbf{\bibinfo{volume}{69}},
  \bibinfo{pages}{033003} (\bibinfo{year}{2004}),
  \href{http://arXiv.org/abs/arXiv:hep-ph/0309218v1}{{
  [arXiv:hep-ph/0309218v1]}}.

\bibitem[{\citenamefont{Georgi and Glashow}(1974)}]{GUT}
\bibinfo{author}{\bibfnamefont{H.}~\bibnamefont{Georgi}} \bibnamefont{and}
  \bibinfo{author}{\bibfnamefont{S.~L.} \bibnamefont{Glashow}},
  \bibinfo{journal}{Phys. Rev. Lett.} \textbf{\bibinfo{volume}{32}},
  \bibinfo{pages}{438} (\bibinfo{year}{1974});
\bibinfo{author}{\bibfnamefont{H.}~\bibnamefont{Georgi}},
  \bibinfo{author}{\bibfnamefont{H.~R.} \bibnamefont{Quinn}}, \bibnamefont{and}
  \bibinfo{author}{\bibfnamefont{S.}~\bibnamefont{Weinberg}},
  \bibinfo{journal}{Phys. Rev. Lett.} \textbf{\bibinfo{volume}{33}},
  \bibinfo{pages}{451} (\bibinfo{year}{1974}).

\bibitem[{\citenamefont{Buras et~al.}(1978)\citenamefont{Buras, Ellis,
  Gaillard, and Nanopoulos}}]{Buras1978}
\bibinfo{author}{\bibfnamefont{A.~J.} \bibnamefont{Buras}},
  \bibinfo{author}{\bibfnamefont{J.~R.} \bibnamefont{Ellis}},
  \bibinfo{author}{\bibfnamefont{M.~K.} \bibnamefont{Gaillard}},
  \bibnamefont{and} \bibinfo{author}{\bibfnamefont{D.~V.}
  \bibnamefont{Nanopoulos}}, \bibinfo{journal}{Nucl. Phys. B}
  \textbf{\bibinfo{volume}{135}}, \bibinfo{pages}{66} (\bibinfo{year}{1978}).

\bibitem[{\citenamefont{Gell-Mann et~al.}(1979)\citenamefont{Gell-Mann, Ramond,
  and Slansky}}]{Seesaw}
\bibinfo{author}{\bibfnamefont{M.}~\bibnamefont{Gell-Mann}},
  \bibinfo{author}{\bibfnamefont{P.}~\bibnamefont{Ramond}}, \bibnamefont{and}
  \bibinfo{author}{\bibfnamefont{R.}~\bibnamefont{Slansky}}, in
  \emph{\bibinfo{booktitle}{Supergarvity: Proceedings of the Supergravity
  Workshop at Stony Brook}}, edited by
  \bibinfo{editor}{\bibfnamefont{P.}~\bibnamefont{van Niuwenhuizen}}
  \bibnamefont{and} \bibinfo{editor}{\bibfnamefont{D.~Z.}
  \bibnamefont{Freedman}} (\bibinfo{publisher}{North-Holland, Amsterdam},
  \bibinfo{year}{1979}), p. \bibinfo{pages}{315};
\bibinfo{author}{\bibfnamefont{T.}~\bibnamefont{Yanagida}}, in
  \emph{\bibinfo{booktitle}{Proceedings of the Workshop on Unified Theory and
  Baryon Number in the Universe}}, edited by
  \bibinfo{editor}{\bibfnamefont{O.}~\bibnamefont{Sawada}} \bibnamefont{and}
  \bibinfo{editor}{\bibfnamefont{A.}~\bibnamefont{Sugamoto}}
  (\bibinfo{publisher}{Tsukuba, Japan}, \bibinfo{year}{1979}),
  p.~\bibinfo{pages}{95};
\bibinfo{author}{\bibfnamefont{R.~N.}~\bibnamefont{Mohapatra}} \bibnamefont{and}
  \bibinfo{author}{\bibfnamefont{G.}~\bibnamefont{Senjanovic}},
  \bibinfo{journal}{Phys. Rev. Lett.} \textbf{\bibinfo{volume}{44}},
  \bibinfo{pages}{912} (\bibinfo{year}{1980}).

\bibitem[{\citenamefont{Hung}(2007)}]{Hung2007}
\bibinfo{author}{\bibfnamefont{P.~Q.} \bibnamefont{Hung}},
  \bibinfo{journal}{Phys. Lett. B} \textbf{\bibinfo{volume}{649}},
  \bibinfo{pages}{275} (\bibinfo{year}{2007}),
  \href{http://arXiv.org/abs/arXiv:hep-ph/0612004v4}{{
  [arXiv:hep-ph/0612004v4]}}.

\bibitem[{\citenamefont{Zubkov}(2007)}]{Zubkov2007}
\bibinfo{author}{\bibfnamefont{M.~A.} \bibnamefont{Zubkov}},
  \bibinfo{journal}{Phys. Lett. B} \textbf{\bibinfo{volume}{649}},
  \bibinfo{pages}{91} (\bibinfo{year}{2007}),
  \href{http://arXiv.org/abs/arXiv:hep-ph/0609029v4}{{
  [arXiv:hep-ph/0609029v4]}}.

\bibitem[{\citenamefont{Buras et~al.}(2004)\citenamefont{Buras, Hung, Tran,
  Poschenrieder, and Wyszomirski}}]{Buras2004}
\bibinfo{author}{\bibfnamefont{A.~J.} \bibnamefont{Buras}},
  \bibinfo{author}{\bibfnamefont{P.~Q.} \bibnamefont{Hung}},
  \bibinfo{author}{\bibfnamefont{N.~K.} \bibnamefont{Tran}},
  \bibinfo{author}{\bibfnamefont{A.}~\bibnamefont{Poschenrieder}},
  \bibnamefont{and}
  \bibinfo{author}{\bibfnamefont{E.}~\bibnamefont{Wyszomirski}},
  \bibinfo{journal}{Nucl. Phys. B} \textbf{\bibinfo{volume}{699}},
  \bibinfo{pages}{253} (\bibinfo{year}{2004}),
  \href{http://arXiv.org/abs/arXiv:hep-ph/0406048v1}{{
  [arXiv:hep-ph/0406048v1]}}.

\bibitem[{\citenamefont{Rubakov and Shaposhnikov}(1983)}]{Rubakov1983}
\bibinfo{author}{\bibfnamefont{V.~A.} \bibnamefont{Rubakov}} \bibnamefont{and}
  \bibinfo{author}{\bibfnamefont{M.~E.} \bibnamefont{Shaposhnikov}},
  \bibinfo{journal}{Phys. Lett. B} \textbf{\bibinfo{volume}{125}},
  \bibinfo{pages}{136} (\bibinfo{year}{1983}).

\bibitem[{\citenamefont{Georgi et~al.}(2001)\citenamefont{Georgi, Grant, and
  Haliu}}]{Georgi2001}
\bibinfo{author}{\bibfnamefont{H.}~\bibnamefont{Georgi}},
  \bibinfo{author}{\bibfnamefont{A.~K.} \bibnamefont{Grant}}, \bibnamefont{and}
  \bibinfo{author}{\bibfnamefont{G.}~\bibnamefont{Hailu}},
  \bibinfo{journal}{Phys. Rev. D} \textbf{\bibinfo{volume}{63}},
  \bibinfo{pages}{064027} (\bibinfo{year}{2001}),
  \href{http://arXiv.org/abs/arXiv:hep-ph/0007350v2}{{
  [arXiv:hep-ph/0007350v2]}}.

\bibitem[{\citenamefont{Dienes and Hossenfelder}(2006)}]{Split}
\bibinfo{author}{\bibfnamefont{K.~R.} \bibnamefont{Dienes}} \bibnamefont{and}
  \bibinfo{author}{\bibfnamefont{S.}~\bibnamefont{Hossenfelder}},
  \bibinfo{journal}{Phys. Rev. D} \textbf{\bibinfo{volume}{74}},
  \bibinfo{pages}{065013} (\bibinfo{year}{2006}),
  \href{http://arXiv.org/abs/arXiv:hep-ph/0607112v1}{{
  [arXiv:hep-ph/0607112v1]}};
\bibinfo{author}{\bibfnamefont{M.}~\bibnamefont{Lindner}},
  \bibinfo{author}{\bibfnamefont{M.}~\bibnamefont{Ratz}}, \bibnamefont{and}
  \bibinfo{author}{\bibfnamefont{M.~A.} \bibnamefont{Schmidt}},
  \bibinfo{journal}{J. High Energy Phys.} \textbf{\bibinfo{volume}{2005}},
  \bibinfo{pages}{081} (\bibinfo{year}{2005}),
  \href{http://arXiv.org/abs/arXiv:hep-ph/0506280v2}{{
  [arXiv:hep-ph/0506280v2]}};
\bibinfo{author}{\bibfnamefont{Z.}~\bibnamefont{Surujon}},
  \bibinfo{journal}{Phys. Rev. D} \textbf{\bibinfo{volume}{73}},
  \bibinfo{pages}{016008} (\bibinfo{year}{2006}),
  \href{http://arXiv.org/abs/arXiv:hep-ph/0507036v2}{{
  [arXiv:hep-ph/0507036v2]}};
\bibinfo{author}{\bibfnamefont{G.}~\bibnamefont{Moreau}} \bibnamefont{and}
  \bibinfo{author}{\bibfnamefont{J.~I.} \bibnamefont{Silva-Marcos}},
  \bibinfo{journal}{J. High Energy Phys.} \textbf{\bibinfo{volume}{2006}},
  \bibinfo{pages}{048} (\bibinfo{year}{2006}),
  \href{http://arXiv.org/abs/arXiv:hep-ph/0507145v1}{{
  [arXiv:hep-ph/0507145v1]}};
\bibinfo{author}{\bibfnamefont{Y.}~\bibnamefont{Grossman}},
  \bibinfo{author}{\bibfnamefont{R.}~\bibnamefont{Harnik}},
  \bibinfo{author}{\bibfnamefont{G.}~\bibnamefont{Perez}},
  \bibinfo{author}{\bibfnamefont{M.~D.} \bibnamefont{Schwartz}},
  \bibnamefont{and} \bibinfo{author}{\bibfnamefont{Z.}~\bibnamefont{Surujon}},
  \bibinfo{journal}{Phys. Rev. D} \textbf{\bibinfo{volume}{71}},
  \bibinfo{pages}{056007} (\bibinfo{year}{2005}),
  \href{http://arXiv.org/abs/arXiv:hep-ph/0507145v1}{{
  [arXiv:hep-ph/0507145v1]}};
\bibinfo{author}{\bibfnamefont{G.}~\bibnamefont{Barenboim}} \bibnamefont{and}
  \bibinfo{author}{\bibfnamefont{N.~E.} \bibnamefont{Mavromatos}},
  \bibinfo{journal}{J. High Energy Phys.} \textbf{\bibinfo{volume}{2005}},
  \bibinfo{pages}{034} (\bibinfo{year}{2005}),
  \href{http://arXiv.org/abs/arXiv:hep-ph/0404014v4}{{
  [arXiv:hep-ph/0404014v4]}};
\bibinfo{author}{\bibfnamefont{J.~A.} \bibnamefont{Aguilar-Saavedra}},
  \bibinfo{author}{\bibfnamefont{G.~C.} \bibnamefont{Branco}},
  \bibnamefont{and} \bibinfo{author}{\bibfnamefont{F.~R.}
  \bibnamefont{Joaquim}}, \bibinfo{journal}{Phys. Rev. D}
  \textbf{\bibinfo{volume}{69}}, \bibinfo{pages}{073004}
  (\bibinfo{year}{2004}),
  \href{http://arXiv.org/abs/arXiv:hep-ph/0310305v2}{{
  [arXiv:hep-ph/0310305v2]}};
\bibinfo{author}{\bibfnamefont{A.}~\bibnamefont{Coulthurst}},
  \bibinfo{author}{\bibfnamefont{K.~L.} \bibnamefont{McDonald}},
  \bibnamefont{and} \bibinfo{author}{\bibfnamefont{B.~H.~J.}
  \bibnamefont{McKellar}}, \bibinfo{journal}{Phys. Rev. D}
  \textbf{\bibinfo{volume}{75}}, \bibinfo{pages}{045018}
  (\bibinfo{year}{2007}{\natexlab{a}}),
  \href{http://arXiv.org/abs/arXiv:hep-ph/0611164v2}{{
  [arXiv:hep-ph/0611164v2]}};
\bibinfo{author}{\bibfnamefont{A.}~\bibnamefont{Coulthurst}},
  \bibinfo{author}{\bibfnamefont{J.}~\bibnamefont{Doukas}}, \bibnamefont{and}
  \bibinfo{author}{\bibfnamefont{K.~L.} \bibnamefont{McDonald}}
  (\bibinfo{year}{2007}{\natexlab{b}}),
  \href{http://arXiv.org/abs/arXiv:hep-ph/0702285v1}{{
  arXiv:hep-ph/0702285v1}}.

\bibitem[{\citenamefont{Harari et~al.}(1978)\citenamefont{Harari, Haut, and
  Weyers}}]{Dem}
\bibinfo{author}{\bibfnamefont{H.}~\bibnamefont{Harari}},
  \bibinfo{author}{\bibfnamefont{H.}~\bibnamefont{Haut}}, \bibnamefont{and}
  \bibinfo{author}{\bibfnamefont{J.}~\bibnamefont{Weyers}},
  \bibinfo{journal}{Phys. Lett. B} \textbf{\bibinfo{volume}{78}},
  \bibinfo{pages}{459} (\bibinfo{year}{1978});
\bibinfo{author}{\bibfnamefont{Y.}~\bibnamefont{Chikashige}},
  \bibinfo{author}{\bibfnamefont{G.}~\bibnamefont{Gelmini}},
  \bibinfo{author}{\bibfnamefont{R.~P.} \bibnamefont{Peccei}},
  \bibnamefont{and}
  \bibinfo{author}{\bibfnamefont{M.}~\bibnamefont{Roncadelli}},
  \bibinfo{journal}{Phys. Lett. B} \textbf{\bibinfo{volume}{94}},
  \bibinfo{pages}{499} (\bibinfo{year}{1980});
\bibinfo{author}{\bibfnamefont{H.}~\bibnamefont{Fritzsch}}, in
  \emph{\bibinfo{booktitle}{Proceedings of Europhysics Topical Conference on
  Flavor Mixing in Weak Interactions, Erice, Italy}}, edited by
  \bibinfo{editor}{\bibfnamefont{L.~L.} \bibnamefont{Chau}}
  (\bibinfo{publisher}{Plenum, New York}, \bibinfo{year}{1984}), p.
  \bibinfo{pages}{717};
\bibinfo{author}{\bibfnamefont{C.}~\bibnamefont{Jarlskog}}, in
  \emph{\bibinfo{booktitle}{Proceedings of the International Symposium on
  Production and Decay of Heavy Flavors, Heidelberg, Germany}}, edited by
  \bibinfo{editor}{\bibfnamefont{K.~R.} \bibnamefont{Schubert}}
  \bibnamefont{and} \bibinfo{editor}{\bibfnamefont{R.}~\bibnamefont{Waldi}}
  (\bibinfo{publisher}{DESY, Hamburg}, \bibinfo{year}{1986}), p.
  \bibinfo{pages}{331};
\bibinfo{author}{\bibfnamefont{P.}~\bibnamefont{Kaus}} \bibnamefont{and}
  \bibinfo{author}{\bibfnamefont{S.}~\bibnamefont{Meshkov}},
  \bibinfo{journal}{Mod. Phys. Lett. A} \textbf{\bibinfo{volume}{3}},
  \bibinfo{pages}{1251} (\bibinfo{year}{1988});
\bibinfo{author}{\bibfnamefont{Y.}~\bibnamefont{Koide}},
  \bibinfo{journal}{Phys. Rev. D} \textbf{\bibinfo{volume}{39}},
  \bibinfo{pages}{1391} (\bibinfo{year}{1989}).

\bibitem[{\citenamefont{Fritzsch and Xing}(2000)}]{Fritzsch2000}
\bibinfo{author}{\bibfnamefont{H.}~\bibnamefont{Fritzsch}} \bibnamefont{and}
  \bibinfo{author}{\bibfnamefont{Z.~Z.} \bibnamefont{Xing}},
  \bibinfo{journal}{Prog. Part. Nucl. Phys.} \textbf{\bibinfo{volume}{45}},
  \bibinfo{pages}{1} (\bibinfo{year}{2000}),
  \href{http://arXiv.org/abs/arXiv:hep-ph/9912358v2}{{
  [arXiv:hep-ph/9912358v2]}}.

\bibitem[{\citenamefont{{Yao} et~al.}(2006)\citenamefont{{Yao}, {Amsler},
  {Asner}, {Barnett}, {Beringer}, {Burchat}, {Carone}, {Caso}, {Dahl},
  {D'Ambrosio} et~al.}}]{PDBook}
\bibinfo{author}{\bibfnamefont{W.-M.} \bibnamefont{{Yao}}},
  \bibinfo{author}{\bibfnamefont{C.}~\bibnamefont{{Amsler}}},
  \bibinfo{author}{\bibfnamefont{D.}~\bibnamefont{{Asner}}},
  \bibinfo{author}{\bibfnamefont{R.}~\bibnamefont{{Barnett}}},
  \bibinfo{author}{\bibfnamefont{J.}~\bibnamefont{{Beringer}}},
  \bibinfo{author}{\bibfnamefont{P.}~\bibnamefont{{Burchat}}},
  \bibinfo{author}{\bibfnamefont{C.}~\bibnamefont{{Carone}}},
  \bibinfo{author}{\bibfnamefont{C.}~\bibnamefont{{Caso}}},
  \bibinfo{author}{\bibfnamefont{O.}~\bibnamefont{{Dahl}}},
  \bibinfo{author}{\bibfnamefont{G.}~\bibnamefont{{D'Ambrosio}}},
  \bibnamefont{et~al.}, \bibinfo{journal}{{Journal of Physics G}}
  \textbf{\bibinfo{volume}{33}}, \bibinfo{pages}{1+} (\bibinfo{year}{2006}),
  \urlprefix\url{http://pdg.lbl.gov}.

\bibitem[{\citenamefont{Branco et~al.}(1990)\citenamefont{Branco, Silva-Marcos,
  and Rebelo}}]{NonDem}
\bibinfo{author}{\bibfnamefont{G.~C.} \bibnamefont{Branco}},
  \bibinfo{author}{\bibfnamefont{J.~I.} \bibnamefont{Silva-Marcos}},
  \bibnamefont{and} \bibinfo{author}{\bibfnamefont{M.~N.}
  \bibnamefont{Rebelo}}, \bibinfo{journal}{Phys. Lett. B}
  \textbf{\bibinfo{volume}{237}}, \bibinfo{pages}{446} (\bibinfo{year}{1990});
\bibinfo{author}{\bibfnamefont{H.}~\bibnamefont{Fritzsch}} \bibnamefont{and}
  \bibinfo{author}{\bibfnamefont{J.}~\bibnamefont{Plankl}},
  \bibinfo{journal}{Phys. Lett. B} \textbf{\bibinfo{volume}{237}},
  \bibinfo{pages}{451} (\bibinfo{year}{1990});
\bibinfo{author}{\bibfnamefont{H.}~\bibnamefont{Fritzsch}},
  \bibinfo{journal}{Phys. Lett. B} \textbf{\bibinfo{volume}{289}},
  \bibinfo{pages}{92} (\bibinfo{year}{1992}).

\bibitem[{\citenamefont{Hung and Seco}(2003)}]{HungSeco}
\bibinfo{author}{\bibfnamefont{P.~Q.} \bibnamefont{Hung}} \bibnamefont{and}
  \bibinfo{author}{\bibfnamefont{M.}~\bibnamefont{Seco}},
  \bibinfo{journal}{Nucl. Phys. B} \textbf{\bibinfo{volume}{653}},
  \bibinfo{pages}{123} (\bibinfo{year}{2003}),
  \href{http://arXiv.org/abs/arXiv:hep-ph/0111013v8}{{
  [arXiv:hep-ph/0111013v8]}};
\bibinfo{author}{\bibfnamefont{P.~Q.} \bibnamefont{Hung}},
  \bibinfo{author}{\bibfnamefont{M.}~\bibnamefont{Seco}}, \bibnamefont{and}
  \bibinfo{author}{\bibfnamefont{A.}~\bibnamefont{Soddu}},
  \bibinfo{journal}{Nucl. Phys. B} \textbf{\bibinfo{volume}{692}},
  \bibinfo{pages}{83} (\bibinfo{year}{2004}),
  \href{http://arXiv.org/abs/arXiv:hep-ph/0311198v1}{{
  [arXiv:hep-ph/0311198v1]}}.

\bibitem[{\citenamefont{Pontecorvo}(1967)}]{PMNS}
\bibinfo{author}{\bibfnamefont{B.}~\bibnamefont{Pontecorvo}},
  \bibinfo{journal}{Zh. Eksp. Teor. Fiz.} \textbf{\bibinfo{volume}{53}},
  \bibinfo{pages}{1717} (\bibinfo{year}{1967}) [Sov. Phys. JETP \textbf{26}, 984
(1968)];
\bibinfo{author}{\bibfnamefont{Z.}~\bibnamefont{Maki}},
  \bibinfo{author}{\bibfnamefont{M.}~\bibnamefont{Nakagawa}}, \bibnamefont{and}
  \bibinfo{author}{\bibfnamefont{S.}~\bibnamefont{Sakata}},
  \bibinfo{journal}{Prog. Theor. Phys.} \textbf{\bibinfo{volume}{28}},
  \bibinfo{pages}{870} (\bibinfo{year}{1962}).

\bibitem[{\citenamefont{Peskin and Takeuchi}(1990)}]{Peskin}
\bibinfo{author}{\bibfnamefont{M.~E.} \bibnamefont{Peskin}} \bibnamefont{and}
  \bibinfo{author}{\bibfnamefont{T.}~\bibnamefont{Takeuchi}},
  \bibinfo{journal}{Phys. Rev. Lett.} \textbf{\bibinfo{volume}{65}},
  \bibinfo{pages}{964} (\bibinfo{year}{1990});
\bibinfo{author}{\bibfnamefont{M.~E.} \bibnamefont{Peskin}} \bibnamefont{and}
  \bibinfo{author}{\bibfnamefont{T.}~\bibnamefont{Takeuchi}},
  \bibinfo{journal}{Phys. Rev. D} \textbf{\bibinfo{volume}{46}},
  \bibinfo{pages}{381} (\bibinfo{year}{1992}).

\bibitem[{\citenamefont{Georgi}(1991)}]{Georgi1991}
\bibinfo{author}{\bibfnamefont{H.}~\bibnamefont{Georgi}},
  \bibinfo{journal}{Nucl. Phys. B} \textbf{\bibinfo{volume}{363}},
  \bibinfo{pages}{301} (\bibinfo{year}{1991}).

\bibitem[{\citenamefont{Dugan and Randall}(1991)}]{Dugan1991}
\bibinfo{author}{\bibfnamefont{M.~J.} \bibnamefont{Dugan}} \bibnamefont{and}
  \bibinfo{author}{\bibfnamefont{L.}~\bibnamefont{Randall}},
  \bibinfo{journal}{Phys. Lett. B} \textbf{\bibinfo{volume}{264}},
  \bibinfo{pages}{154} (\bibinfo{year}{1991}).

\bibitem[{\citenamefont{Frampton et~al.}(2000)\citenamefont{Frampton, Hung, and
  Sher}}]{Frampton2000}
\bibinfo{author}{\bibfnamefont{P.~H.} \bibnamefont{Frampton}},
  \bibinfo{author}{\bibfnamefont{P.~Q.} \bibnamefont{Hung}}, \bibnamefont{and}
  \bibinfo{author}{\bibfnamefont{M.}~\bibnamefont{Sher}},
  \bibinfo{journal}{Phys. Rep.} \textbf{\bibinfo{volume}{330}},
  \bibinfo{pages}{263} (\bibinfo{year}{2000}),
  \href{http://arXiv.org/abs/arXiv:hep-ph/9903387v2}{{
  [arXiv:hep-ph/9903387v2]}}.

\bibitem[{\citenamefont{He et~al.}(2001)\citenamefont{He, Polansky, and
  Su}}]{He2001}
\bibinfo{author}{\bibfnamefont{H.-J.} \bibnamefont{He}},
  \bibinfo{author}{\bibfnamefont{N.}~\bibnamefont{Polonsky}}, \bibnamefont{and}
  \bibinfo{author}{\bibfnamefont{S.}~\bibnamefont{Su}}, \bibinfo{journal}{Phys.
  Rev. D} \textbf{\bibinfo{volume}{64}}, \bibinfo{pages}{053004}
  (\bibinfo{year}{2001}).

\end{thebibliography}
\end{document}